\newif\ifdraft\draftfalse 
\newif\ifsoon\soontrue    
\newif\iffull\fullfalse   
\newif\ifsmall\smalltrue  
\newif\ifbackref\backreffalse 
\newif\ifcamera\cameratrue 
\newif\ifanon\anonfalse     
\newif\iflater\laterfalse 
\newif\ifdesperateforspace\desperateforspacefalse

\makeatletter \@input{texdirectives} \makeatother

\documentclass[\ifcamera\else preprint,nocopyrightspace\fi]{sigplanconf}

\usepackage{preamble}

\begin{document}

\toappear{}


\title{\ifcamera\else\Huge\fi Beginner's Luck}
\subtitle{\ifcamera\else\LARGE\fi A Language for Property-Based Generators
\ifanon\vspace*{-1.5cm}\fi}

\ifanon
\authorinfo{}{}{}
\else
\authorinfo
{
  Leonidas Lampropoulos\textsuperscript{1} \qquad
  Diane Gallois-Wong\textsuperscript{2,3} \qquad
  C\u{a}t\u{a}lin Hri\c{t}cu\textsuperscript{2} \\[0.5em]
  John Hughes\textsuperscript{4} \qquad
  Benjamin C. Pierce\textsuperscript{1} \qquad
  Li-yao Xia\textsuperscript{2,3} \qquad
\\[1em]
}{
  \textsuperscript{1}University of Pennsylvania\ifcamera, USA \fi\qquad
  \textsuperscript{2}INRIA Paris\ifcamera, France\fi\qquad
  \textsuperscript{3}ENS Paris\ifcamera, France\fi\qquad
  \textsuperscript{4}Chalmers University\ifcamera, Sweden\fi
\vspace*{-1cm}
}{}
\fi

\maketitle








\begin{abstract}
Property-based random testing {\em \`a la} QuickCheck requires building
efficient generators for well-distributed random data satisfying complex logical
predicates, but writing these generators can be difficult and error prone.  We
propose 
a domain-specific language in which generators are conveniently
expressed by decorating predicates with lightweight annotations to control both
the distribution of generated values and the amount of constraint solving that
happens before each variable is instantiated.  This language, called {\em Luck},
makes generators easier to write, read, and maintain.

We give Luck a formal semantics and prove several fundamental
properties, including the soundness and completeness of random
generation with respect to a standard predicate semantics.  We
evaluate Luck on common examples from the property-based testing
literature and on two significant case studies, showing that it can be
used in complex domains with comparable bug-finding effectiveness and
a significant reduction in testing code size compared to handwritten
generators.
\end{abstract}

\ifcamera

\sloppy

\leo{Any others?}

\fussy
\fi

\section{Introduction}
\label{sec:intro}

Since being popularized by QuickCheck~\cite{ClaessenH00}, property-based
random testing has become a standard technique for improving software
quality in a wide variety of programming languages~\cite{Arts2008,
  Lindblad07, Hughes07, Pacheco:2007} and for streamlining interaction with
proof assistants~\cite{ChamarthiDKM11, Bulwahn12, OwreAFP2006,
  DybjerHT03, itp2015}.

When using a property-based random testing tool, one writes {\em
  properties} in the form of executable predicates. For example, a
natural property to test for a list \lk{reverse} function is that, for any
list 
\lk{xs}, reversing \lk{xs} twice yields \lk{xs} again.
In QuickCheck notation:
\begin{verbatim}
  prop_reverse xs  =  (reverse (reverse xs) == xs)
\end{verbatim}
To test this property, QuickCheck generates random lists until either it finds a
counterexample or a \iffull predetermined \else certain \fi number of tests succeed.

An appealing feature of QuickCheck is that it offers a library of property
combinators resembling standard logical operators.  For example, a property
of the form \lk{p ==> q}, built using the implication combinator
\lk{==>}, will be tested automatically by generating {\em
  valuations} (assignments of random values, of appropriate type, to the
free variables of \lk{p} and \lk{q}), discarding those valuations that
fail to satisfy \lk{p}, and checking whether any of the ones that remain
are counterexamples to \lk{q}.

QuickCheck users soon learn that this default gene\-rate-and-test
approach sometimes does not give satisfactory results.  In particular, if the
precondition \lk{p} is satisfied by relatively few values of the appropriate
type, then most of the random inputs that QuickCheck generates will be
discarded, so that \lk{q} will seldom be exercised.  Consider, for example,
testing a simple property of a school database system: that every student in a
list of \lk{registeredStudents} should be taking at least one
course,
\begin{verbatim}
  prop_registered studentId =
    member studentId registeredStudents ==>
    countCourses studentId > 0
\end{verbatim}
where, as usual:
\begin{verbatim}
  member x [] = False
  member x (h:t) = (x == h) || member x t
\end{verbatim}
If the space of possible student ids is large (e.g., because they are
represented as 
machine integers), then a randomly generated id is very unlikely to be a
member of 
\lk{registeredStudents}, so almost all test cases will be discarded.

To enable effective testing in such cases, the QuickCheck user can provide a
{\em property-based generator} for inputs satisfying \lk{p}---here, a generator that
always returns student ids drawn from the members of {\tt registeredStudents}.
Indeed, QuickCheck provides a library of combinators for defining such
generators.  These combinators also allow fine control over the {\em
distribution} of generated values---a crucial feature in
practice~\cite{ClaessenH00, TestingNI, swarm-testing}.

Property-based generators generators work well for small to medium-sized examples,
but writing them can become challenging as \lk{p} gets more
complex---sometimes turning into a research contribution in
its own right! For example, papers have been written about \iffull random \fi
generation techniques for well-typed
lambda-terms \cite{PalkaAST11,Yakushev10, FetscherCPHF15,
Tarau15} and for ``indistinguishable'' machine states that can be used for
finding 
bugs in information-flow monitors~\cite{TestingNI, testing_ni_jfp}.
Moreover, if we use QuickCheck to test an {\em invariant} property
(\EG type preservation), then the same condition will appear in both the
precondition and the conclusion of the property, requiring that we
express this condition both as a boolean predicate \lk{p} and as a generator
whose outputs all 
satisfy \lk{p}. These two artifacts must then be kept in sync,
which can become both a maintenance issue and a rich source of
confusion in the testing process. These difficulties are not hypothetical: Hri\c{t}cu \ETAL's
machine-state generator~\cite{TestingNI} is over 1500 lines of tricky
Haskell, while Pa{\l}ka \ETAL's generator for well-typed
lambda-terms~\cite{PalkaAST11} is over 1600 even trickier ones.  To enable
effective property-based random testing of complex software artifacts,
we need a better way of writing predicates and corresponding
generators.

A natural idea is to derive an efficient generator for a given predicate
\lk{p} directly from \lk{p} itself.  Indeed, two variants of this idea,
with complementary strengths and weaknesses, have been explored by
others---one based on local choices 
and backtracking, one on general constraint solving. 
Our language, Luck, synergistically combines these two approaches.

The first approach can be thought of as a kind of incremental
generate-and-test: rather than generating completely random valuations and
then testing them against \lk{p}, we instead walk over the structure of
\lk{p} and instantiate each unknown variable \lk{x} at the first point
where we meet a constraint involving \lk{x}.
In the \lk{member} example above, on each recursive call, we
make a random choice between the branches of the $\LuckOr{}{}$.  If we choose
the left, we instantiate \lk{x} to the head of the list; otherwise we
leave \lk{x} unknown and continue with the recursive call to \lk{member} on the
tail.  This has the effect of traversing the list of
registered students and picking one of its elements.
%
%
This process resembles {\em narrowing} from functional logic
programming~\cite{Antoy94aneeded, Hanus97, Lindblad07, TolmachA03}.  It is
attractively lightweight, admits natural control over distributions (as we will
see in the next section), and has been used successfully~\cite{FischerK07,
ChristiansenF08, ReichNR11,GligoricGJKKM10}, even in challenging domains such as
generating well-typed programs to test compilers~\cite{ClaessenFLOPS14,
FetscherCPHF15}.

However, choosing a value for an unknown when we encounter the {\em first}
constraint on it risks making choices that do not
satisfy {\em later} constraints, forcing us to backtrack and make a
different choice when the problem is discovered. For example,
consider the \lk{notMember} predicate:
\begin{verbatim}
  notMember x []     =  True
  notMember x (h:t)  =  (x /= h) && notMember x t
\end{verbatim}
Suppose we wish to generate values for \lk{x} such that \lk{notMember x ys} for
some predetermined list \lk{ys}.  When we first encounter the constraint \lk{x /=
h}, we generate a value for \lk{x} that is not equal to the known value \lk{h}.
We then proceed to the recursive call of \lk{notMember}, where we {\em check}
that the chosen \lk{x} does not appear in the rest of the list.  Since the
values in the rest of the list are not taken into account when choosing \lk{x},
this may force us to backtrack if our choice of \lk{x} was unlucky.  If the
space of possible values for \lk{x} is not much bigger than \iffull  the length
of \fi \lk{ys}---say, just \iffull twice as big\else $2\times$\fi---\iffull
then \fi we will backtrack 50\% of the time.
Worse yet, if \lk{notMember} is used to define another
predicate---e.g., \lk{distinct}, which tests whether each element of an input
list is different from all the others---and we want to generate 
a list satisfying
\lk{distinct}, then \lk{notMember}'s 50\% chance of backtracking
will be compounded on each recursive call of \lk{distinct}, leading to
unacceptably low rates of successful generation.

The second existing approach uses a {\em constraint solver}
to generate a diverse set of valuations satisfying a
predicate.%
\footnote{Constraint solvers
can, of course, be used to {\em directly} search for
counterexamples to a property of interest
by software model checking~\cite[etc.]{BlanchetteN10, AlloyBook,
  BallLR11, JhalaM09}.
We are interested here in the rather different task of
quickly generating a large number of diverse inputs, so that we can
thoroughly test systems like compilers whose state spaces are too
large to be exhaustively explored.  }
This approach has been widely investigated, both for generating inputs directly
from predicates~\cite{CarlierDG10, SeidelVJ15, GotliebICST09, KoksalKS11} and
for symbolic-execution-based testing~\cite{DART:2005, CUTE:2005, KLEE:2008:OSDI,
AvgerinosRCB14, TorlakB14}, which additionally uses the system under test to
guide generation of inputs that exercise different control-flow paths. For
\lk{notMember}, gathering a set of disequality constraints
on \lk{x} before choosing its value avoids any backtracking.

However, {\em pure} constraint-solving approaches do not give us everything we
need. They do not provide effective control over the distribution
of generated valuations.  At best, they might guarantee a {\em uniform} (or near
uniform) distribution~\cite{Chakraborty2014}, but this is typically not the
distribution we want in practice (see~\autoref{sec:examples}).
Moreover, the overhead of maintaining and solving constraints can make these
approaches significantly less efficient than the more lightweight, local approach
of needed narrowing when the latter does not lead to backtracking, as
for instance in \lk{member}.

The complementary strengths and weaknesses of local instantiation and global
constraint solving suggest a hybrid approach, where limited constraint
propagation, under explicit user control, is used to refine the domains (sets of
possible values) of unknowns before instantiation. Exploring this approach 
is the goal of this paper.
Our main contributions are:

\begin{itemize}
\item We propose a new domain-specific language, \Luck, for writing
generators via lightweight annotations on predicates, combining the
strengths of the local-instantiation and constraint-solving
approaches to generation. Section \autoref{sec:examples} illustrates Luck's novel
features using binary search trees as an example.

\item To place Luck's design on a firm formal foundation, we define a
core calculus and establish key properties, including the soundness and
completeness of its probabilistic generator semantics with respect to a
straightforward interpretation of expressions as predicates
(\autoref{sec:semantics}).

\item We provide a prototype interpreter (\autoref{sec:source}) including a
simple implementation of the constraint-solving primitives used by the
generator semantics. %
We do not use an off-the shelf constraint solver because we want to experiment
with a per-variable uniform sampling approach (\autoref{sec:examples}) which
is not 
supported by modern solvers.  In addition, using such a solver would require
translating Luck expressions---datatypes, pattern matching, etc.---into a form
that it can handle. We leave this for future work.


\item We evaluate Luck's expressiveness on a collection of common examples
from the random testing literature (\autoref{sec:casestudies}) and on
two significant case studies; the latter demonstrate that \Luck can be used (1) to
find bugs 
in a widely used compiler (GHC) by randomly generating well-typed
lambda terms and (2) to help design information-flow abstract machines by
generating ``low-indistinguishable'' machine states.  Compared to hand-written
generators, these experiments show comparable bug-finding
effectiveness (measured in test cases generated per counterexample
found) and a significant reduction in the size of testing
code.
The interpreted Luck generators run an order of magnitude slower than compiled
QuickCheck versions ($8$ to $24$ times per test), but many
opportunities for optimization remain.
\end{itemize}
Sections \autoref{sec:relwork} and~\autoref{sec:concl} discuss related work and
future directions.
This paper is accompanied by several auxiliary materials:
(1) a Coq formalization of the narrowing semantics of Luck and machine-checked
    proofs of its properties 
    (available at \url{https://github.com/QuickChick/Luck}) (\autoref{sec:narrow});
(2) the prototype Luck interpreter and a battery of example programs, including
    all the ones we used for evaluation (also
    at \url{https://github.com/QuickChick/Luck}) (\autoref{sec:casestudies});
(3) an extended version of the paper with full definitions and paper proofs
  for the whole semantics (\url{https://arxiv.org/abs/1607.05443}).

\section{\Luck by Example}
\label{sec:examples}

\autoref{fig:bst} shows a recursive Haskell predicate \lk{bst} that checks
whether a given tree with labels strictly between \lk{low}
and \lk{high} satisfies the standard binary-search tree (BST)
invariant~\cite{Okasaki99}.  It is followed by a QuickCheck
generator \lk{genTree}, which generates BSTs with a given maximum
depth, controlled by the \lk{size} parameter. 
This generator first
checks whether \lk{low + 1 >= high}, in which case it returns the only
valid BST satisfying this constraint---the \lk{Empty} one.
Otherwise, it uses QuickCheck's \lk{frequency} combinator, which takes
a list of pairs of positive integer weights and associated
generators and randomly selects one of the generators using the
probabilities specified by the weights. In this example,
$\frac{1}{size + 1}$ of the time it creates an \lk{Empty} tree,
while $\frac{size}{size + 1}$ of the time it returns
a \lk{Node}. The \lk{Node} generator is specified using monadic
syntax: first it generates an integer \lk{x} that is strictly
between \lk{low} and \lk{high}, and then the left and right
subtrees \lk{l} and \lk{r} by calling \lk{genTree} recursively; 
finally it returns \lk{Node x l r}.

\begin{figure}
\begin{small}

{\it Binary tree datatype (in both Haskell and Luck):}
\begin{verbatim}
  data Tree a = Empty | Node a (Tree a) (Tree a)
\end{verbatim}

\vspace{.7ex}

{\it Test predicate for BSTs (in Haskell):}
\begin{verbatim}
  bst :: Int -> Int -> Tree Int -> Bool
  bst low high tree =
    case tree of
      Empty -> True
      Node x l r -> 
        low < x && x < high
        && bst low x l && bst x high r
\end{verbatim}

\vspace{.7ex}

{\it QuickCheck generator for BSTs (in Haskell):}
\begin{verbatim}
  genTree :: Int -> Int -> Int -> Gen (Tree Int)
  genTree size low high
    | low + 1 >= high = return Empty
    | otherwise =
        frequency [(1, return Empty),
                   (size, do
                     x <- choose (low + 1, high - 1)
                     l <- genTree (size `div` 2) low x
                     r <- genTree (size `div` 2) x high
                     return (Node x l r))]
\end{verbatim}

\vspace{.7ex}

{\it Luck generator (and predicate) for BSTs:}
\begin{verbatim}
  sig bst :: Int -> Int -> Int -> Tree Int -> Bool
  fun bst size low high tree =
    if size == 0 then tree == Empty
    else case tree of
         | 1    % Empty -> True
         | size % Node x l r ->
            ((low < x && x < high) !x)
            && bst (size / 2) low x l
            && bst (size / 2) x high r
\end{verbatim}
\end{small}
\caption{Binary Search Tree tester and two generators}
\label{fig:bst}
\end{figure}

The generator for BSTs allows us to efficiently test conditional properties
of the form ``if \lk{bst t} then {\em $\langle$some other property of}
\lk{t}$\rangle$,'' but it raises some new issues of its own.  First, even for
this simple example, getting the generator right is a bit tricky (for
instance because of potential off-by-one errors in generating \lk{x}), and
it is not immediately obvious that the set of trees 
generated by the generator is exactly the set accepted by the predicate.
Worse, we now need to maintain two similar but distinct artifacts and
keep them in sync.\iffull{} (We can't just throw away the predicate and keep the
generator because we often need them both, for example to test properties
like ``the \lk{insert} function applied to a BST and a value returns a
BST.'')\leo{Not sure this parenthetical is helpful.}\ch{I find it quite helpful}\fi{}  As predicates and generators become more complex, these issues
can become quite problematic (e.g.,~\cite{TestingNI}).  

Enter Luck.
The bottom of \autoref{fig:bst} shows a Luck program that represents {\em
  both} a BST predicate {\em and} a generator for random BSTs.
Modulo variations in concrete syntax, the Luck code follows the Haskell {\tt
bst} predicate quite closely.  The significant differences are: (1) the {\em
sample-after expression} \lk{!x}, which controls when node labels are generated,
and (2) the \lk{size} parameter, which is used, as in the QuickCheck generator,
to annotate the branches of the \lk{case} with relative weights.
Together, these enable us to give the program both a natural interpretation as a
predicate (by simply ignoring weights and sampling expressions) and an efficient
interpretation as a generator of random trees with the same distribution as the
QuickCheck version.  For example, evaluating the top-level query \lk{bst 10 0 42
u = True}---i.e., ``generate values \lk{t} for the {unknown} \lk{u} such
that \lk{bst 10 0 42 t} evaluates to \lk{True}''---will yield random binary
search trees of size up to 10 with node labels strictly between 0 and 42, with
the same distribution as the QuickCheck generator \lk{genTree 10 0 42}.


An {\em unknown} in Luck is a special kind of value, similar to logic variables
found in logic programming languages and unification variables used by
type-inference algorithms.  Unknowns are typed, and each is associated with a
domain of possible values from its type. Given an expression $e$ mentioning some
set $U$ of unknowns, our goal is to generate {\em valuations} over these
unknowns (maps from $U$ to concrete values) by iteratively refining the
unknowns' domains, so that, when any of these valuations is substituted into
$e$, the resulting concrete term evaluates to a desired value (e.g., \lk{True}).

Unknowns can be introduced both explicitly, as in the
top-level query above (see also \autoref{sec-toplevel}), and implicitly, as in
the generator semantics of \lk{case} expressions.
In the \lk{bst} example, when the \lk{Node} branch is chosen, the
pattern variables \lk{x}, \lk{l}, and \lk{r} are replaced by fresh unknowns,
which are then instantiated by evaluating the constraint \lk{low < x \&\& x <
high} and the recursive calls to \lk{bst}.

Varying the placement of unknowns in the top-level \lk{bst} query yields
different behaviors. For instance, if we change the query to \lk{bst
10 ul uh u = True}, replacing the \lk{low} and \lk{high} parameters with
unknowns \lk{ul} and \lk{uh}, the domains of these unknowns will be refined
during tree generation and the result will be a generator for random valuations
$(\lk{ul}\mapsto\lk{i},\, \lk{uh}\mapsto\lk{j},\, \lk{u}\mapsto\lk{t})$
where \lk{i} and \lk{j} are lower and upper bounds on the node labels in
\lk{t}.

Alternatively, we can evaluate the top-level query \lk{bst 10 0 42 t = True},
replacing \lk{u} with a concrete tree \lk{t}.  In this case, Luck will return a
trivial valuation only if \lk{t} is a binary search tree; otherwise it will
report that the query is unsatisfiable.  A less useful possibility is that we
provide explicit values for \lk{low} and
\lk{high} but choose them with $\lk{low} > \lk{high}$, e.g., \lk{bst 10 6 4
  u = True}. Since there are no satisfying valuations for \lk{u} other than
\lk{Empty}, Luck will now generate  only \lk{Empty} trees.
%

A {\em sample-after expression} of the form \lk{e !x} is used to control
instantiation of 
unknowns. Typically, \lk{x} will be an unknown \lk{u}, and evaluating \lk{e !u}
will cause \lk{u} to be instantiated to a concrete value (after
evaluating \lk{e} to refine the domains of all of the unknowns
in \lk{e}).  If \lk{x} reduces to a value rather than an unknown, we
similarly instantiate any unknowns appearing within this value.

As a concrete example, consider the compound inequality constraint \lk{0 < x \&\& x <
  4}.  A generator based on pure narrowing (as in~\cite{GligoricGJKKM10}),
would instantiate \lk{x} when the evaluator meets the first constraint where
it appears, namely \lk{0 < x} (assuming left-to-right evaluation order).  We can
mimic this behavior in Luck by writing \lk{((0 < x) !x) \&\& (x <
  4)}. However, picking a value for \lk{x} at this point ignores the
constraint \lk{x < 4}, which can lead to backtracking. If, for instance, the
domain from which we are choosing values for \lk{x} is 32-bit integers, then
the probability that a random choice satisfying \lk{0 < x} will also satisfy
\lk{x < 4} is minuscule.  It is better in this case to write
\lk{(0 < x \&\& x < 4) !x}, instantiating \lk{x} after the entire
conjunction has been evaluated and all the constraints on the domain of
\lk{x} recorded and thus avoiding backtracking completely. Finally, if we do
not include a sample-after expression for \lk{x} here at all, we can further
refine its domain with constraints later on, at the cost of dealing with a more
abstract representation of it internally in the meantime.  Thus, sample-after
expressions give Luck users explicit control over the tradeoff between the
expense of possible backtracking---when unknowns are instantiated early---and
the expense of maintaining constraints on unknowns---so that they can be
instantiated late (\EG so that \lk{x} can be instantiated after the
recursive calls to \lk{bst}).

Sample-after expressions choose random values with {\em uniform} probability
from the domain associated with each unknown.
While this behavior is sometimes useful,
effective property-based
  random testing often requires fine control over the distribution of
  generated test cases.  
%
Drawing inspiration from the QuickCheck combinator library for building
complex generators, and particularly
\lk{frequency} (which we saw in \lk{genTree} (\autoref{fig:bst})),
Luck also allows weight annotations on the branches of a {\tt case} expression
which have a \lk{frequency}-like effect. In the Luck version
of \lk{bst}, for example, the unknown \lk{tree} is either instantiated to
an \lk{Empty} tree $\frac{1}{1+size}$ of the time or partially instantiated to
a \lk{Node} (with fresh unknowns for \lk{x} and the left and right subtrees)
$\frac{size}{1+size}$ of the time.

Weight annotations give the user control over the probabilities of local
choices.  These do not necessarily correspond to a specific posterior
probability, but the QuickCheck community has established techniques for
guiding the user in tuning local weights to obtain good testing.  For
example, the user can wrap properties inside a \lk{collect x} combinator;
during testing, QuickCheck will gather information on \lk{x}, grouping equal
values to provide an estimate of the posterior distribution that is being
sampled.  The \lk{collect} combinator is an effective tool for
adjusting \lk{frequency} weights and dramatically increasing
bug-finding rates (\EG~\cite{TestingNI}). The Luck implementation
provides a similar primitive.

One further remark on uniform sampling:
While {\em locally} instantiating unknowns uniformly from their domain is a
useful default, generating {\em globally} uniform distributions of test cases is
usually not what we want, as this often leads to inefficient testing in
practice.  A simple example comes from the information flow control experiments
of \Hritcu~\ETAL~\cite{TestingNI}.  There are two ``security levels,'' called
{\em labels}, \lk{Low} and {High}, and pairs of integers and labels are
considered ``indistinguishable'' to a \lk{Low} observer if the labels are equal
and, if the labels are \lk{Low}, so are the integers. In Haskell:
\begin{verbatim}
  indist (v1,High) (v2,High)  =  True
  indist (v1,Low ) (v2,Low)   =  v1 == v2 
  indist _          _         =  False
\end{verbatim}
If we use 32-bit integers, then for
every \lk{Low} indistinguishable pair there are $2^{32}$ \lk{High} ones!
Thus, choosing a uniform distribution over indistinguishable pairs means
that we will essentially never generate pairs with \lk{Low} labels.
\iffull
Clearly, such a distribution cannot provide effective testing; indeed, 
\Hritcu~\ETAL{} found that the best distribution was actually somewhat
skewed in favor of \lk{Low} labels.  
\fi

However, in other areas where random sampling is used, it is sometimes important
to be able to generate globally uniform distributions; if desired, this effect
can be achieved in Luck by emulating {\em Boltzmann
samplers}~\cite{DuchonFLS04}.  This technique fits naturally in Luck, providing
an efficient way of drawing samples from combinatorial structures of approximate
size \lk{n}---in time linear in \lk{n}---where any two objects with the same
size have an equal probability of being generated.
\iffull Details can be found in Section \ref{sec:relwork}.\else Details can be found in the
extended version. \fi

\section{Semantics of Core Luck}
\label{sec:semantics}

We next present a core calculus for Luck---a minimal subset into which the
examples in the previous section can in principle be desugared (though our
implementation does not do this).  The core  omits primitive booleans
and integers and replaces datatypes with binary sums, products, and
iso-recursive types.

We begin in \autoref{sec:core} with the syntax and standard {\em predicate
semantics} of the core. (We call it the ``predicate'' semantics because, in our
examples, the result of evaluating a top-level expression will typically be a
boolean, though this expectation is not baked into the formalism.)  We then
build up to the full generator semantics in three steps.  
First, we give an interface to a {\em constraint solver} (\autoref{sec:constr}),
abstracting over the primitives required to implement our semantics.
Then we define a probabilistic {\em narrowing semantics}, which enhances the
local-instantiation approach to random generation with QuickCheck-style
distribution control (\autoref{sec:narrow}).
Finally, we introduce a {\em matching semantics}, building on the narrowing
semantics, that unifies constraint solving and narrowing into a single evaluator
(\autoref{sec:gen}).
\iffull We \else In the long version, we \fi also show how integers and
booleans can be encoded and how the semantics applies to the binary search
tree example\iffull (\autoref{sec:example-application})\fi.
The key properties of the generator semantics (both narrowing and matching
versions) are soundness and completeness with respect to the predicate semantics
(\autoref{sec:prop}); informally, whenever we use a Luck program to generate a
valuation that satisfies some predicate, the valuation will satisfy the boolean
predicate semantics (soundness), and it will generate {every} possible
satisfying valuation with non-zero probability (completeness).

\subsection{Syntax, Typing, and Predicate Semantics}
\label{sec:core}

The syntax of Core Luck is given in \autoref{fig:CoreLuck}.  Except for
the last line in the definitions of values and expressions, it is a standard
simply typed call-by-value lambda calculus with sums, products, and iso-recursive types. We 
include recursive lambdas for convenience in examples, although in principle
they could  
be encoded using recursive types.

\begin{figure}
\begin{grammar}
  v ::= () | (v, v) | \inl{T}{v} | \inr{T}{v}
     | \RecT{f}{x}{e}{T_1}{T_2} | \fold{T}{v}
     | u

  e ::= x | () | \ErecT{f}{x}{e}{T_1}{T_2} | (e ~ e)
     | (e, e) | \EcaseofP{e}{x}{y}{e}
     | \Einl{T}{e} | \Einr{T}{e}  | \Ecaseof{e}{x}{e}{x}{e}
     | \Efold{T}{e} | \Eunfold{T}{e}
     | u | {e \leftarrow (e, e)} | ~!e | \Etil{e}{e}

  \Tnf ::= X | 1 | \Tnf + \Tnf | \Tnf \times \Tnf| \mu X. ~ \Tnf

  T ::= X | 1 | T + T | T \times T | \mu X. ~ T | T \rightarrow T

  \Gamma ::= \emptyset | \Gamma, x : T
\end{grammar}
\caption{Core Luck Syntax}
\label{fig:CoreLuck}
\end{figure}

Values include unit, pairs of values, sum constructors ($\iinl$ and
$\iinr$) applied to values (and annotated with types, to eliminate ambiguity),
first class \iffull(potentially){}\fi recursive functions (\ii{rec}),
\ii{fold}-annotated values indicating where an
iso-recursive type should be ``folded,'' and {\em unknowns} drawn from an
infinite set.  The standard expression forms include variables, unit, functions,
function applications, pairs with a single-branch pattern-matching construct for
deconstructing them, value tagging ($\iinl$ and $\iinr$), pattern matching on
tagged values, and \ii{fold}/\ii{unfold}.  The nonstandard additions are
unknowns ($u$), \emph{instantiation} ($e \leftarrow (e_1, e_2)$), \emph{sample} ($!e$)
and \emph{after} ($\Etil{e_1}{e_2}$) expressions.

The ``after'' operator, written with a backwards semicolon, evaluates both
$e_1$ and $e_2$ in sequence. 
However, unlike the standard sequencing operator $e_1;e_2$, the result of
$\Etil{e_1}{e_2}$ is the result of $e_1$; the
expression $e_2$ is evaluated just for its
side-effects. For example, the sample-after expression \lk{e !x} of the previous
section is desugared to a combination of sample and after: $\Etil{e}{!x}$.
If we evaluate this snippet in a context where $x$ is bound to
some unknown $u$, 
then the expression $e$ is evaluated first, refining the domain 
of $u$ (amongst other unknowns); 
then the sample expression $!u$ is evaluated for its side effect,
instantiating $u$ to a uniformly generated value from its domain; and
finally the result of $e$ is returned as the result of the whole expression.
A reasonable way to implement $\Etil{e_1}{e_2}$ using standard lambda
abstractions would be as $(\lambda ~ x . ~ (\lambda \_.~ x)~ e_2) ~ e_1$.
However, there is a slight difference in the semantics of this encoding
compared to our intended semantics---we will return to this point in \autoref{sec:gen}.

Weight annotations like the ones in the \lk{bst} example
\iffull
\bcp{show the annotated case expression that we're encoding here?}
\fi
 can be
desugared using {\em instantiation expressions}.  For example, 
assuming a standard encoding of binary search trees ($\ii{Tree} = \mu X.~ 1 +
int \times X \times X$) and naturals, plus syntactic
sugar for constant naturals:
\[ 
\Ecaseof
 {(\Einst
  {\Eunfold{\ii{Tree}}{\mathit{tree}}}
  {1}{\mathit{size}})}
 {x}{\dots}{y}{\dots}
\]
%


\iffull
\begin{figure}
{
\[
\begin{array}{c}
\\ 
\inference[\TVar~]
{x : T \in \Gamma}
 {\Gamma \vdash x : T}\quad
\inference[\TUnit~]
{}
 {\Gamma \vdash () : 1}\\
\\ 
\inference[\TAbs~]
{\Gamma, x : T_1, f : T_1 \rightarrow T_2 \vdash e_2 : T_2}
{\Gamma \vdash \RecT{f}{x}{e_2}{T_1}{T_2} : T_1 \rightarrow T_2}\\
\\ 
\inference[\TApp~]
{\Gamma \vdash e_0 : T_1 \rightarrow T_2 & \Gamma \vdash e_1 : T_1}
{\Gamma \vdash (e_0~e_1) : T_2}\\
\\ 
\inference[\TPair~]
{\Gamma \vdash e_1 : T_1 & \Gamma \vdash e_2 : T_2}
{\Gamma \vdash (e_1, e_2) : (T_1 \times T_2)}\\
\\ 
\inference[\TCasePair~]
{\Gamma \vdash e : (T_1 \times T_2) \\
 \Gamma, x : T_1, y : T_2 \vdash e' : T}
{\Gamma \vdash \case e ~ \ii{of} ~ (x,y) \rightarrow e': T}\\
\\ 
\inference[\TInl~]
{\Gamma \vdash e : T_1}
{\Gamma \vdash \inl{T_1 \mathord{+} T_2}{e} : T_1 + T_2}\\
\\ 
\inference[\TInr~]
{\Gamma \vdash e : T_2}
{\Gamma \vdash \inr{T_1 \mathord{+} T_2}{e} : T_1 + T_2}\\
\\ 
\inference[\TCase~]
{\Gamma \vdash e : T_1 + T_2 \\
 \Gamma, x : T_1 \vdash e_1 : T & \Gamma, y : T_2 \vdash e_2 : T}
{\Gamma \vdash \case e ~ \of (\ii{inl}~x \rightarrow e_1) ~ (\ii{inr}~y \rightarrow e_2) : T}\\
\\ 
\inference[\TFold~]
{U = \mu X. ~ T_1 & \Gamma \vdash e_1 : T_1[U/X]}
{\Gamma \vdash \fold{U}{e_1} : U}\\
\\ 
\inference[\TUnfold~]
{U = \mu X. ~ T_1 & \Gamma \vdash e_1 : U}
{\Gamma \vdash \unfold{U}{e_1} : T_1[U/X]}\\
\end{array}
\]
}
\caption{Standard Typing Rules}
\label{fig:StandardTyping}
\end{figure}
\fi

Most of the typing rules are standard (these can be found in \iffull
\autoref{fig:StandardTyping}.  \else the extended version of the paper.\fi)
The four non-standard rules are given in \autoref{fig:Typing2}.  
Unknowns are typed: each will be
associated with a domain (set of values) drawn from 
a type $\Tnf$ that does not contain arrows. Luck does not support
constraint solving over functional domains (which would require something like
higher-order unification), and the restriction of unknowns to non-functional
types reflects this.  To remember the types of unknowns, we extend the
typing context 
to include a component $U$, a map from unknowns to non-functional
types. When the variable typing environment $\Gamma = \emptyset$,
we write $U \vdash e : T$ as a shorthand for $\emptyset ; U \vdash e :T$.
\iffull
The rules for the standard constructs in \autoref{fig:StandardTyping}
are as expected (adding $U$ everywhere).
\fi
An unknown $u$ has
type $\Tnf$ if $U(u) = \Tnf$. If $e_1$ and $e_2$ are well typed,
then $\Etil{e_1}{e_2}$ shares the type of $e_1$. An instantiation expression
$e \leftarrow (e_l, e_r)$ is well typed if $e$ has sum type $\Tnf_1 + \Tnf_2$
and $e_l$ and $e_r$ are natural numbers.
A sample expression $!e$ has the (non-functional) type $\Tnf$
when $e$ has type $\Tnf$.

\begin{figure}
{
\[
\begin{array}{c}
\\ 
\inference[\TUnknown~]
{U(u) = \Tnf}
{\Gamma; U \vdash u : \Tnf}
\quad  
\inference[\TTil~]
{\Gamma; U \vdash e_1 : T_1  & \Gamma ; U \vdash e_2 : T_2}
{\Gamma; U \vdash \Etil{e_1}{e_2} : T_1 }\\
\\ 
\inference[\TBang~]
{\Gamma; U \vdash e : \Tnf }
{\Gamma; U \vdash !e : \Tnf}
\quad 
\inference[\TInst~]
{\Gamma; U \vdash e : \Tnf_1 + \Tnf_2 \\
 \Gamma; U \vdash e_l : nat & \Gamma \vdash e_r : nat}
{\Gamma; U \vdash e \leftarrow (e_l, e_r) : \Tnf_1 + \Tnf_2}\\
\\
nat := \mu X. ~ 1 + X
\end{array}
\]
}
\caption{Typing Rules for Nonstandard Constructs}
\label{fig:Typing2}
\end{figure}


The predicate semantics for Core Luck, written $e \Downarrow v$,
are defined as a big-step operational semantics.  We assume that $e$ is
closed with 
respect to ordinary variables and free of unknowns.
The rules for the standard constructs are
unsurprising
\iffull
(\autoref{fig:standard}).
\else
(see the extended version).
\fi
The only non-standard rules are the ones for narrow, sample and after
expressions, which are essentially ignored (\autoref{fig:Standard2}). With the
predicate semantics we can implement a naive generate-and-test method
for generating valuations satisfying some predicate by generating arbitrary
well-typed valuations and filtering out those for which the predicate does
not evaluate to \lk{True}. 

\iffull
\begin{figure}
{
\[
\begin{array}{c}
\text{\leo{Do we want explicit (not) is value}}\\
\text{\leo{preconditions everywhere to clutter things?}}\\
\\ 
\inference[\PVal~]
  {\mathit{is\_value ~ v}}
  { v \Downarrow v}\\
\\ 
\inference[\PApp~]
 {e_0 \Downarrow (\RecT{f}{x}{e_2}{T_1}{T_2}) \\
 e_1 \Downarrow v_1 \\
 e[(\RecT{f}{x}{e_2}{T_1}{T_2})/f, v_1/x] \Downarrow v}
 {(e_0 ~ e_1) \Downarrow v} \\
\\ 
\inference[\PPair~]
  {e_1 \Downarrow v_1 ~~~~~ e_2 \Downarrow v_2}
  {(e_1, e_2) \Downarrow (v_1, v_2)}\\
\\ 
\inference[\PCasePair~]
  {e \Downarrow (v_1, v_2) \\
   e'[v_1/x, v_2/y] \Downarrow v}
  {\case e ~ \ii{of} (x,y) \rightarrow e' \Downarrow v}\\
\\ 
\inference[\PInl~]
  {e \Downarrow v}
  {\inl{T}{e} \Downarrow \inl{T}{v}} \\
\\ 
\inference[\PInr~]
  {e \Downarrow v}
  {\inr{T}{e} \Downarrow \inr{T}{v}} \\
\\ 
\inference[\PCaseInl~]
  {e \Downarrow \inl{T}{v} ~~~~~ e_1[v/x] \Downarrow v_1}
  {\Ecaseof{e}{x}{e_1}{y}{e_2} \Downarrow v_1} \\
\\ 
\inference[\PCaseInr~]
  {e \Downarrow \inr{T}{v} ~~~~~ e_2[v/y] \Downarrow v_2}
  {\Ecaseof{e}{x}{e_1}{y}{e_2} \Downarrow v_2} \\
\\ 
\inference[\PFold~]
{e \Downarrow v}
{\fold{S}{e} \Downarrow \fold{S}{v}}\\
\\ 
\inference[\PUnfold~]
{e \Downarrow \fold{T}{v}}
{\unfold{T}{e} \Downarrow v}\\
\end{array}
\]
}
\caption{Predicate Semantics for Standard Core Luck Constructs}
\label{fig:standard}
\end{figure}
\fi

\begin{figure}
{
\[
\begin{array}{c}
\inference[\PInst~]
{e \Downarrow v & e_1 \Downarrow v_1 & e_2 \Downarrow v_2 \\
  \sem{v_1} > 0 & \sem{v_2} > 0}
{   e \leftarrow (e_1, e_2) \Downarrow v}
\quad
\inference[\PBang~]{e \Downarrow v}{!e \Downarrow v}\\
\\
\inference[\PTil~]{e_1 \Downarrow v_1 & e_2 \Downarrow v_2}
                  {\Etil{e_1}{e_2} \Downarrow v_1}\\
\\
\begin{array}{rcl}
\sem{\fold{nat}(\inl{1 \mathord{+} nat}{()})} & = & 0\\
\sem{\fold{nat}(\inr{1 \mathord{+} nat}{v})} & = & 1 + \sem{v}\\
\end{array}
\end{array}
\]
}
\caption{Predicate Semantics for Nonstandard Constructs}
\label{fig:Standard2}
\vspace{5mm}
\end{figure}

\subsection{Constraint Sets}
\label{sec:constr}

The rest of this section develops an alternative probabilistic generator
semantics for Core Luck.  This semantics will use {\em constraint sets}
$\kappa \in \cset$ to describe the possible values that unknowns can take.
For the moment, we leave the implementation of constraint sets open (the one
used by our prototype interpreter is described in \autoref{sec:source}),
simply requiring that they support the following operations:
\[
\begin{array}{lcl}
\sem{\cdot} & :: & \cset -> \ii{Set} ~ \ii{Valuation} \\
U &::& \cset -> \ii{Map} ~ \mathcal{U} ~ \Tnf\\
\ii{fresh} &::& \cset \rightarrow \Tnf^{*} \rightarrow (\cset\times \mathcal{U}^{*})\\
\ii{unify} &::& \cset \rightarrow \ii{Val} \rightarrow \ii{Val} \rightarrow \cset\\
\ii{SAT}   &::& \cset \rightarrow Bool\\
{}[\cdot] &::& \cset \rightarrow \mathcal{U} \rightarrow \ii{Maybe} ~ \ii{Val}
\\
\ii{\Csamplename} &::& \cset \rightarrow \mathcal{U} \rightarrow \cset^{*}
\end{array}
\]
Here we describe these operations informally, deferring technicalities until
after we have presented the generator semantics (\autoref{sec:prop}).


A constraint set $\kappa$ denotes a set of valuations ($\sem{\kappa}$),
representing the solutions to the constraints. Constraint sets also carry type
information about existing unknowns: $U(\kappa)$ is a mapping from $\kappa$'s
unknowns to types. A constraint set $\kappa$ is {\em well typed}
($\vdash \kappa$) if, for every valuation $\sigma$ in the denotation of $\kappa$
and every unknown $u$ bound in $\sigma$, the type map $U(\kappa)$ contains $u$
and $\emptyset;U(\kappa)\vdash\sigma(u):U(\kappa)(u)$.
\iffull
$$\forall (\sigma \in \sem{\kappa}) (u \in \sigma). ~~
u \in U(\kappa) \wedge
\emptyset;U(\kappa)\vdash\sigma(u):U(\kappa)(u) $$
\diane{Now that the text above has been rephrased (the type judgment now appears clearly), maybe we don't need this anymore even in full version}
\fi

Many of the semantic rules will need to introduce fresh unknowns.  The
$\ii{fresh}$ function 
takes as inputs a constraint set $\kappa$ and a sequence of (non-functional) types of length $k$;
it draws the next $k$ unknowns (in some deterministic order) from the infinite
set $\mathcal{U}$ and extends $U(\kappa)$ with the respective bindings.

The main way constraints are introduced during evaluation is unification. Given
a constraint set $\kappa$ and two values, each potentially containing unknowns,
$\ii{unify}$ updates $\kappa$ to preserve only those valuations in which the
values match.

$\ii{SAT}$ is a total predicate that holds on constraint sets whose denotation
contains at least one valuation.  The totality requirement implies that our
constraints must be decidable.

The value-extraction function $\kappa[u]$ returns an optional (non-unknown)
value: if in the denotation of $\kappa$, all valuations map $u$ to the same
value $v$, then that value is returned (written $\just{v}$); otherwise 
nothing (written $\emptyset$).

The $\ii{sample}$ operation is used to implement sample expressions
($!e$): given a constraint set $\kappa$ and an unknown $u \in U(\kappa)$, it
returns a list of constraint sets representing all possible concrete choices for $u$,
in all of which $u$ is completely determined---that is $\forall \kappa \in
(\ii{sample}~\kappa~u). ~ \exists v. ~ \kappa[u] = \just{v}$. To allow for
reasonable implementations of this interface, we maintain an invariant that the
input unknown to $\ii{sample}$ will always have a finite denotation; thus, the
resulting list is also finite.

\subsection{Narrowing Semantics}
\label{sec:narrow}

As a first step toward a semantics for Core Luck that incorporates both
constraint solving 
and local instantiation, we define a simpler {\em narrowing} semantics.
This semantics is of some interest in its own right, in that it extends
traditional ``needed narrowing'' with explicit probabilistic instantiation
points, but its role here is as a subroutine of the matching semantics
in \autoref{sec:gen}. 

The narrowing evaluation judgment takes as inputs an expression $e$ and a
constraint set $\kappa$.  As in the predicate semantics, evaluating $e$ returns
a value $v$, but now it also depends on a constraint set $\kappa$ and returns a
new constraint set $\kappa'$.  The latter is intuitively a refinement
of\bcp{``is guaranteed to be smaller than''?}\leo{What is
  ``smaller''?}\bcp{I'd intuitively assume pointwise subset.  But in any
case we did say ``intuitively''...}
$\kappa$---\IE evaluation will only remove valuations.
$$\narrow{e}{\kappa}{t}{q}{v}{\kappa'}$$
The semantics is annotated with a representation of the sequence of random
choices made during evaluation, in the form of a \emph{trace} $t$. A trace is a
sequence of {\em choices}:
integer pairs $(m,n)$ with
$0 \leq m < n$, where $n$ 
denotes the number of possibilities chosen among and $m$ is the index of the
one actually 
taken.  We write $\emptylist$ for the empty trace and $t\concat t'$ for the
concatenation of two traces.
We also annotate the judgment with the probability $q$ of making the choices
represented in the trace.
Recording traces is useful after the fact in calculating the total
probability of some given outcome of evaluation (which may be reached by
many different derivations).  Traces play no role in determining how
evaluation proceeds.
\iffull
We model probability distributions using rational numbers $q \in
(0,1] \cap \mathbb{Q}$, for simplicity in the Coq formalization.
\fi

We maintain the invariant that the input constraint set $\kappa$ is well typed
and that the input expression $e$ is well typed with respect to an empty
variable context and the unknown context $U(\kappa)$.
Another invariant is that every constraint set $\kappa$ that appears as input to
a judgment is satisfiable and the restriction of its denotation to the unknowns
in $e$ is finite. These invariants are established at the top-level (see
\autoref{sec-toplevel}).  The finiteness invariant ensures the
output of $\ii{sample}$ will always be a finite collection (and therefore the
probabilities involved will be positive rational numbers.
Moreover, they guarantee termination of constraint
solving, as we will see in
\autoref{sec:gen}. 
Finally, we assume that the type of every expression has been determined by
an initial type-checking phase. We write $e^T$ to show that $e$ has type
$T$. This information is used in the semantic rules to provide types for fresh
unknowns.

\begin{figure}
{
\[
\begin{array}{c}
\MoveRuleLabel{-2.9em}{0.25em}{\NVal~}
\inference
{	v=() ~\lor~ v=(\ErecT{f}{x}{e'}{T_1}{T_2}) ~\lor~ \isunknown{v}	} 
{	\narrow{v}{\kappa}{\emptylist}{1}{v}{\kappa}	}
\\
\\
\MoveRuleLabel{-2.9em}{0.25em}{\NPair~}
\inference
{	
	\narrow{e_1}{\kappa}{t_1}{q_1}{v_1}{\kappa_1}
&	\narrow{e_2}{\kappa_1}{t_2}{q_2}{v_2}{\kappa_2}
} 
{	\narrow{(e_1,e_2)}{\kappa}{t_1\concat t_2}{q_1*q_2}{(v_1,v_2)}{\kappa_2}
}
\\
\\
\MoveRuleLabel{-4.8em}{1.3em}{\NCasePairP~}
\inference
{	
	\narrow{e}{\kappa}{t}{q}{(v_1,v_2)}{\kappa_a}
\\	
	\narrow{\substs{e'}{x}{v_1}{y}{v_2}}{\kappa_a}{t'}{q'}{v}{\kappa'}
} 
{	\narrow{\EcaseofP{e}{x}{y}{e'}}{\kappa}{t\concat t'}{q*q'}{v}{\kappa'}
}
\\
\\
\MoveRuleLabel{-2.25em}{1.3em}{\NCasePairU~}
\inference
{	
	\narrow{e}{\kappa}{t}{q}{u}{\kappa_a}
\\
	\Cfresh{\kappa_a}{[\Tnf_1,\Tnf_2]}{[u_1,u_2]}{\kappa_b}
\\	
	\Cunify{\kappa_b}{(u_1,u_2)}{u}{\kappa_c}
\\	
	\narrow{\substs{e'}{x}{u_1}{y}{u_2}}{\kappa_c}{t'}{q'}{v}{\kappa'}
} 
{	\narrow{\EcaseofP{e^{\Tnf_1\times \Tnf_2}}{x}{y}{e'}}{\kappa}{t\concat t'}{q*q'}{v}{\kappa'}
}
\\
\\
\MoveRuleLabel{-2.5em}{0.25em}{\NInl~}
\inference
{	\narrow{e}{\kappa}{t}{q}{v}{\kappa'}
} 
{	\narrow{\Einl{T_1 \Tsum T_2}{e}}{\kappa}{t}{q}{\Einl{T_1 \Tsum T_2}{v}}{\kappa'}	}
\\
\\
\iffull
\MoveRuleLabel{-2.5em}{0.25em}{\NInr~}
\inference
{	\narrow{e}{\kappa}{t}{q}{v}{\kappa'}
}
{	\narrow{\Einr{T_1 \Tsum T_2}{e}}{\kappa}{t}{q}{\Einr{T_1 \Tsum T_2}{v}}{\kappa'}	}
\\
\\
\fi
\MoveRuleLabel{2em}{1.3em}{\NCaseInl~}
\inference
{	
	\narrow{e}{\kappa}{t}{q}{\Einl{T}{v_l}}{\kappa_a}
\\	
	\narrow{\subst{e_l}{x_l}{v_l}}{\kappa_a}{t'}{q'}{v}{\kappa'}
}
{	\narrow{\Ecaseof{e}{x_l}{e_l}{x_r}{e_r}}{\kappa}{t\concat t'}{q*q'}{v}{\kappa'}
}
\\
\\
\iffull
\MoveRuleLabel{2em}{1.3em}{\NCaseInr~}
\inference
{	
	\narrow{e}{\kappa}{t}{q}{\Einr{T}{v_r}}{\kappa_a}
\\	
	\narrow{\subst{e_r}{x_r}{v_r}}{\kappa_a}{t'}{q'}{v}{\kappa'}
}
{	\narrow{\Ecaseof{e}{x_l}{e_l}{x_r}{e_r}}{\kappa}{t\concat t'}{q*q'}{v}{\kappa'}
}
\\
\\
\fi
\hspace{-0.75em}
\MoveRuleLabel{4.25em}{1.3em}{\NCaseU~}
\inference
{	
	\narrow{e}{\kappa}{t_1}{q_1}{u}{\kappa_a}
\\	
	\Cfresh{\kappa_a}{[\Tnf_l, \Tnf_r]}{[u_l, u_r]}{\kappa_0}
\\	
	\Cunify{\kappa_0}{u}{(\Einl{\Tnf_l\mathord{+}\Tnf_r}{u_l})}{\kappa_l}
&	
	\Cunify{\kappa_0}{u}{(\Einr{\Tnf_l\mathord{+}\Tnf_r}{u_r})}{\kappa_r}
\\
        \Cchoose{1}{\kappa_l}{\subst{e_l}{x_l}{v_l}}
               {1}{\kappa_r}{\subst{e_r}{x_r}{v_r}}
               {t_2}{q_2}{i}
\\
        \narrow{\subst{e_i}{x_i}{u_i}}{\kappa_i}{t_3}{q_3}{v}{\kappa'}
}
{	\narrow{\Ecaseof{e^{\Tnf_l\mathord{+}\Tnf_r}}{x_l}{e_l}{x_r}{e_r}}{\kappa}
               {t_1\concat t_2 \concat t_3}{q_1*q_2*q_3}{v}{\kappa'}
}
\\
\\
\MoveRuleLabel{2em}{2.5em}{\NApp~}
\inference
{
	\narrow{e_0}{\kappa}{t_0}{q_0}{(\ErecT{f}{x}{e_2}{T_1}{T_2})}{\kappa_a}
\\
        \narrow{e_1}{\kappa_a}{t_1}{q_1}{v_1}{\kappa_b}
\\	
	\narrow{\substs{e_2}{f}{(\ErecT{f}{x}{e_2}{T_1}{T_2})}{x}{v_1}}{\kappa_b}{t_2}{q_2}{v}{\kappa'}
}
{	\narrow{(e_0\;e_1)}{\kappa}{t_0\concat t_1\concat t_2}{q_0*q_1*q_2}{v}{\kappa'}	}
\iffull
\\
\\
\MoveRuleLabel{-2.7em}{0.25em}{\NFold~}
\inference
{	\narrow{e}{\kappa}{t}{q}{v}{\kappa'}
} 
{	\narrow{\Efold{T}{e}}{\kappa}{t}{q}{\Efold{T}{v}}{\kappa'}	}
\\
\\
\MoveRuleLabel{-4.7em}{0.25em}{\NUnfoldF~}
\inference
{	\narrow{e}{\kappa}{t}{q}{\Efold{T}{v}}{\kappa'}	} 
{	\narrow{\Eunfold{T}{e}}{\kappa}{t}{q}{v}{\kappa'}	}
\\
\\
\MoveRuleLabel{-4.7em}{0.25em}{\NUnfoldU~}
\inference
{	
	\narrow{e}{\kappa}{t}{q}{u}{\kappa_a}
\\
	\Cfresh{\kappa_a}{\subst{T}{X}{\mu X.T}}{u'}{\kappa_b}
\\
	\Cunify{\kappa_b}{u}{(\Efold{\mu X.T}{u'})}{\kappa'}
} 
{	\narrow{\Eunfold{\mu X.T}{e}}{\kappa}{t}{q}{u'}{\kappa'}	}
\fi
\end{array}
\]
}
\caption{Narrowing Semantics of Standard Core Luck Constructs}

\label{fig:LuckNarrowStandard}
\end{figure}


\begin{figure}
{ 
\[
\begin{array}{c}
\MoveRuleLabel{-2.9em}{0.25em}{\NTil~~~}
\inference
{	
	\narrow{e_1}{\kappa}{t_1}{q_1}{v_1}{\kappa_1}
&	\narrow{e_2}{\kappa_1}{t_2}{q_2}{v_2}{\kappa_2}
} 
{	\narrow{\Etil{e_1}{e_2}}{\kappa}{t_1\concat t_2}{q_1*q_2}{v_1}{\kappa_2}
}
\\
\\
\MoveRuleLabel{-3.2em}{0.25em}{\NBang~}
\inference
{	
	\narrow{e}{\kappa}{t}{q}{v}{\kappa_a}
&
	\CsampleV{v}{\kappa_a}{t'}{q'}{\kappa'}
}
{	\narrow{!e}{\kappa}{t\concat t'}{q*q'}{v}{\kappa'}
}
\\
\\
\hspace{-0.8em}
\MoveRuleLabel{5.5em}{1.3em}{\NInst~}
\inference
{	
	\narrow{e}{\kappa}{t}{q}{v}{\kappa_a}
\\	
	\narrow{e_1}{\kappa_a}{t_1}{q_1}{v_1}{\kappa_b}
&	\narrow{e_2}{\kappa_b}{t_2}{q_2}{v_2}{\kappa_c}
\\
	\CsampleV{v_1}{\kappa_c}{t_1'}{q_1'}{\kappa_d}
&	\CsampleV{v_2}{\kappa_d}{t_2'}{q_2'}{\kappa_e}
\\
	\Cnatdenote{\kappa_e}{v_1}{n_1} & n_1>0
&	\Cnatdenote{\kappa_e}{v_2}{n_2} & n_2>0
\\	
	\Cfresh{\kappa_e}{[\Tnf_1, \Tnf_2]}{[u_1, u_2]}{\kappa_0}
\\
	\Cunify{\kappa_0}{v}{(\Einl{\Tnf_1\mathord{+}\Tnf_2}{u_1})}{\kappa_l}
&
	\Cunify{\kappa_0}{v}{(\Einr{\Tnf_1\mathord{+}\Tnf_2}{u_2})}{\kappa_r}
\\
        \Cchoose{n_1}{\kappa_l}{v}{n_2}{\kappa_r}{v}
               {t'}{q'}{i}
}
{	
	\narrow{\Einst{e^{\Tnf_1\mathord{+}\Tnf_2}}
		{e^{\nat}_1}{e^{\nat}_2}}{\kappa}
		{t\concat t_1\concat t_2\concat t_1'\concat t_2'\concat t'}
        {q*q_1*q_2*q_1'*q_2'*q'}{v}{\kappa_i}
}
\end{array}
\]
}
\caption{Narrowing Semantics for Non-Standard Expressions}
\label{fig:LuckNarrowNew}
\end{figure}


\begin{figure}
\[
\begin{array}{c}
\inference[]
 {
   \ii{SAT}(\kappa_1) & \ii{SAT}(\kappa_2) 
 }
 {
   \Cchoose{n}{\kappa_1}{e_1}{m}{\kappa_2}{e_2}
          {\listsingleton{\chose{0}{2}}}{n / (n+m)} l
          
 }
\quad
{
\inference[]
 {
   \neg \ii{SAT}(\kappa_1) & \ii{SAT}(\kappa_2)
 }
 {
   \Cchoose{n}{\kappa_1}{e_1}{m}{\kappa_2}{e_2}
          {\emptylist}{1}{r}
 }
}\\
\\
\inference[]
 {
   \ii{SAT}(\kappa_1) & \ii{SAT}(\kappa_2) 
 }
 {
   \Cchoose{n}{\kappa_1}{e_1}{m}{\kappa_2}{e_2}
          {\listsingleton{\chose{1}{2}}}{m/(n+m)} r
 }
\quad
{
\inference[]
 {
   \ii{SAT}(\kappa_1) & \neg\ii{SAT}(\kappa_2)
 }
 {
   \Cchoose{n}{\kappa_1}{e_1}{m}{\kappa_2}{e_2}
          {\emptylist}{1}{l}
 }
}
\end{array}
\]
\caption{Auxiliary relation \Cchoosename}
\label{fig:choose}
\end{figure}

\begin{figure}
\[
\begin{array}{c}
\inference
{
	\Csampleres{u}{\kappa}{S}
&	S[m]=\kappa'
}
{	\CsampleV{u}{\kappa}{\listsingleton{(m,|S|)}}{1/|S|}{\kappa'}	}
\\
\\
\inference
{	}
{	\CsampleV{()}{\kappa}{\emptylist}{1}{\kappa}	}
\quad
\inference
{	\CsampleV{v}{\kappa}{t}{q}{\kappa'}	}
{	\CsampleV{(\Efold{T}{v})}{\kappa}{t}{q}{\kappa'}	}
\\
\\
\inference
{	\CsampleV{v}{\kappa}{t}{q}{\kappa'}	}
{	\CsampleV{(\Einl{T}{v})}{\kappa}{t}{q}{\kappa'}	}
\quad
\inference
{	\CsampleV{v}{\kappa}{t}{q}{\kappa'}	}
{	\CsampleV{(\Einr{T}{v})}{\kappa}{t}{q}{\kappa'}	}
\\
\\
\inference
{	\CsampleV{v_1}{\kappa}{t_1}{q_1}{\kappa_1}
&	\CsampleV{v_2}{\kappa_1}{t_2}{q_2}{\kappa'}
}
{	\CsampleV{(v_1,v_2)}{\kappa}{t_1\concat t_2}{q_1*q_2}{\kappa'}	}
\end{array}
\]
\caption{Auxiliary relation \CsampleVname}
\label{fig:sampleV}
\end{figure}


The narrowing semantics is given in \autoref{fig:LuckNarrowStandard} for the
standard constructs\iffull\else{} (omitting {\em fold/unfold} and {\bf N-R} and
{\bf N-Case-R} rules analogous to the {\bf N-L} and {\bf N-Case-L} rules
shown)\fi{} and in \autoref{fig:LuckNarrowNew} for instantiation
expressions; \autoref{fig:sampleV} and \autoref{fig:choose} give some auxiliary
definitions. Most of the rules are intuitive. A common pattern is sequencing two
narrowing judgments $\narrow{e_1}{\kappa}{t_1}{q_1}{v}{\kappa_1}$ and
$\narrow{e_2}{\kappa_1}{t_2}{q_2}{v}{\kappa_2}$.  The constraint-set result of
the first narrowing judgment ($\kappa_1$) is given as input to the second,
while traces and 
probabilities are accumulated by concatenation ($t_1 \concat t_2$) and
multiplication ($q_1 * q_2$).
%
We now explain the rules in detail.
\leo{Andrew:- Although you talk informally about back-tracking in the implementation, it would be helpful to state 
more clearly up front which semantics are non-deterministic and where (e.g. first rule of sampleV in Fig. 7).}

\explainrule{\NVal} is
the base case of the evaluation relation, handling values that are not
handled by other rules by returning them as-is.
%
No choices are made, so the
probability of the result is 1 and the trace is empty.

\explainrule{\NPair}:
To evaluate $(e_1, e_2)$ given a constraint set $\kappa$, we sequence the
derivations for $e_1$ and $e_2$.

\explainrules{\NCasePairP, \NCasePairU}:
To evaluate the pair elimination expression $\case e ~\ii{of} ~
(x,y) \rightarrow ~ e'$ in a constraint set $\kappa$, we first
evaluate $e$ in $\kappa$. Typing ensures that the resulting value is
either a pair or an unknown.
If it is a pair (\NCasePairP), we substitute its components for
$x$ and $y$ in $e'$ and continue evaluating.
If it is an unknown $u$
of type $\Tnf_1\times \Tnf_2$ (\NCasePairU), we first use $\Tnf_1$ and
$\Tnf_2$ as types for fresh unknowns $u_1$, $u_2$ 
and remember the constraint that the pair $(u_1, u_2)$ must unify with
$u$. We then proceed as above, this time substituting $u_1$ and $u_2$
for $x$ and $y$.

(The first pair rule might appear unnecessary since, even in the case where
the scrutinee evaluates to a pair, we could generate unknowns, unify, and
substitute, as in \NCasePairU. However, unknowns in Luck only range over
non-functional types $\Tnf$, so this trick does not work when the type of
the $e$ contains arrows.)

The \NCasePairU{} rule also shows how the finiteness invariant is preserved:
when we generate the unknowns $u_1$ and $u_2$, their domains are
unconstrained, but before we substitute them into an expression used
as ``input'' to a subderivation, we unify them with the result of a
narrowing derivation, which already has a finite representation
in~$\kappa_a$.

\iffull
\explainrules{\NInl, \NInr}:
\else
\explainrule{\NInl}:
\fi
To evaluate $\inl{T_1 \mathord{+} T_2}{e}$, we evaluate $e$ and tag the
resulting value with $\inl{T_1\mathord{+}T_2}$\!\!, with the resulting
constraint set, trace, and probability unchanged.  $\inr{T_1 \mathord{+}
T_2}{e}$ is handled similarly
\iffull \else(the rule is elided) \fi.

\explainrules{\NCaseInl,\iffull \NCaseInr,\fi \NCaseU}:
As in the pair elimination rule, we first evaluate the discriminee $e$ to a
value, which must have one of the shapes $\Einl{T}{v_l}$, $\Einr{T}{v_r}$, or
$u\in\Ukn$, thanks to typing.  The cases for $\Einl{T}{v_l}$ (rule \NCaseInl)
and $\Einr{T}{v_r}$ \iffull (rule \NCaseInr) \else (elided) \fi are similar to
\NCasePairP: $v_l$ 
or $v_r$ can be directly substituted for $x_l$ or $x_r$ in $e_l$ or
$e_r$. 
The unknown case (\NCaseU) is similar to \NCasePairU{} but a
bit more complex.
Once again $e$ shares with the unknown $u$ a type $\Tnf_l+\Tnf_r$
that does not contain any arrows, so we can generate fresh unknowns
$u_l$, $u_r$ with types $\Tnf_l$, $\Tnf_r$.
We unify $\Einl{\Tnf_l+\Tnf_r}{v_l}$ with $u$ to get the
constraint set $\kappa_l$ and $\Einr{\Tnf_l+\Tnf_r}{v_r}$ with $u$ to
get $\kappa_r$.
%
%
We then use the auxiliary relation $\Cchoosename$ (\autoref{fig:choose}), which
takes two integers $n$ and $m$ (here equal to $1$) as well as two constraint
sets (here $\kappa_l$ and $\kappa_r$), 
to select either $l$ or $r$.
If exactly one of $\kappa_l$ and $\kappa_r$ is satisfiable, then
\Cchoosename{} will return the corresponding index  
with probability $1$ and an empty trace (because no random choice were made).
If both are satisfiable, then the resulting index is randomly chosen.  Both
outcomes are equiprobable (because of the $1$ arguments to $\Cchoosename$), so
the probability is one half in each case.  This uniform binary choice is
recorded in the trace $t_2$ as either $\chose{0}{2}$ or $\chose{1}{2}$.
Finally, we evaluate the expression corresponding to the chosen index, with the
corresponding unknown substituted for the variable.
The satisfiability checks enforce the invariant that constraint sets are
satisfiable, which in turn ensures that $\kappa_l$ and $\kappa_r$
cannot both be unsatisfiable at the same time, since there must exist at
least one
valuation in $\kappa_0$ that maps $u$ to a value (either $\iinl$ or $\iinr$)
which ensures that the corresponding unification will succeed.

\explainrule{\NApp}:
To evaluate an application $(e_0 ~ e_1)$, we first evaluate $e_0$ to
$\RecT{f}{x}{e_2}{T_1}{T_2}$ (since unknowns only range over arrow-free types
$\Tnf$, the result cannot be an unknown) and its argument $e_1$ to a value
$v_1$. We then evaluate the appropriately substituted body, 
$\substs{e_2}{f}{(\ErecT{f}{x}{e_2}{T_1}{T_2})}{x}{v_1}$, and combine the various
probabilities and traces appropriately.

Rule \NTil{} is similar to \NPair; however, the value result of the derivation
is that of the first narrowing evaluation, implementing the reverse form of
sequencing described in the introduction of this section.

\iffull
Rule \NFold{} is similar to \NInl. \NUnfoldF{} and \NUnfoldU{} are similar
to (though simpler than) \NCasePairP{} and \NCasePairU{}.
\fi

\explainrule{\NBang}:
To evaluate $!e$ we evaluate $e$ to a value $v$, then use the auxiliary
relation \CsampleVname{} (\autoref{fig:sampleV}) to completely instantiate $v$,
walking down the structure of $v$. When unknowns are encountered, $\ii{sample}$
is used to produce a list of constraint sets $S$; with probability
$\frac{1}{|S|}$ (where $|S|$ is the size of the list) we can select the $m$th
constraint set in $S$, for each $0 \leq m < |S|$.

Rule \NInst{} is similar to \NCaseU.
The main difference is the ``weight'' arguments $e_1$ and $e_2$.
These are evaluated to values $v_1$ and $v_2$, and \CsampleVname{} is called
to ensure that they are fully instantiated in all subsequent constraint
sets, in particular in $\kappa_e$.
The relation $\Cnatdenote{\kappa_e}{v_1}{n_1}$ walks down the structure of the
value $v_1$ (like $\CsampleVname$) and calculates the unique
natural number $n_1$ 
corresponding to $v_1$. Specifically, when the input value is an unknown,
$\Cnatdenote{\kappa}{u}{n}$ holds if $\kappa[u] = v'$ and $\sem{v} = n$, where
the notation $\sem{v}$ is defined in \autoref{fig:Standard2}.
%
The rest of the rule is the same as \NCaseU, except that the computed weights
$n_1$ and $n_2$ are given as arguments to \Cchoosename{} in order to shape the distribution
accordingly.

Using the narrowing semantics, we can implement a more efficient method for
generating valuations than the naive generate-and-test described in
Section~\autoref{sec:core}: instead of generating arbitrary valuations we
only lazily instantiate a subset of unknowns as we encounter them.  This
method has the additional advantage that, if a generated valuation yields an
unwanted result, the implementation can backtrack to the point of the
latest choice, which can drastically improve
performance~\cite{ClaessenFLOPS14}.


Unfortunately, using the narrowing semantics in this way can lead to a lot
of backtracking.
\newcommand{\andand}{\mathrel{\mbox{\small \&\&}}}
To see why, consider three unknowns, $u_1, u_2$, and
$u_3$, and a constraint set $\kappa$ 
where each unknown has type \lk{Bool} (\IE $1 + 1$) and the domain
associated with each contains both \lk{True} and 
\lk{False} ($\inl{1+1}()$ and $\inr{1+1}()$). Suppose we want to generate
valuations for these three unknowns such that the conjunction $u_1 \andand u_2
\andand u_3$ holds, where $e_1 \andand e_2$ is shorthand for
$\Ecaseof{e_1}{x}{e_2}{y}{\False}$. If we attempt to evaluate the expression $u_1
\andand u_2 \andand u_3$ using the narrowing semantics, we first apply
the \NCaseU{} rule with $e = u_1$. That
means that $u_1$ will be unified with either $\iinl$ or $\iinr$ (applied to a
fresh unknown) with equal probability, leading to a \lk{False} result for the
entire expression $50\%$ of the time.  If we choose to unify $u_1$ with an
$\iinl$, then we apply the \NCaseU{} rule again, returning either \lk{False} or
$u_3$ (since unknowns are values---rule \NVal) with equal
probability. Therefore, we will have generated a desired valuation only $25\%$
of the time; we will need to backtrack $75\%$ of the time.

The problem here is that the narrowing semantics is agnostic to the desired
result of the whole computation---we only find out at the very end that we
need to backtrack.  But we can do better...

\subsection{Matching Semantics}
\label{sec:gen}

In this section we present a {\em matching} semantics that takes as an
additional input a {\em pattern} (a value not containing lambdas but possibly
containing unknowns)
\iffull
$$ p ::= () ~ | ~ (p, p) ~ | ~ \inl{\Tnf}{p} ~ | ~ \inr{\Tnf}{p} ~ | ~ \fold{\Tnf}{p} ~ | ~ u$$
\fi
and propagates this pattern backwards to guide the generation
process. By allowing our semantics to look ahead in this way, we can often avoid case
branches that lead to non-matching results.

The matching judgment is again a variant of big-step evaluation; it has the form
$$\geneval {e} {p} {\kappa} {q} {t} {\kappa^?}$$
where \iffull the pattern \fi $p$ can mention the unknowns in $U(\kappa)$ and
where the metavariable $\kappa^?$ stands for an {\em optional} constraint set
($\emptyset$ or $\just{\kappa}$) returned by matching.  Returning an option
allows us to calculate the {probability} of backtracking by summing the
$q$'s
of all failing derivations.  (The combined
probability of failures and successes may be less than $1$, because some
reduction paths may diverge.)

We keep the invariants from \autoref{sec:narrow}: the input constraint set
$\kappa$ is well typed and so is the input expression $e$ (with respect to an
empty variable context and $U(\kappa)$); moreover $\kappa$ is satisfiable, and
the restriction of its denotation to the unknowns in $e$ is finite.
To these invariants we add that the input pattern $p$ is well typed in
$U(\kappa)$ and that the common type of $e$ and $p$ does not contain any arrows
($e$ can still contain functions and applications internally; these are handled
by calling the narrowing semantics).

\iffull
The following properties are essential to maintaining these invariants.
Whenever the output option has the form $\just{\kappa'}$, then
$\kappa'$ is satisfiable.
This is easily ensured by checking the satisfiability of candidate
constraint sets and outputting $\emptyset$ if they are not
satisfiable.
Moreover, when the output has the form $\just{\kappa'}$, then all the
unknowns of $p$ have finite denotations in $\kappa'$ (despite them not
necessarily having finite denotations in the input constraint set
$\kappa$).
\fi

\iffull
The evaluation relation appears in \autoref{fig:LuckSemanticsStandard} (standard
constructs) and \autoref{fig:LuckSemanticsNew} (novel Luck constructs).
Additional rules concerning failure propagating cases appear
in \autoref{fig:LuckSemanticsFailure}, while match rules that deal with
discriminees containing arrow types appear in \autoref{fig:match-fun}.  Most of
them are largely similar to the narrowing rules, only introducing unifications
with target patterns in key places.  Several of them rely on the narrowing
semantics defined previously.
\else
The rules except for $\ii{case}$ are similar to the narrowing
semantics.  \autoref{fig:LuckSemanticsStandard} shows several; the rest
appear in the extended version.
\fi

\begin{figure}
{ 
\[
\begin{array}{c}
\\ 
{ 
  \MoveRuleLabel{-3.25em}{0.25em}{\GVal~}
  \inference
  {
    v = () ~\lor~ \isunknown{v}  & \Cunify {\kappa} {v} {p} {\kappa'}
  }
  {
    \geneval {v} {p} {\kappa} {1} {\emptylist} 
             {\ii{if}~\ii{SAT}(\kappa')~\ii{then}~\just{\kappa'}~\ii{else}~\emptyset}
  }
}\\
\\ 
{
  \MoveRuleLabel{-3.25em}{0.25em}{\GPair~}
  \inference
  {
    \Cfresh{\kappa}{[\Tnf_1,\Tnf_2]}{[u_1,u_2]}{\kappa'}\\
    \Cunify {\kappa'} {(u_1, u_2)} {p} {\kappa_0} \\
    \geneval {e_1} {u_1} {\kappa_0} {q_1} {t_1} {\just{\kappa_1}} & 
    \geneval {e_2} {u_2} {\kappa_1} {q_2} {t_2} {\kappa_2^?} 
  }
  {
    \geneval {(e_1^{\Tnf_1}, e_2^{\Tnf_2})} {p} {\kappa} {q_1 * q_2} {t_1 \concat t_2} {\kappa_2^?}
  }
}\\
\iffull
\else
\\ 
{
  \MoveRuleLabel{-5.25em}{0.25em}{\GPairFail~}
  \inference
  {
    \Cfresh{\kappa}{[\Tnf_1,\Tnf_2]}{[u_1,u_2]}{\kappa'}\\
    \Cunify {\kappa'} {(u_1, u_2)} {p} {\kappa_0} \\
    \geneval {e_1} {u_1} {\kappa_0} {q_1} {t_1} {\emptyset}
  }
  {
    \geneval {(e_1^{\Tnf_1}, e_2^{\Tnf_2})} {p} {\kappa} {q_1} {t_1} {\emptyset}
  }
}\\
\fi
\iffull
\\ 
{
  \MoveRuleLabel{-2.5em}{2.25em}{\GCasePair~}
  \inference
  {
    \Cfresh{\kappa} {[\Tnf_1, \Tnf_2]} {[u_{1}, u_{2}]} {\kappa_a} \\
    \geneval{e}{(u_1, u_2)}{\kappa_a}{q_1}{t_1}{\just{\kappa_b}}\\
    \geneval{e'[u_1/x, u_2/y]}{p}{\kappa_b}{q_2}{t_2}{\kappa^?}\\
  }
  {
    \geneval{\EcaseofP{e^{\Tnf_1 \times \Tnf_2}}{x}{y}{e'}} {p} {\kappa} {q_1 * q_2} {t_1 \concat t_2} {\kappa^?}
  }
}\\
\\ 
{
  \MoveRuleLabel{-3.5em}{0.25em}{\GInlSat~}
  \inference
  {
    \Cfresh{\kappa}{\Tnf_1}{u}{\kappa_1}\\
    \Cunify{\kappa_1}{(\inl{\Tnf_1 + \Tnf_2}{u})}{p}{\kappa_2}\\
    \ii{SAT}(\kappa_2) & \geneval {e} {u} {\kappa_2} {q} {t} {\kappa^?}
  }
  {
    \geneval {\inl{\Tnf_1 + \Tnf_2}{e}} {p} {\kappa} {q} {t} {\kappa^?}
  }
}\\
\\ 
{
  \MoveRuleLabel{-3.5em}{0.25em}{\GInrSat~}
  \inference
  {
    \Cfresh{\kappa}{\Tnf_2}{u}{\kappa_1}\\
    \Cunify{\kappa_1}{(\inr{\Tnf_1 + \Tnf_2}{u})}{p}{\kappa_2}\\
    \ii{SAT}(\kappa_2) & \geneval {e} {u} {\kappa_2} {q} {t} {\kappa^?}
  }
  {
    \geneval {\inr{\Tnf_1 + \Tnf_2}{e}} {p} {\kappa} {q} {t} {\kappa^?}
  }
}\\
\fi
\\ 
{
  \MoveRuleLabel{-3.25em}{0.25em}{\GApp~}
  \inference
  {
    \narrow {e_0} {\kappa} {t_0} {q_0} {(\Rec{f}{x}{e_2})} {\kappa_0}\\
    \narrow {e_1} {\kappa_0} {t_1} {q_1} {v_1} {\kappa'} \\
    \geneval{e_2[(\Rec{f}{x}{e_2})/f, v_1/x]}{p}{\kappa'}{q_2}{t_2}{\kappa^?}
  }
  {
    \geneval{(e_0\;e_1)}{p}{\kappa}{q_0 * q_1 * q_2}{t_0 \concat t_1 \concat t_2}{\kappa^?}
  }
}\\
\iffull
\\ 
{
  \MoveRuleLabel{-3.5em}{0.25em}{\GFold~}
  \inference
  {
    \Cfresh{\kappa} {\Tnf[\mu X. ~ \Tnf / X]} {u} {\kappa_1}\\
    \Cunify{\kappa_1}{(\fold{\mu X. ~ \Tnf}{u})}{p}{\kappa_2}\\
    \geneval {e} {u} {\kappa_2} {q} {t} {\kappa^?}
  }
  {
    \geneval {\fold{\mu X. ~ \Tnf}{e}} {p} {\kappa} {q} {t} {\kappa^?}
  }
}\\
\\ 
{
  \MoveRuleLabel{-4em}{0.25em}{\GUnfold~}
  \inference
  {
    \geneval{e}{(\fold{\mu X. ~ \Tnf}{p})}{\kappa}{q}{t}{\kappa^?}
  }
  {
    \geneval{\unfold{\mu X. ~ \Tnf}{e}}{p}{\kappa}{q}{t}{\kappa^?}
  }
}\\
\else
\\ 
\MoveRuleLabel{-3.75em}{0.25em}{\GTil~}
\inference
{	
        \geneval{e_1}{p}{\kappa}{q_1}{t_1}{\just{\kappa_1}} &
        \narrow{e_2}{\kappa_1}{t_2}{q_2}{v}{\kappa_2}
} 
{	
        \geneval{\Etil{e_1}{e_2}}{p}{\kappa}{q_1 * q_2}{t_1\concat t_2}{\just{\kappa_2}}
}
\\
\fi
\end{array}
\]
}
\caption{Matching Semantics of \iffull Standard\else Selected\fi{} Core Luck
  Constructs}
\label{fig:LuckSemanticsStandard}
\end{figure}

\iffull
\begin{figure}
{
\[
\begin{array}{c}
\\ 
{
  \MoveRuleLabel{-0.5em}{2.25em}{\GCasePairF~}
  \inference
  {
    T_1 \not \in \Tnf \vee T_2 \not \in \Tnf\\
    \narrow{e}{\kappa}{t_1}{q_1}{(v_1, v_2)}{\kappa_1}\\
    \geneval{e'[v_1/x, v_2/y]}{p}{\kappa_1}{q_2}{t_2}{\kappa_2^?}\\
  }
  {
    \geneval{\EcaseofP{e^{T_1 \times T_2}}{x}{y}{e'}} {p} {\kappa} {q_1 * q_2} {t_1 \concat t_2} {\kappa_2^?}
  }
}\\
\\
{
\MoveRuleLabel{3.2em}{1em}{\GCaseInlF~}
\hspace{-1.2em}  
\inference
  {
    T_1 \not \in \Tnf \vee T_2 \not \in \Tnf\\
    \narrow{e}{\kappa}{t_1}{q_1}{\inl{T_1 \Tsum T_2}{v_1}}{\kappa_1}\\
    \geneval{e_1[v_1/x_l]}{p}{\kappa_1}{q_1'}{t_1'}{\kappa^?}
  }
  {
    \geneval{\Ecaseof{e^{T_1 \Tsum T_2}}{x_l}{e_1}{x_r}{e_2}}{p}{\kappa}{q_1 * q_1'}{t_1 \concat t_1'}{\kappa^?}
  }
}\\
\\
{
\MoveRuleLabel{3.2em}{1em}{\GCaseInrF~}
\hspace{-1.2em}  
\inference
  {
    T_1 \not \in \Tnf \vee T_2 \not \in \Tnf\\
    \narrow{e}{\kappa}{t_1}{q_1}{\inr{T_1 \Tsum T_2}{v_2}}{\kappa_1}\\
    \geneval{e_2[v_2/x_r]}{p}{\kappa_1}{q_1'}{t_1'}{\kappa^?}
  }
  {
    \geneval{\Ecaseof{e^{T_1 \Tsum T_2}}{x_l}{e_1}{x_r}{e_2}}{p}{\kappa}{q_1 * q_1'}{t_1 \concat t_1'}{\kappa^?}
  }
}
\end{array}
\]
}
\caption{Matching Semantics for Function Cases}
\label{fig:match-fun}
\end{figure}
\begin{figure}
{ 
\[
\begin{array}{c}
\\ 
{
  \inference[\GPairFail~]
  {
    \Cfresh{\kappa}{[\Tnf_1,\Tnf_2]}{[u_1,u_2]}{\kappa'}\\
    \Cunify {\kappa'} {(u_1, u_2)} {p} {\kappa_0} \\
    \geneval {e_1} {u_1} {\kappa_0} {q_1} {t_1} {\emptyset}
  }
  {
    \geneval {(e_1^{\Tnf_1}, e_2^{\Tnf_2})} {p} {\kappa} {q_1} {t_1} {\emptyset}
  }
}\\
\\ 
{
  \inference[\GCasePairFail~]
  {
    \Cfresh{\kappa} {[\Tnf_1, \Tnf_2]} {\kappa_0} {[u_{1}, u_{2}]}  \\
    \geneval{e}{(u_1, u_2)}{\kappa_0}{q_1}{t_1}{\emptyset}\\
  }
  {
    \geneval{\case e^{\Tnf_1 \times \Tnf_2} ~ \ii{of} ~ (x,y) \rightarrow e'} {p} {\kappa} 
            {q_1} {t_1} {\emptyset}
  }
}\\
\\ 
\inference[\GTilFail~]
{	
        \geneval{e_1}{p}{\kappa}{q_1}{t_1}{\emptyset}
} 
{	
        \geneval{\Etil{e_1}{e_2}}{p}{\kappa}{q_1}{t_1}{\emptyset}
}
\\
\\ 
{
  \inference[\GInlUnSat~]
  {
    \Cfresh{\kappa}{\Tnf_1}{u}{\kappa_1}\\
    \Cunify{\kappa_1}{(\inl{\Tnf_1 + \Tnf_2}{u})}{p}{\kappa_2}\\
    \neg \ii{SAT}(\kappa_2)
  }
  {
    \geneval {\inl{\Tnf_1 + \Tnf_2}{e}} {p} {\kappa} {1} {\emptylist} {\emptyset}
  }
}\\
\\ 
{
  \inference[\GInrUnSat~]
  {
    \Cfresh{\kappa}{\Tnf_2}{u}{\kappa_1}\\
    \Cunify{\kappa_1}{(\inr{\Tnf_1 + \Tnf_2}{u})}{p}{\kappa_2}\\
    \neg \ii{SAT}(\kappa_2)
  }
  {
    \geneval {\inr{\Tnf_1 + \Tnf_2}{e}} {p} {\kappa} {1} {\emptylist} {\emptyset}
  }
}\\
\end{array}
\]
}
\caption{Failure Propagation for Matching Semantics}
\label{fig:LuckSemanticsFailure}
\end{figure}
\fi

\iffull
\begin{figure}
{ 
\[
\begin{array}{c}
\\ 
\MoveRuleLabel{-3.5em}{0.25em}{\GTil~}
\inference
{	
        \geneval{e_1}{p}{\kappa}{q_1}{t_1}{\just{\kappa_1}} &
        \narrow{e_2}{\kappa_1}{t_2}{q_2}{v}{\kappa_2}
} 
{	
        \geneval{\Etil{e_1}{e_2}}{p}{\kappa}{q_1 * q_2}{t_1\concat t_2}{\just{\kappa_2}}
}
\\
\\ 
{
  \MoveRuleLabel{-3.5em}{0.25em}{\GBang~}
  \inference
  {
    \geneval{e}{p}{\kappa}{q_1}{t_1}{\just{\kappa_1}}\\
    \CsampleV{p}{\kappa_1}{t_2}{q_2}{\kappa'} \\
  }
  {
    \geneval{!e}{p}{\kappa}{q_1 * q_2}{t_1 \concat t_2}{\just{\kappa'}}
  }
}\\
\\ 
{
  \MoveRuleLabel{-5.5em}{0.25em}{\GBangFail~}
  \inference
  {
    \geneval{e}{p}{\kappa}{q_1}{t_1}{\nothing}\\
  }
  {
    \geneval{!e}{p}{\kappa}{q_1}{t_1}{\nothing}
  }
}\\
\\ 
{
  \MoveRuleLabel{3.5em}{1.3em}{\GInst~}
  \inference
  {	
    \geneval{e}{p}{\kappa}{q}{t}{\just{\kappa_a}}\\
    \narrow{e_1}{\kappa_a}{t_1}{q_1}{v_1}{\kappa_b} &
    \narrow{e_2}{\kappa_b}{t_2}{q_2}{v_2}{\kappa_c} \\
    \CsampleV{v_1}{\kappa_c}{t_1'}{q_1'}{\kappa_d} &
    \CsampleV{v_2}{\kappa_d}{t_2'}{q_2'}{\kappa_e} \\
    \Cnatdenote{\kappa_e}{v_1}{n_1} & n_1>0 &
    \Cnatdenote{\kappa_e}{v_2}{n_2} & n_2>0 \\
    \Cfresh{\kappa_e}{[\Tnf_1, \Tnf_2]}{[u_1, u_2]}{\kappa_0}\\
    \Cunify{\kappa_0}{p}{(\Einl{\Tnf_1+\Tnf_2}{u_1})}{\kappa_l}\\
    \Cunify{\kappa_0}{p}{(\Einr{\Tnf_1+\Tnf_2}{u_2})}{\kappa_r}\\
    \Cchoose{n_1}{\kappa_l}{v}{n_2}{\kappa_r}{v}{t'}{q'}{i}
  }
  {
    \geneval{\Einst{e^{\Tnf_1+\Tnf_2}}{e_1^{\nat}}{e_2^{\nat}}}{p}{\kappa}
            {q*q_1*q_2*q_1'*q_2'*q'}
            {t\concat t_1\concat t_2\concat t_1'\concat t_2'\concat t'}
            {\{\kappa_i\}}
  }
}\\
\\ 
{
  \MoveRuleLabel{-2.5em}{1em}{\GInstFail~}
  \inference
  {	
    \geneval{e}{p}{\kappa}{q}{t}{\nothing}\\
  }
  {
    \geneval{\Einst{e^{\Tnf_1+\Tnf_2}}{e_1^{\nat}}{e_2^{\nat}}}{p}{\kappa}{q}{t}{\nothing}
  }
}\\
\end{array}
\]
}
\caption{Matching Semantics of Nonstandard Core Luck Constructs}
\label{fig:LuckSemanticsNew}
\end{figure}
\fi


\explainrule{\GVal}:
To generate valuations for a unit value or an unknown, we unify $v$ and the
target pattern $p$ under the input constraint set $\kappa$.
Unlike \NVal, there is no case for functions, since the expression being
evaluated must have a non-function type.

\explainrules{\GPair, \GPairFail}:
To evaluate $(e_1, e_2)$, where $e_1$ and $e_2$ have types $\Tnf_1$ and
$\Tnf_2$, we first generate fresh unknowns $u_1$ and
$u_2$\iffull{} with these types\fi. We unify the pair $(u_1, u_2)$ with the target
pattern $p$, obtaining a new constraint set $\kappa'$.
We then proceed as in \NPair, evaluating $e_1$ against pattern $u_1$ and $e_2$
against $u_2$, threading constraint sets and accumulating traces and
probabilities.
\GPair{} handles the case where the evaluation of $e_1$ succeeds\iffull, yielding a constraint
set $\{\kappa_1\}$\fi, while \GPairFail{} handles failure: 
if evaluating $e_1$ yields $\emptyset$, the whole computation
immediately yields $\emptyset$ as well; $e_2$ is not
evaluated, and the final trace and probability are $t_1$ and $q_1$.

\iffull
\explainrules{\GCasePair, \GCasePairFail, \GCasePairF}:
If the type of the discriminee $e$ contains function types 
(\GCasePairF), 
we narrow $e$ to a pair and substitute its components as in \NCasePairP, 
but then we evaluate the resulting expression against the original target pattern $p$.
Otherwise $e$ has a type of form $\Tnf_1\times\Tnf_2$ and we proceed as in \NCasePairU{} with a few differences. The unknowns
$u_1$ and $u_2$ are introduced before the evaluation of $e$ to provide a target
pattern $(u_1, u_2)$. If the evaluation succeeds in yielding
$\just{\kappa_b}$ (\GCasePair) we proceed to
substitute $u_1$ and $u_2$ (that now have a finite domain as all pattern
unknowns at the resulting constraint sets). If instead evaluation of $e$ yields
$\nothing$
(\GCasePairFail), the whole computation returns $\nothing$ immediately. 

\explainrules{\GInlSat, \GInrSat, \GInlUnSat, \GInrUnSat}:
To evaluate $\Einl{\Tnf_1+\Tnf_2}{e}$, we generate an unknown $u$ of type $\Tnf_1$ and unify $\Einl{\Tnf_1+\Tnf_2}{u}$ with the target pattern $p$. 
If the constraint set obtained is satisfiable (\GInlSat),
we simply evaluate $e$ against the pattern $u$. 
Otherwise (\GInlUnSat) we immediately return $\emptyset$.
The same goes for $\iinr$.

\fi

\explainrules{\GApp, \GTil}:
To evaluate an application $e_0 ~ e_1$, we use the narrowing semantics to
reduce $e_0$ to  $\Rec{f}{x}{e_2}$ and $e_1$ to a value $v_1$, then
evaluate $e_2[(\Rec{f}{x}{e_2})/f, v_2/x]$ against the original \iffull target
pattern \fi $p$\iffull{} in the matching semantics\fi.
In this rule we cannot use a pattern during the evaluation of $e_1$: we do not
have any candidates!
%
This is the main reason for introducing the sequencing
operator as a primitive $\Etil{e_1}{e_2}$ instead of
encoding it using lambda abstractions. In \GTil{}, we evaluate $e_1$ against
\iffull the target pattern \fi $p$ and then evaluate $e_2$ using narrowing, just for its side
effects. If we used lambdas to encode sequencing, $e_1$ would be narrowed
instead, which is not what we want.

\iffull
\explainrules{\GFold, \GUnfold}:
\GFold{} is similar to \GPair, only simpler. 
To evaluate $\Eunfold{\mu X. ~ \Tnf}{e}$ with pattern $p$, \GUnfold{} simply evaluates $e$ with the pattern $\Efold{\mu X. ~ \Tnf}{p}$.


\explainrules{\GBang, \GBangFail}:
This rule is very similar to \NBang. We first evaluate $e$ against pattern
$p$. If that succeeds we proceed to use the same auxiliary
relation \CsampleVname{} as in \NBang{} (defined
in \autoref{fig:sampleV}). Otherwise, the whole computation returns $\nothing$.

\explainrules{\GInst, \GInstFail}:
Like in \GBang, we propagate the pattern $p$ and evaluate $e$ against it. After
checking that the resulting constraint set option is not $\nothing$, we proceed
exactly as in \NInst.

\explainrules{\GCaseInlF, \GCaseInrF}:
If the type of the discriminee $e$ contains function types (meaning it cannot be written as $\Tnf_1+\Tnf_2$), we proceed as in \NCaseInl{} and \NCaseInr, except in the final evaluation we match the expression against $p$.
\fi

The interesting rules are the ones for $\ii{case}$ when the type of the
scrutinee does not contain functions. For these rules, we can actually use the
patterns to guide the generation that occurs during the evaluation of the
scrutinee as well.
We model the behavior of constraint solving\bcp{??}: instead of choosing which branch to
follow with some probability (50\% in \NCaseU), we evaluate both branches, just
like a constraint solver would exhaustively search the entire
domain.\bcp{I (still) find that sentence confusing / unhelpful.}

Before looking at the rules in detail, we need to extend the constraint set 
interface with two new functions:
\begin{center}
\begin{tabular}{lcl}
\ii{rename} &$::$& $\mathcal{U}^{*} \rightarrow \cset \rightarrow \cset$\\
\ii{union} &$::$& $\cset \rightarrow \cset \rightarrow \cset$
\end{tabular}
\end{center}

\noindent
The $\ii{rename}$ operation freshens a constraint set by 
replacing all the unknowns in a given sequence with freshly generated
ones\iffull{} (of the same type)\fi.
The $\ii{union}$ of two constraint sets intuitively denotes the union of their
corresponding denotations. 

\begin{figure}
\[
\begin{array}{@{}c@{}}
{
\MoveRuleLabel{3.2em}{1em}{\GCaseOne~}
\hspace{-1.2em}  
\inference
  {
    \Cfresh{\kappa}{[\Tnf_1, \Tnf_2]}{[u_1,u_2]}{\kappa_0}\\
    \geneval{e}{(\inl{\Tnf_1 + \Tnf_2}{u_1})}{\kappa_0}{q_1}{t_1}{\just{\kappa_1}}\\
    \geneval{e}{(\inr{\Tnf_1 + \Tnf_2}{u_2})}{\kappa_0}{q_2}{t_2}{\just{\kappa_2}}\\
    \geneval{e_1[u_1/x_l]}{p}{\kappa_1}{q_1'}{t_1'}{\kappa_a^?}\qquad
    \geneval{e_2[u_2/y_r]}{p}{\kappa_2}{q_2'}{t_2'}{\kappa_b^?}\\
	\kappa^? = \ii{combine} ~ \kappa_0 ~ \kappa_a^? ~ \kappa_b^? \\
  }
  {
    \genevalone{\Ecaseof{e^{\Tnf_1 + \Tnf_2}}{x_l}{e_1}{y_r}{e_2}}{p}{\kappa}
  }
}
\\
    \genevaltwo{q_1 * q_2 * q_1' * q_2'} {t_1 \concat t_2 \concat t_1' \concat t_2'} {\kappa^?}
\\
\\
\begin{array}{@{}l@{}ll}
  \text{where }&
  \ii{combine} ~ \kappa ~ \nothing ~ \nothing = \nothing\\
  &\ii{combine} ~ \kappa ~ \just{\kappa_1} ~ \nothing = \just{\kappa_1} \\
  &\ii{combine} ~ \kappa ~ \nothing ~ \just{\kappa_2} = \just{\kappa_2}\\
  &\ii{combine} ~ \kappa ~ \just{\kappa_1} ~ \just{\kappa_2} = &\\
  &\multicolumn{2}{l}{
    \qquad\quad  \{\ii{union} ~ \kappa_1 ~ 
    (rename~(U(\kappa_1) \hbox{-} U(\kappa))~\kappa_2)\}}\\ 
\end{array}
\\
\\
\\
{
\MoveRuleLabel{4.8em}{1em}{\GCaseTwo~}
  \inference
  {
    \Cfresh{\kappa}{[\Tnf_1, \Tnf_2]}{[u_1,u_2]}{\kappa_0}\\
    \geneval{e}{(\inl{\Tnf_1 + \Tnf_2}{u_1})}{\kappa_0}{q_1}{t_1}{\nothing}\\
    \geneval{e}{(\inr{\Tnf_1 + \Tnf_2}{u_2})}{\kappa_0}{q_2}{t_2}{\just{\kappa_2}}\\
    \geneval{e_2[u_2/y]}{p}{\kappa_2}{q_2'}{t_2'}{\kappa_b^?}\\
  }
  {
    \genevalone{\Ecaseof{e^{\Tnf_1+\Tnf_2}}{x}{e_1}{y}{e_2}}{p}{\kappa}
    \genevaltwo{q_1 * q_2 * q_2'} {t_1 \concat t_2 \concat t_2'} {\kappa_b^?}
  }
}
\iffull
\\
\\
{
\MoveRuleLabel{4.2em}{1.3em}{\GCaseThree~}
\hspace{-1.2em}  
  \inference
  {
    \Cfresh{\kappa}{[\Tnf_1, \Tnf_2]}{[u_1,u_2]}{\kappa_0}\\
    \geneval{e}{(\inl{\Tnf_1 + \Tnf_2}{u_1})}{\kappa_0}{q_1}{t_1}{\just{\kappa_1}}\\
    \geneval{e}{(\inr{\Tnf_1 + \Tnf_2}{u_2})}{\kappa_0}{q_2}{t_2}{\nothing}\\
    \geneval{e_1[u_1/x]}{p}{\kappa_1}{q_1'}{t_1'}{\kappa_a^?}\\
  }
  {
    \genevalone{\Ecaseof{e^{\Tnf_1+\Tnf_2}}{x}{e_1}{y}{e_2}}{p}{\kappa}
    \genevaltwo{q_1 * q_2 * q_1'} {t_1 \concat t_2 \concat t_1'} {\kappa_a^?}
  }
}
\\
\\
{
\MoveRuleLabel{4.2em}{1.3em}{\GCaseFour~}
\hspace{-1.2em}  
  \inference
  {
    \Cfresh{\kappa}{[\Tnf_1, \Tnf_2]}{[u_1,u_2]}{\kappa_0}\\
    \geneval{e}{(\inl{\Tnf_1 + \Tnf_2}{u_1})}{\kappa_0}{q_1}{t_1}{\nothing}\\
    \geneval{e}{(\inr{\Tnf_1 + \Tnf_2}{u_2})}{\kappa_0}{q_2}{t_2}{\nothing}\\
  }
  {
    \genevalone{\Ecaseof{e^{\Tnf_1+\Tnf_2}}{x}{e_1}{y}{e_2}}{p}{\kappa}
    \genevaltwo{q_1 * q_2} {t_1 \concat t_2} {\nothing}
  }
}
\fi
\end{array}
\]
\[
\]
\caption{Matching Semantics for Constraint-Solving {\em case}}
\label{fig:Case-CST}
\end{figure}

\iffull
The four \ii{case} rules with function-free types appear in \autoref{fig:Case-CST}.
\else
Two of the rules appear in \autoref{fig:Case-CST}.
(A third
is symmetric to \GCaseTwo;
a fourth handles failures.)
\fi
We independently evaluate $e$ against both an $\iinl$ pattern and
an $\iinr$ pattern.
%
If both of them yield failure, then the whole evaluation yields failure 
(\iffull \GCaseFour\else elided\fi).
If exactly one  succeeds,
we evaluate just the corresponding branch 
(\GCaseTwo{} or \iffull \GCaseThree\else the other elided rule\fi).
If both succeed (\GCaseOne), we evaluate both branch bodies
and combine the results with \ii{union}.
We use \ii{rename} to avoid conflicts, since we may generate the same fresh
unknowns while independently computing $\kappa_a^?$ and $\kappa_b^?$.


If desired, the user can ensure that only one branch will be executed by using
an instantiation expression before the \ii{case} is reached.
Since $e$ will then begin with a concrete constructor, only one of the
evaluations of $e$ against the patterns $\iinl$ and 
$\iinr$ will succeed, and only the corresponding branch will be executed.

The \GCaseOne{} rule is the second place where the need for finiteness of the
restriction of $\kappa$ to the input expression $e$ arises. In order for the
semantics to terminate in the presence of (terminating) recursive calls, it is
necessary that the domain be finite. To see this, consider a simple recursive predicate that
holds for every number:
\[
\ErecTL{f}{u}{\Ecaseof{\Eunfold{nat}{u}}{x}{\lk{True}}{y}{(f ~ y)}}{nat}{bool}
\]
Even though $f$ terminates in the predicate semantics for every input $u$, if we
allow a constraint set to map $u$ to the infinite domain of all natural numbers,
the matching semantics will not terminate.
While this
finiteness restriction feels a bit 
unnatural, we have not found it to be a problem in practice---see \autoref{sec:source}.
%

\subsection{Example} 
\label{sec:example-application}


\newcommand{\EsumR}{\Einr}
\newcommand{\Ebang}[1]{!#1}
\newcommand{\myand}{\andand}
\newcommand{\choice}[3]{(#1,#2)}
\newcommand{\MAfter}{\GTil}
\newcommand{\ekp}[3]{#3 \; \Shortleftarrow \; #1\Dashv#2}

To show how all this works, let's trace the main steps of the matching derivations of two
given expressions against the pattern $\True$ in a given constraint set.  We
will also extract probability distributions about optional constraint sets from these
derivations.

We are going to evaluate
$A:=\Etil{(0<u\myand u<4)}{!u}$
and
$B:=(\Etil{0<u}{!u})\myand u<4$
against the pattern $\True$ in a constraint set $\kappa$, 
in which $u$ is independent from other unknowns and its possible values are $0,...,9$.
Similar expressions were introduced as examples in \autoref{sec:examples};
the results we obtain here confirm the intuitive explanation given there.

Recall that the conjunction expression $e_1\myand e_2$ is shorthand
for ${\Ecaseof{e_1}{a}{e_2}{b}{\False}}$, and that we are using a standard Peano
encoding of naturals: $\nat=\mu X.~1+X$.  We elide
folds for brevity.  The inequality $a<b$ can be encoded as $lt~a~b$, where:
\[
\begin{array}{c}
\hspace{-0.5em}
lt=\Erecone{f}{x}{\nat}{\nat->\bool} \Erecone{g}{y}{\nat}{\bool}\\
\begin{array}{cll}
  \EcaseofDisc{y} & \EcaseL{\_}{\False} \\
                  & \EcaseR{y_R}{\EcaseofDisc{x} & \EcaseL{\_}{\True}\\
                  &                              & \EcaseR{x_R}{f~x_R~y_R}}\\
\end{array}
\end{array}
\]

%
Many rules 
introduce fresh unknowns, many of which are irrelevant: 
they might be directly equivalent to some other unknown, 
or there might not exist any reference to them.
We abusively use the same variable for two constraint sets which differ only 
in the addition of a few irrelevant variables to one of them.

\paragraph{Evaluation of $A$}

We first derive $\geneval{(0<u)}{\True}{\kappa}{1}{\emptylist}{\{\kappa_0\}}$.
Since in the desugaring of $0<u$ as an application $lt$ is already in \ii{rec}
form and both $0$ and $u$ are values, the constraint set after the narrowing
calls of \GApp~will stay unchanged. We then evaluate
$\Ecaseof{u}{\_}{\False}{y_R}{...}$. Since the domain of $u$ contains both zero
and non-zero elements, unifying $u$ with $\Einl{1+\nat}{u_1}$ and
$\Einr{1+\nat}{u_2}$ (\GVal) will produce some non-empty constraint
sets. Therefore, rule \GCaseOne~applies. Since the body of the left hand side of
the match is \False, the result of the left derivation in \GCaseOne~is
$\emptyset$ and in the resulting constraint set $\kappa_0$ the domain of $u$
is $\{1,...,9\}$.

Next, we turn to $\geneval{(0<u\myand u<4)}{\True}{\kappa}
{1}{\emptylist}{\{\kappa_1\}}$, where, by a similar argument following the
recursion, the domain of $u$ in $\kappa_1$ is $\{1,2,3\}$.  There are 3 possible
narrowing-semantics derivations for $\Ebang{u}$:
(1) $\narrow{!u}{\kappa_1}{\listsingleton{\choice{0}{3}{\frac{1}{3}}}}{1/3}{u}{\kappa^A_1}$,
(2) $\narrow{!u}{\kappa_1}{\listsingleton{\choice{1}{3}{\frac{1}{3}}}}{1/3}{u}{\kappa^A_2}$,
and
(3) $\narrow{!u}{\kappa_1}{\listsingleton{\choice{2}{3}{\frac{1}{3}}}}{1/3}{u}{\kappa^A_3}$,
where the domain of $u$ in $\kappa^A_i$ is $\{i\}$.
(We have switched to narrowing-semantics judgments because of the rule $\MAfter$.)
Therefore all the possible derivations
for $A=\Etil{(0<u\myand u<4)}{!u}$ matching $\True$ in $\kappa$ are:
$$\geneval{A}{\True}{\kappa}
 {1/3}{\listsingleton{\choice{i-1}{3}{\frac{1}{3}}}}{\{\kappa^A_i\}}
\qquad \text{for }i\in\{1,2,3\}
$$
From the set of possible derivations, we can extract a probability distribution:
for each resulting optional constraint set, we sum the probabilities
of each of the traces that lead to this result. 
Thus the probability distribution associated
with $\genevalone{A}{\True}{\kappa}$ is
$$[\{\kappa^A_1\}\mapsto\frac{1}{3};\quad 
\{\kappa^A_2\}\mapsto\frac{1}{3};\quad 
\{\kappa^A_3\}\mapsto\frac{1}{3}].
$$

\paragraph{Evaluation of $B$}

The evaluation of $0<u$ is the same as before, after which we narrow $!u$
directly in $\kappa_0$ and there are 9 possibilities:
$\narrow{!u}{\kappa_0}{\listsingleton{\choice{i-1}{9}{\frac{1}{9}}}}{1/9}{u}{\kappa^B_i}$ for each
$i\in\{1,...,9\}$, where the domain of $u$ in $\kappa^B_i$ is $\{i\}$. 
Then we evaluate 
$\ekp{u<4}{\kappa^B_i}{\True}$:
if $i$ is 1, 2 or 3 this yields $\{\kappa^B_i\}$;
if $i>3$ this yields a failure $\nothing$. Therefore the
possible derivations for $B=(\Etil{0<u}{!u})\myand u<4$ are:
$$\begin{aligned}
&\geneval{B}{\True}{\kappa}
 {1/9}{\listsingleton{\choice{i-1}{9}{\frac{1}{9}}}}{\{\kappa^B_i\}}
\qquad &\text{for }i\in\{1,2,3\}\\
&\geneval{B}{\True}{\kappa}
 {1/9}{\listsingleton{\choice{i-1}{9}{\frac{1}{9}}}}{\nothing}
\qquad &\text{for }i\in\{4,...,9\}
\end{aligned}
$$
We can again compute the corresponding probability distribution:
$$[\{\kappa^B_1\}\mapsto\frac{1}{9};\quad 
\{\kappa^B_2\}\mapsto\frac{1}{9};\quad 
\{\kappa^B_3\}\mapsto\frac{1}{9};\quad 
\nothing\mapsto\frac{2}{3}]$$
Note that if we were just recording the probability of an execution 
and not its trace,
we would not know that there are six distinct executions leading to $\nothing$ with probability $\frac{1}{9}$,
so we would not be able to compute its total probability.

The probability associated with $\nothing$ ($0$ for $A$, $2/3$ for $B$) is the
probability of backtracking.  As stressed in \autoref{sec:examples}, $A$ is much
better than $B$ in terms of backtracking---i.e., it is more efficient in this
case to instantiate $u$ only after all the constraints on its domain have been
recorded. For a more formal treatment of backtracking strategies in Luck using
Markov Chains, see~\cite{diane-report}.

\subsection{Properties}
\label{sec:prop}
%

We close our discussion of Core Luck by
\iffull
formally stating and proving some key properties.
\else
summarizing some key properties;
more details and proofs can be found in the extended version.
\fi
Intuitively, we show that, when we evaluate an expression $e$ against a pattern $p$ in the
presence of a constraint set $\kappa$, 
we can only remove valuations from the denotation of $\kappa$ ({\em
decreasingness}),
any derivation in the generator semantics corresponds to an execution in the
predicate semantics ({\em soundness}),
and every valuation that matches $p$ will be found in the denotation of the
resulting constraint set of some derivation ({\em completeness}).

Since we have two flavors of generator semantics, narrowing and matching, we
also present these properties in two steps. First, we present the properties
for the narrowing semantics; their proofs have been verified using
Coq\iffull---we just sketch them here\fi.  Then we present the properties
for the matching semantics; for these, we have only paper proofs, but 
these proofs are quite similar
to the narrowing ones (\iffull\else details are in the extended
version; \fi the only real difference is the case rule).


We begin by giving the formal specification of constraint sets\iffull and a
few helpful lemmas that derive from it \fi.
We introduce one extra abstraction, the \emph{domain} of a constraint
set $\kappa$, written $\dom{\kappa}$. This domain corresponds to the
unknowns in a constraint set that actually have bindings in
$\sem{\kappa}$. For example, when we generate a fresh unknown $u$ from
$\kappa$, $u$ does not appear in the domain of $\kappa$; it only
appears in the denotation after we use it in a unification. The domain
of $\kappa$ is a subset of the set of keys of $U(\kappa)$.

When we write that for a valuation and constraint set $\sigma \in \sem{\kappa}$,
it also implies that the unknowns that have bindings in $\sigma$ are exactly the
unknowns that have bindings in $\sem{\kappa}$, i.e., in $\dom{\kappa}$. We use
the overloaded notation $\sigma|_{x}$ to denote the restriction of $\sigma$ to
$x$, where $x$ is either a set of unknowns or another valuation.
\iffull
The following straightforward lemma relates the two restrictions: 
\footnote{All the definitions in this section are implicitly universally quantified
over the free variables appearing the formulas.}
if we restrict a valuation $\sigma'$ to the domain of a constraint set $\kappa$,
the resulting valuation is equivalent to restricting $\sigma'$ to any valuation
$\sigma \in \sem{\kappa}$.

\lemma
\label{lemma:restrict}
 $\sigma \in \kappa \;\Rightarrow\; \sigma'|_{\dom{\kappa}} \equiv \sigma'|_{\sigma}$
\fi

\iffull
\paragraph{Ordering}

We introduce an ordering on constraint sets: two constraints sets
are \emph{ordered} ($\kappa_1 \leq \kappa_2$) if
$\dom{\kappa_2} \subseteq \dom{\kappa_1}$ and for all valuations
$\sigma \in \sem{\kappa_1}$, $\sigma |_{\dom{\kappa_2}} \in \sem{\kappa_2}$.
Right away we can prove that $\leq$ is reflexive and transitive, using
\autoref{lemma:restrict} and basic set properties.
\fi

\paragraph{Specification of fresh}
\[
\Cfresh{\kappa}{T}{u}{\kappa'} 
  \,\Rightarrow
  \left\{
    \begin{array}{l} 
      u ~{\not\in}~ U(\kappa)\\
      U(\kappa') = U(\kappa) \oplus (u \mapsto T)\\
      \sem{\kappa'} = \sem{\kappa}
    \end{array}
  \right.
\]

\noindent
Intuitively, when we generate a fresh unknown $u$ of type $T$ from
$\kappa$, $u$ is really fresh for $\kappa$, meaning $U(\kappa)$ does
not have a type binding for it. The resulting constraint set $\kappa'$
has an extended unknown typing map, where $u$ maps to $T$ and its
denotation remains unchanged. That means that $\dom{\kappa'}
= \dom{\kappa}$.
\iffull
Based on this specification, we can easily prove that $\kappa'$ is
smaller than $\kappa$, the generated unknowns are not contained in any
valuation in $\sem{\kappa}$ and that $\kappa'$ is well typed.

\lemma[fresh\_ordered]
\label{lemma:fresh_ordered}
\[
 \Cfresh{\kappa}{T}{u}{\kappa'} \;\Rightarrow\; \kappa' \leq \kappa
\]

\lemma[fresh\_for\_valuation]
\label{lemma:fresh_for_valuation}
 \[
   \Cfresh{\kappa}{T}{u}{\kappa'} \;\Rightarrow\;
   \forall \sigma. ~\sigma \in \sem{\kappa} \,\Rightarrow\, u \notin \sigma
\]

\lemma[fresh\_types]
\label{lemma:fresh_types}
\[
 \Cfresh{\kappa}{T}{u}{\kappa'} \;\Rightarrow\;
 (\vdash \kappa \,\Rightarrow\, ~ \vdash \kappa')
\]
\fi

\paragraph{Specification of sample}

\[
\kappa' \in \Csample{u}{\kappa}
  \,\Rightarrow
  \left\{
    \begin{array}{l} 
       U(\kappa') = U(\kappa) \\
       \ii{SAT}(\kappa') \\
       \exists v. ~ \sem{\kappa'} =
         \{\,\sigma \mid \sigma \in \sem{\kappa},\, \sigma(u) = v\,\}\\
      \end{array}
  \right.
\]

\noindent
When we sample $u$ in a constraint set $\kappa$ and obtain a list, for every
member constraint set $\kappa'$, the typing map of $\kappa$ remains
unchanged and all of the valuations that remain in the denotation of $\kappa'$
are the ones that mapped to some specific value $v$ in $\kappa$. Clearly, the
domain of $\kappa$ remains unchanged. We also require a completeness property
from \ii{sample}, namely that if we have a valuation $\sigma \in \sem{\kappa}$
where $\sigma(u) = v$ for some $u, v$, then 
$\sigma$ is in some member $\kappa'$ of the result:\bcp{Could save a line if
needed by reformatting this on one line:}
\[ 
\left.
  \begin{array}{l}
    \sigma(u) = v  \\
    \sigma \in \sem{\kappa}
  \end{array}
\right\} \Rightarrow 
\exists\kappa'. 
\left \{
  \begin{array}{l}
    \sigma \in \sem{\kappa'}\\
     \kappa' \in \Csample{u}{\kappa}
  \end{array}
\right.
\]

\iffull
\noindent
We can prove similar lemmas as in \ii{fresh}: ordering and preservation. 

In addition, we can show that if some unknown is a singleton in
$\kappa$, it remains a singleton in $\kappa'$. This is necessary for
the proof of narrowing expressions.

\lemma[sample\_ordered]
\label{lemma:sample_ordered}
\[
 \kappa' \in \Csample{u}{\kappa} \;\Rightarrow\; \kappa' \leq \kappa
\]

\lemma[sample\_types]
\label{lemma:sample_types}
\[
 \kappa' \in \Csample{u}{\kappa} \;\Rightarrow\; (\vdash \kappa \,\Rightarrow\, ~ \vdash \kappa')
\]

\lemma[sample\_preserves\_singleton]
\label{lemma:sample_preserves_singleton}
\[
\begin{array}{l}
  \kappa[u'] \neq \nothing ~\wedge~ \kappa' \in \Csample{u}{\kappa}
  \;\Rightarrow\; \kappa'[u'] = \kappa[u']
\end{array}
\]

Finally, we can lift all of these properties to \CsampleVname{} by simple induction, using
this spec to discharge the base case.
\fi

\paragraph{Specification of unify}
\begin{gather*}
  U(\luckunify{\kappa}{v_1}{v_2}) = U(\kappa) \\
  \sem{\luckunify{\kappa}{v_1}{v_2}} =
    \{\,\sigma\in\sem{\kappa} \mid \sigma(v_1) = \sigma(v_2)\,\}
\end{gather*}

\noindent
When we unify in a constraint set $\kappa$ two (well-typed for
$\kappa$) values $v_1$ and $v_2$, the typing map remains unchanged
while the denotation of the result contains just the valuations from
$\kappa$ that when substituted into $v_1$ and $v_2$ make them
equal. The domain of $\kappa'$ is the union of the domain of $\kappa$
and the unknowns in $v_1, v_2$.

\iffull
Once again, we can prove ordering and typing lemmas.

\lemma[unify\_ordered]
\label{lemma:unify_ordered}
\[
  \luckunify{\kappa}{v_1}{v_2} \leq \kappa
\]

\lemma[unify\_types]
\label{lemma:unify_types}
\[
\left.
  \begin{array}{l}
    \typeC{\kappa}{v_1}{\Tnf}\\
    \typeC{\kappa}{v_2}{\Tnf}\\
    \vdash \kappa
  \end{array}
\right\} \Rightarrow\enskip \vdash \luckunify{\kappa}{v_1}{v_2}
\]
\fi

\paragraph*{Properties of the Narrowing Semantics}

\iffull
With the above specification of constraint sets, we can proceed to
proving our main theorems for the narrowing semantics: decreasingness,
soundness and completeness. \fi
The first theorem, {\em decreasingness} states that we never add new valuations to
our constraint sets; our semantics can only refine the denotation of the input
$\kappa$.

\thm[Decreasingness]
\label{thm:narrow_decreasing}
\[
\narrow{e}{\kappa}{t}{q}{v}{\kappa'} \;\Rightarrow\; \kappa' \leq \kappa
\]
\iffull
\Proof
By induction on the derivation of narrowing, using the lemmas about ordering for
fresh (\autoref{lemma:fresh_ordered}), sample (\autoref{lemma:sample_ordered})
and unify (\autoref{lemma:unify_ordered}), followed by repeated applications of
the transitivity of $\leq$.
\qed
\fi

\iffull
\medskip

Before we reach the other main theorems we need to prove preservation
for the narrowing semantics; to do that we first need to prove that
the typing map of constraint sets only increases when narrowing.

\lemma[Narrowing Effect on Type Environments]
\label{lemma:narrowing_types}
\[
\narrow{e}{\kappa}{t}{q}{v}{\kappa'} \;\Rightarrow\; U(\kappa')|_{U(\kappa)} \equiv U(\kappa)
\]
\Proof
By induction on the derivation.

\pcase{\NVal} 
$U(\kappa)|_{U(\kappa)} \equiv U(\kappa)$ by the definition of restriction.

\pcase{\NPair}
By the inductive hypothesis we have\linebreak
$U(\kappa_1)|_{U(\kappa)} \equiv U(\kappa)$ and
$U(\kappa_2)|_{U(\kappa_1)} \equiv U(\kappa_1)$.
The result follows by transitivity.

\pcase{\NCasePairP}
Similar to \NPair.

\pcase{\NCasePairU}
By the inductive hypothesis we have\linebreak
$U(\kappa_a)|_{U(\kappa)} \equiv U(\kappa)$ and
$U(\kappa')|_{U(\kappa_c)} \equiv U(\kappa_c)$.
By transitivity, we only need to show that
$U(\kappa_c)|_{U(\kappa_a)} \equiv U(\kappa_a)$. This follows by
transitivity of restrict (through $U(\kappa_b)$), and the
specifications of $\ii{fresh}$ and $\ii{unify}$. 

\pcases{\NInl, \NInr} 
The induction hypothesis gives us\linebreak
$U(\kappa')|_{U(\kappa)} \equiv U(\kappa)$,
which is exactly what we want to prove.

\pcases{\NCaseInl, \NCaseInr, \NApp}
Similar to \NCasePairP.

\pcases{\NCaseU-*}
For each of the four cases derived by inlining $\ii{choose}$, 
we proceed exactly like \NCasePairU.

\pcases{\NInst-*}
For each of the four cases derived by inlining $\ii{choose}$ the
result follows as in \NCasePairU, with additional uses of transitivity
to accommodate the narrowing derivations for $e_1$ and $e_2$. 

\pcase{\NBang}
Directly from the induction hypothesis, as in \NInl.

\qed

We also need to prove a form of context invariance for unknowns: we
can substitute a typing map $U$ with a supermap $U'$ in a typing
relation.

\lemma[Unknown Invariance]
\label{lemma:unknown_invariance}
\[
\left.
\begin{array}{l}
  \typedU{U}{\Gamma}{e}{T} \\
   U' |_U \equiv U 
\end{array}
\right\}
\Rightarrow\;
  \typedU{U'}{\Gamma}{e}{T}
\]
\Proof
By induction on the typing derivation for $e$. The only interesting
case is the one regarding unknowns: we know for some unknown $u$ that
$U(u) = T$ and that $U' |_U \equiv U$ and want to prove that
$\typed{U'}{\Gamma}{u}{T}$. By the \TUnknown{} rule we just need to
show that $U'(u) = T$, which follows the definition of $\equiv$ and
$|_{\cdot}$ for maps.
\qed

\smallskip

We can now prove preservation: if a constraint set $\kappa$ is
well typed and an expression $e$ has type $T$ in $U(\kappa)$ and the
empty context, then if we narrow $\ewithkR{e}{\kappa}$ to obtain
$\ewithk{v}{\kappa'}$, $\kappa'$ will be well typed and $v$ will also have
the same type $T$ in $U(\kappa')$.

\thm[Preservation]
\label{thm:narrow_preservation}
\[
\left.
\begin{array}{l}
  \narrow{e}{\kappa}{t}{q}{v}{\kappa'}\\
  \typeCE{\kappa}{e}{T}\\
  \vdash \kappa 
\end{array}
\right\}
\Rightarrow
\left\{
\begin{array}{l}
  \typeCE{\kappa'}{v}{T}\\
  \vdash{\kappa'}
\end{array}
\right.
\]

\Proof
Again, we proceed by induction on the narrowing derivation.

\pcase{\NVal}
Since $v=e$ and $\kappa' = \kappa$, the result follows immediately
from the hypothesis.

\pcase{\NPair}
We have 
$$\narrow{e_1}{\kappa}{t_1}{q_1}{v_1}{\kappa_1}
\tand
\narrow{e_2}{\kappa_1}{t_2}{q_2}{v_2}{\kappa_2}.$$
The inductive hypothesis for the first derivation gives us that
\prsvIH[.]{T'}{\kappa}{e_1}{\kappa_1}{v_1}
Similarly, the second inductive hypothesis gives us that
\prsvIH[.]{T'}{\kappa_1}{e_2}{\kappa_2}{v_2}
The typing assumption of the theorem states that 
$$\typeCE{\kappa}{(e_1, e_2)}{T}.$$
We want to show that 
$$\typeCE{\kappa_2}{(v_1,v_2)}{T} \tand \vdash \kappa_2.$$

By inversion on the typing relation for $(e_1, e_2)$ we know that there exist
$T_1, T_2$ such that
$$T = T_1 \times T_2 
\tand \typeCE{\kappa}{e_1}{T_1} 
\tand \typeCE{\kappa}{e_2}{T_2}.$$
We first instantiate the first inductive hypothesis
on $T_1$ which gives us $\prsvE{\kappa_1}{v_1}{T_1}$. Then, to instantiate the
second one on $T_2$ and obtain $\prsvE{\kappa_2}{v_2}{T_2}$, we need to show that
$\typeCE{\kappa_1}{e_2}{T_2}$, which follows by
combining \autoref{lemma:narrowing_types} and Unknown Invariance
(\autoref{lemma:unknown_invariance}). The same combination also gives us
$\typeCE{\kappa_2}{e_1}{T_1}$. The result follows by the \TPair{} constructor.

\pcase{\NCasePairP}
We have 
$$\narrow{e}{\kappa}{t}{q}{(v_1,v_2)}{\kappa_a}
\tand 
\narrow{\substs{e'}{x}{v_1}{y}{v_2}}{\kappa_a}{t'}{q'}{v}{\kappa'}.$$
By the inductive hypothesis, 
\prsvIH{T'}{\kappa}{e}{\kappa_a}{(v_1,v_2)}
and
\prsvIH[.]{T'}{\kappa_a}{e'[v_1/x,v_2/y]}{\kappa'}{v}
The typing assumption gives us
$$\typeCE{\kappa}{\EcaseofP{e}{x}{y}{e'}}{T}.$$ 
Inversion on this typing relation means that there exist types $T_1, T_2$ such
that 
$$\typeCE{\kappa}{e}{T_1 \times T_2}
\tand 
\typed{\kappa}{(y \mapsto T_2, x \mapsto T_1)}{e'}{T}$$
We want to prove that
$$\prsvE{\kappa'}{v}{T}.$$

We instantiate the first inductive hypothesis on $T_1 \times T_2$ and use
inversion on the resulting typing judgment for $(v_1, v_2)$, which yields
$$\vdash \kappa_a 
\tand \typeCE{\kappa_a}{v_1}{T_1} 
\tand \typeCE{\kappa_a}{v_2}{T_2}.$$
By the second inductive hypothesis, we only need to show that
$$\prsv{\kappa_a}{e'[v_1/x, v_2/y]}{T}$$
Applying the Substitution Lemma twice, Unknown Invariance
and \autoref{lemma:narrowing_types} yields the desired result.
 
\pcase{\NCasePairU}
Similarly to the previous case, we have 
$$\narrow{e}{\kappa}{t}{q}{u}{\kappa_a}
\tand
\narrow{\substs{e'}{x}{u_1}{y}{u_2}}{\kappa_c}{t'}{q'}{v}{\kappa'},$$
where $\Cfresh{\kappa_a}{[\Tnf_1,\Tnf_2]}{[u_1,u_2]}{\kappa_b}$

\noindent
and $\Cunify{\kappa_b}{(u_1,u_2)}{u}{\kappa_c}$
and $\typeCE{\kappa}{e}{\Tnf_1 \times \Tnf_2}.$

\noindent
As in the previous case we have two inductive hypotheses
\prsvIH{T'}{\kappa}{e}{\kappa_a}{u}
and
\prsvIH[.]{T'}{\kappa_c}{e'[u_1/x,u_2/y]}{\kappa'}{v}
The same typing assumption,
$$\typeCE{\kappa}{\EcaseofP{e}{x}{y}{e'}}{T},$$ 
can be inverted to introduce $T_1$ and $T_2$ such that
$$\typeCE{\kappa}{e}{T_1 \times T_2}
\tand 
\typed{\kappa}{(y \mapsto T_2, x \mapsto T_1)}{e'}{T}.$$
By type uniqueness, \leo{Mention debruijn for mu?} $\Tnf_1 = T_1$ and 
$\Tnf_2 = T_2$.
Once again, we want to prove that
$$\prsvE{\kappa'}{v}{T}.$$

Like in the previous case, by the first inductive hypothesis instantiated on
$T_1 \times T_2$ we get that $\kappa_a$ is well typed and $u$ has type
$T_1 \times T_2$ in $\kappa_a$. By the specification of $\ii{fresh}$
(\autoref{lemma:fresh_types}) we get that $\kappa_b$ is well typed and that
$U(\kappa_b) = U(\kappa_a) \oplus u_1 \mapsto T_1 \oplus u_2 \mapsto T_2$, where
$u_1$ and $u_2$ do not appear in $U(\kappa_a)$. By the specification of
$\ii{unify}$ we know that $U(\kappa_c) = U(\kappa_b)$ and
\autoref{lemma:unify_types} means that $\kappa_c$ is well typed. Finally, we
instantiate the second inductive hypothesis on $T$, using the Substitution Lemma
and Unknown Invariance to prove its premise.

\pcases{\NInl, \NInr, \NFold}
Follows directly from the induction hypothesis after inversion of the
typing derivation for $e$.

\pcases{\NCaseInl, \NCaseInr, \NUnfoldF}
Similar to \NCasePairP.

\pcases{\NUnfoldU, \NCaseU-*}
The unknown case for unfold as well as the four cases derived by inlining
$\ii{choose}$ are similar to \NCasePairU.

\pcase{\NApp}
We have
\begin{gather*}
  \narrow{e_0}{\kappa}{t_0}{q_0}{v_0}{\kappa_a},\quad
  \narrow{e_1}{\kappa_a}{t_1}{q_1}{v_1}{\kappa_b},\\
  \narrow{\substs{e_2}{f}{v_0}{x}{v_1}}{\kappa_b}{t_2}{q_2}{v}{\kappa'},
\end{gather*}
as well as the corresponding inductive hypotheses,
where $v_0$ is of the form $(\ErecT{f}{x}{e_2}{T_1}{T_2})$.

The preservation typing assumption states that $\kappa$ is well typed and that
$\typeCE{\kappa}{(e_0 ~ e_1)}{T_2'}$. Inverting this typing relation gives us
that
$$\typeCE{\kappa}{e_0}{T_1' \rightarrow T_2'}
\tand
\typeCE{\kappa}{e_1}{T_1'}.$$

\noindent
By the first inductive hypothesis,
\prsvIH[.]{T'}{\kappa}{e_0}{\kappa_a}{v_0}
Instantiating this hypothesis on $T_1' \rightarrow T_2'$ and inverting the
resulting typing relation gives us that $T_1' = T_1$ and $T_2' = T_2$.
The remainder of the proof is similar to the second part of \NCasePairP.

\pcase{N-Bang}
We have
$$\narrow{e}{\kappa}{t}{q}{v}{\kappa_a}
\tand
\CsampleV{v}{\kappa_a}{t'}{q'}{\kappa'}.$$
By the inductive hypothesis, 
\prsvIH[.]{T'}{\kappa}{e}{\kappa_a}{v}

The typing premise of preservation states that $e$ has type $\Tnf$ in
$U(\kappa)$, and we can instantiate the inductive hypothesis with $\Tnf$.  By
the specification of \Csamplename, $U(\kappa') = U(\kappa)$ and the typing lemma
for \Csamplename{} yields that $\kappa'$ is well typed which concludes the
proof.

\pcase{N-Inst-*}

Each of the four cases derived by inlining \ii{choose} are similar.
The typing premise is $\typeC{\kappa}{\Einst{e}{e_1}{e_2}}{T},$
while we know that $\typeCE{\kappa}{e}{\Tnf_1 \times \Tnf_2}$.  Inverting the
premise and using type uniqueness allows us to equate $T$ with
$\Tnf_1 \times \Tnf_2$ and also gives us that $e_1$ and $e_2$ have type
$\ii{nat}$ in $\kappa$.

We have that $\narrow{e}{\kappa}{t}{q}{v}{\kappa_a}$ and the corresponding
inductive hypothesis:
\prsvIH[.]{T'}{\kappa}{e}{\kappa_a}{v}
We instantiate it to $\Tnf_1 \times \Tnf_2$. Using the narrowing of types and
Unknown Invariance, we get that $e_1$ and $e_2$ have type $\ii{nat}$ in
$\kappa_a$.

We proceed similarly for the derivations 
$$\narrow{e_1}{\kappa_a}{t_1}{q_1}{v_1}{\kappa_b}
\tand
\narrow{e_2}{\kappa_b}{t_2}{q_2}{v_2}{\kappa_c},$$
using the induction hypothesis and propagating type information
using \autoref{lemma:narrowing_types} and Unknown Invariance, to obtain that
$\kappa_c$ is well typed, $v$ has type $\Tnf_1 \times \Tnf_2$, and $v_1$ and
$v_2$ have type $\ii{nat}$.

Continuing with the flow of the rule, 
$$\CsampleV{v_1}{\kappa_c}{t_1'}{q_1'}{\kappa_d}
\tand \CsampleV{v_2}{\kappa_d}{t_2'}{q_2'}{\kappa_e},$$
and the specification for \Csamplename{} lifted to \CsampleVname{} yield
\linebreak $U(\kappa_e) = U(\kappa_d) = U(\kappa_c)$ and $\kappa_e$ is well
typed.

We can then apply the specification of \ii{fresh} to the generation of the
unknowns $u_1$ and $u_2$,
$$\Cfresh{\kappa_e}{[\Tnf_1, \Tnf_2]}{[u_1, u_2]}{\kappa_0},$$ which means
$U(\kappa_0) = U(\kappa_e) \oplus u_1 \mapsto \Tnf_1 \oplus u_2 \mapsto \Tnf_2$
and $\vdash \kappa_0$. Therefore, $U(\kappa_0) |_{U(\kappa_e)} \equiv
U(\kappa_e)$ and by Unknown Invariance the type of $v$ carries over to
$U(\kappa_0)$. In addition, $\inl{\Tnf_1 + \Tnf_2}{u_1}$ and $\inr{\Tnf_1
+ \Tnf_2}{u_2}$ have the same type as $v$ in $\kappa_0$.

The two unifications operate on $\kappa_0$,
$$\Cunify{\kappa_0}{v}{(\Einl{\Tnf_1\mathord{+}\Tnf_2}{u_1})}{\kappa_l}$$
and $$\Cunify{\kappa_0}{v}{(\Einr{\Tnf_1\mathord{+}\Tnf_2}{u_2})}{\kappa_r},$$
and the specification of unify applies to give us that\linebreak
$U(\kappa_0) = U(\kappa_l) = U(\kappa_r)$, as well as $\vdash \kappa_l$ and
$\vdash \kappa_r$. Thus, for both $\kappa_l$ and $\kappa_r$, $v$ has type
$\Tnf_1 \times \Tnf_2$. Since \ii{choose} will pick one of $\kappa_l$ or
$\kappa_r$ to return, this concludes the proof.

\fi

\iffull
\medskip 

With preservation for the narrowing semantics proved, we only need one very
simple lemma about the interaction between variable and valuation substitution 
in expressions:

\lemma[Substitution Interaction]
\label{lemma:substU_subst}
\[
\left.
\begin{array}{l}
  \sigma(e) = e'\\
  \sigma(v) = v'
\end{array}
\right\} \Rightarrow \sigma(e[v/x]) = e'[v'/x]
\]
\Proof
The result follows by induction on $\sigma(e) = e'$ and case splitting on
whether $x$ is equal to any variable encountered.
\qed

\medskip
\fi

\noindent
Soundness and completeness can be visualized as follows:
\[
\begin{tikzcd}
e_p \arrow{r}{\Downarrow} 
    & v_p \\
e \Dashv \kappa \arrow{u}{\sigma \in \sem{\kappa}} \arrow[]{r}{\Downarrow_q^t} 
    & v \vDash \kappa' \arrow[swap]{u}{\sigma' \in \sem{\kappa'}} 
\end{tikzcd}
\]
\noindent
Given the bottom and right sides of the diagram, soundness guarantees that we
can fill in the top and left. That is, any narrowing derivation
$\narrow{e}{\kappa}{q}{t}{v}{\kappa'}$ directly corresponds to some derivation
in the predicate semantics, with the additional assumption that all the unknowns
in $e$ are included in the domain of the input constraint set $\kappa$ (which
can be replaced by a stronger assumption that $e$ is well typed in $\kappa$).
\iffull

Before we formally state and prove soundness, we need two technical lemmas
about unknown inclusion in domains.

\lemma[Narrow Result Domain Inclusion]
\label{lemma:narrow_domain}
\[
\left.
\begin{array}{l}
\narrow{e}{\kappa}{t}{q}{v}{\kappa'}\\
(\forall u. ~ u \in e \Rightarrow u \in \ii{dom}(\kappa))\\
\end{array}
\right\}
\Rightarrow\;
\forall u. ~ u \in v \Rightarrow u \in \ii{dom}(\kappa')
\]
\Proof
Straightforward induction on the narrowing derivation.
\qed

\lemma[Narrow Increases Domain]
\label{lemma:narrow_increases_domain}
\[
\left.
\begin{array}{l}
  \narrow{e}{\kappa}{t}{q}{v}{\kappa'}\\
  u \in \ii{dom}(\kappa)
\end{array}
\right\}
\Rightarrow\;
u \in \ii{dom}(\kappa')
\]
\Proof
Straightforward induction on the narrowing derivation,
using the specifications of \ii{fresh}, \ii{unify} and \ii{sample}.
\qed

We also need a sort of inverse to Substitution Interaction:

\lemma[Inverse Substitution Interaction]
\label{lemma:SubVal_subst}
\[
\left.
\begin{array}{c}
  \sigma(e_2) = e_2'\\
  \sigma(e_1[e_2/x]) = e_1''
\end{array}
\right\}
 \Rightarrow\;
\exists e_1'. ~ \sigma(e_1) = e_1'
\]
\Proof
By induction on $e_1$, inversion of the substitution relation and case analysis
on variable equality when necessary.
\qed

\medskip

\fi

\iffull
We can now move on to stating and proving soundness for the narrowing semantics.
\fi

\thm[Soundness]
\label{thm:narrow_soundness}
\[
  \left.
    \begin{array}{l}
      \narrow{e}{\kappa}{q}{t}{v}{\kappa'} \\
       \sigma'(v) = v_p \;\wedge\; \sigma' \in \sem{\kappa'} \\
      \forall u. ~ u \in e \Rightarrow u \in \dom{\kappa}
    \end{array}
  \right\}
  \Rightarrow\;\exists \sigma ~ e_p.
  \left\{
    \begin{array}{l}
      \sigma' |_\sigma \equiv \sigma \\
      \sigma \in \sem{\kappa} \\
      \sigma(e) = e_p \\
      e_p \Downarrow v_p
    \end{array}
  \right.
\]
\iffull
\Proof
By induction on the narrowing derivation.

\pcase{\NVal}
In the base case $e=v$ and therefore the soundness witnesses are trivially
$\sigma'$ and $v_p$.

\pcase{\NPair}
We know that 
$$\narrow{e_1}{\kappa}{t_1}{q_1}{v_1}{\kappa_1} 
\tand \narrow{e_2}{\kappa_1}{t_2}{q_2}{v_2}{\kappa'}.$$
By inversion of the substitution $\sigma'(v)$ we know that $v=(v_1, v_2)$,
$\sigma(v_1) = v_{p_1}$ and $\sigma(v_2) = v_{p_2}$.

By the induction hypothesis for $e_{p_2}$, we get that
\soundIHd{\sigma'}{\kappa'}{v_2}{v_{p_2}}{e_2}{\kappa_1}{\sigma_1}{e_{p_2}}
Its only premise that is not an assumption can be discharged using the lemma
about narrowing increasing domain (\autoref{lemma:narrow_increases_domain}).

The induction hypothesis for $e_{p_1}$ states that
\soundIHd{\sigma_1}{\kappa_1}{v_1}{v_{p_1}}{e_1}{\kappa}{\sigma}{e_{p_1}}
Proving that $\sigma_1(v_1) = v_{p_1}$ is where the requirement that unknowns be
bound in the input $\kappa$ comes into play: since $\sigma_1$ is a restriction
of $\sigma'$, they assign the same values to all unknowns that $\sigma_1$
assigns a value to. Using \autoref{lemma:narrow_domain}, we can show that all
unknowns in $v_1$ are included in the domain of $\kappa_1$ and therefore
$\sigma_1(v_1) = \sigma'(v_1) = v_{p_1}.$

The final witnesses for the \NPair{} case are $\sigma$ and $(e_{p_1},
e_{p_2})$. The conclusion follows using the transitivity of restrict and
the \PPair{} constructor.

\pcase{\NCasePairP}
We know that 
$$\narrow{e}{\kappa}{t}{q}{(v_1,v_2)}{\kappa_a}
\tand
\narrow{e''}{\kappa_a}{t'}{q'}{v}{\kappa'},$$
where $e'' = \substs{e'}{x}{v_1}{y}{v_2}$.
The inductive hypothesis for the second derivation gives us
\soundIHd{\sigma'}{\kappa'}{v}{v_p}{e''}{\kappa_a}{\sigma_1}{e_p''}

After discharging the premises using \autoref{lemma:narrow_domain}
and\linebreak \autoref{lemma:narrow_increases_domain}, we can investigate the shape of the
$e_p''$ witness. First note that, because of the domain inclusions, there exist
$v_{p_1}, v_{p_2}$ such that $\sigma_1(v_1) = v_{p_1}$ and $\sigma_1(v_2) =
v_{p_2}$. But then, by applying Inverse Substitution Interaction
(\autoref{lemma:SubVal_subst}) we know that there exists $e_p'$ such that
$\sigma_1(e') = e_p$. 

By the inductive hypothesis for $e$, we get that 
\soundIHd[\\]{\sigma_1}{\kappa_a}{(v_1, v_2)}{(v_{p_1}, v_{p_2})}{e}{\kappa}{\sigma}{e_p}
This allows us to provide witnesses for the entire case: $\sigma$ and
$\EcaseofP{e_p}{x}{y}{e_p'}$. Using the transitivity of restrict and
the \PCasePair{} rule, we only need to show that $e_p'' = e_p'[v_{p_1}/x,
v_{p_2}/y]$, which follows from two applications of the (normal) Substitution
Interaction lemma (\autoref{lemma:substU_subst}).

\pcase{\NCasePairU}
This case is largely similar with \NCasePairP, with added details for dealing
with \ii{fresh} and \ii{unify}.
We have
$$\narrow{e}{\kappa}{t}{q}{u'}{\kappa_a}
\tand \narrow{e''}{\kappa_c}{t'}{q'}{v}{\kappa'},$$
where
\vspace{-0.5em}
\begin{gather*}
  e'' = \substs{e'}{x}{u_1}{y}{u_2},\\[-0.2em]
  \Cfresh{\kappa_a}{[\Tnf_1,\Tnf_2]}{[u_1,u_2]}{\kappa_b},\\[-0.2em]
  \Cunify{\kappa_b}{(u_1,u_2)}{u'}{\kappa_c}.
\end{gather*}

By the inductive hypothesis for the second derivation, we get 
\soundIHd{\sigma'}{\kappa'}{v}{v_p}{e''}{\kappa_c}{\sigma_1}{e_p''}
Discharging the inclusion premise is slightly less trivial in this case, since
it requires using the specifications of fresh and unify to hand the case where
$u = u_1$ or $u = u_2$. 

Then, by the definition of $\kappa_c,$ we know that $\sigma_1$ is in the
denotation of $\ii{unify}~{\kappa_b} ~ {(u_1, u_2)} ~ {u'}$. But, by the
specification of unify, $\sigma_1
|_{\dom{\kappa_b}} \in \sem{\kappa_b}$. Since \ii{fresh} preserves the domains
of constraints sets, that also means that $\sigma_1
|_{\dom{\kappa_a}} \in \sem{\kappa_a}$. In the following, let $\sigma_1'
= \sigma_1 |_{\dom{\kappa_a}}$; then, since $u \in \dom{\kappa_a}$
by \autoref{lemma:narrow_domain}, there exists some value $v_u$ such that
$\sigma_1'(u) = v_u$.

We can now use the inductive hypothesis for the first narrowing derivation that
states that: 
\soundIHd{\sigma_1'}{\kappa_a}{u}{v_u}{e}{\kappa}{\sigma}{e_p}

We conclude as in \NCasePairP{}, with $\sigma$ and $\EcaseofP{e_p}{x}{y}{e_p'}$
as the witnesses, where $e_p'$ is obtained as in the previous case by
investigating the shape of $e_p''$.

\pcases{\NInl, \NInr, \NFold}
These cases follow similarly to \NPair.

\pcases{\NCaseInl, \NCaseInr, \NUnfoldF, \NTil}
These cases follow similarly to \NCasePairP.

\pcases{\NCaseU, \NUnfoldU}
These cases follow similarly to \NCasePairU.

\pcase{\NApp}
We have
\begin{gather*}
  \narrow{e_0}{\kappa}{t}{q}{v_0}{\kappa_a},\quad
  \narrow{e_1}{\kappa_a}{t}{q}{v_1}{\kappa_b}, \\[-0.2em]
  \narrow{e_2'}{\kappa_b}{t'}{q'}{v}{\kappa'}\rlap{,}
\end{gather*}
where $v_0$ is of the form $(\ErecT{f}{x}{e_2}{T_1}{T_2})$ and
$e_2' = \substs{e_2}{f}{v_0}{x}{v_1}$.

By the inductive hypothesis for the third derivation we get that
\soundIHd{\sigma'}{\kappa'}{v}{v_p}{e_2'}{\kappa_b}{\sigma_2}{e_{p_2}'}

As in \NCasePairP, we can prove that there exists $v_{p_0}$ and $v_{p_1}$
such that $\sigma_2(v_0) = v_{p_0}$ and $\sigma_2(v_1) = v_{p_1}$.

By the inductive hypothesis for the evaluation of the argument we get that there
exist $\sigma_1$ and $e_{p_1}$ such that $\sigma_2
|_{\sigma_1} \equiv \sigma_1$ and $\sigma_1 \in \sem{\kappa_a}$ and
$\sigma(e_1) = e_{p_1}$ and $e_{p_1} \Downarrow v_{p_1}$.

Since $\sigma_1$ is a restriction of $\sigma_2$ and because of the inclusion
hypotheses, $\sigma_1$ also maps the lambda to $v_{p_0}$, which allows us to use
the last inductive hypothesis:
\vspace{-0.5em}
\soundIHd{\sigma_1}{\kappa_a}{v_0}{v_{p_0}}{e_0}{\kappa}{\sigma}{e_{p_0}}

By inverting the substitution for $\sigma_1$ in the lambda expression, we know
that there exists $e_{p_2}$, such that
\[v_{p_0} = (\ErecT{f}{x}{e_{p_2}}{T_1}{T_2}) \tand \sigma_1(e_2) = e_{p_2}.\]
The witnesses needed for soundness are $\sigma$ and
$(e_{p_0} ~ e_{p_1})$. After using the transitivity of restrict and the \PApp{}
constructor, the only goal left to prove is that:
$$e_{p_2}[v_{p_0}/f, v_{p_1}/x] \Downarrow v_p.$$

Applying Substitution Interaction twice concludes the proof.
\leo{Carefully applying? there is a potential loop in rewriting here. Maybe not too important}

\pcase{\NBang}
We know that 
$$\narrow{e}{\kappa}{t}{q}{v}{\kappa_a}
\tand \CsampleV{v}{\kappa_a}{t'}{q'}{\kappa'}.$$

By the specification of \ii{sample}, since $\sigma' \in \sem{\kappa'}$ we know
that $\sigma' \in \sem{\kappa_a}$ and the result follows directly from the
induction hypothesis.

\pcase{\NInst}
The 4 derived cases from inlining \ii{choose} flow similarly, so without loss of
generality let's assume that the first \ii{choose} rule was used.
We know a lot of things from the narrowing derivation:
\begin{gather*}
  \narrow{e}{\kappa}{t}{q}{v}{\kappa_a}, \\
  \narrow{e_1}{\kappa_a}{t_1}{q_1}{v_1}{\kappa_b},\quad
  \narrow{e_2}{\kappa_b}{t_2}{q_2}{v_2}{\kappa_c}, \\
  \CsampleV{v_1}{\kappa_c}{t_1'}{q_1'}{\kappa_d},\quad
  \CsampleV{v_2}{\kappa_d}{t_2'}{q_2'}{\kappa_e}, \\
  \Cnatdenote{\kappa_e}{v_1}{n_1},\quad n_1>0,\quad
  \Cnatdenote{\kappa_e}{v_2}{n_2},\quad n_2>0, \\
  \Cfresh{\kappa_e}{[\Tnf_1, \Tnf_2]}{[u_1, u_2]}{\kappa_0}, \\
  \Cunify{\kappa_0}{v}{(\Einl{\Tnf_1\mathord{+}\Tnf_2}{u_1})}{\kappa_l}, \\
  \Cunify{\kappa_0}{v}{(\Einr{\Tnf_1\mathord{+}\Tnf_2}{u_2})}{\kappa_r}.
\end{gather*}

By the specification of \ii{unify} for the definition of $\kappa_l$ we know that
$\sigma'|_{\dom{\kappa_0}} \in \sem{\kappa_0}$. By using the specification of
\ii{fresh} we can obtain that $\sigma'|_{\dom{\kappa_e}} \in \sem{\kappa_e}$.
Inversion of $\Cnatdenote{\kappa_e}{v_1}{n_1}$ yields that there exists
$v_{p_1}$ such that $\kappa_e[v_1] = v_{p_1}$ (where we lift the $\kappa[\cdot]$
notation to values) and similarly for $v_{p_2}$
(using \autoref{lemma:sample_preserves_singleton} to preserve the first
result). But that means, for all $\sigma \in \sem{\kappa_e}$ (including
$\sigma'|_{\dom{\kappa_e}}$), we have $\sigma(v_i) = v_{p_i}$.

That allows us to use the inductive hypotheses for $e_1$ and $e_2$ yielding
$\sigma_1$, $e_{p_1}$ and $e_{p_2}$ such that $\sigma'
|_{\sigma_1} \equiv \sigma_1$, $\sigma_1(e_i) = e_{p_i}$ and $e_{p_i} \Downarrow
v_{p_i}$.

Finally, we use the last inductive hypothesis to obtain $\sigma$ and $e_p$ as
appropriate and provide $\sigma$ and $\Einst{e_p}{e_{p_1}}{e_{p_2}}$ as
witnesses to the entire case. The result follows immediately.
\qed

\medskip
\fi

Completeness guarantees the opposite direction: given a predicate derivation
$e_p \Downarrow v_p$ and a ``factoring'' of $e_p$ into an expression $e$ and a
constraint set $\kappa$ such that for some valuation $\sigma \in \sem{\kappa}$
substituting $\sigma$ in $e$ yields $e_p$, and under the assumption that
everything is well typed, there is always a nonzero probability of obtaining
some factoring of $v_p$ as the result of a narrowing judgment.

\thm[Completeness]
\label{thm:narrow_completeness}
\[
  \left.
  \begin{array}{l}
    e_p \Downarrow v_p \\
    \sigma(e) = e_p \\
    \sigma \in \sem{\kappa} \;\wedge \prsvB{\kappa}\\
    \prsvA{\kappa}{e}{T}
  \end{array}
  \right\}
  \Rightarrow
  \begin{array}{l}
    \exists v ~ \kappa' ~ \sigma' ~ q ~ t. \\
    \left\{
    \begin{array}{l}
      \sigma' |_\sigma \equiv \sigma \enskip\wedge\enskip
      \sigma' \in \sem{\kappa'} \\
      \sigma'(v) = v_p \\
      \narrow{e}{\kappa}{t}{q}{v}{\kappa'}
    \end{array}
    \right.
  \end{array}
\]
\iffull
\Proof
By induction on the derivation of the predicate semantics judgment.

\pcase{\PVal}
The witnesses for completeness are $v$, $\kappa$, $\sigma$, $1$ and $\emptylist$. The result
holds trivially.

\pcase{\PPair}
We have 
$$e_{p_1} \Downarrow v_{p_1}
\tand 
e_{p_2} \Downarrow v_{p_2}.$$
By inversion on the substitution we have two cases.  In the simple case, $e$ is
some unknown $u$ and $\sigma(u) = (e_{p_1}, e_{p_2})$. But then $e_{p_1}$ and
$e_{p_2}$ must be values, and therefore $e_{p_i} = v_{p_i}$. By the \NVal{}
rule,  $\narrow{u}{\kappa}{\emptylist}{1}{u}{\kappa}$ and the result follows
directly.

In the more interesting case, $e$ is a pair $(e_1, e_2)$ and we know that
$\sigma(e_1) = e_{p_1}$ and $\sigma(e_2) = e_{p_2}$. Inverting the typing
relation gives us
$$\typeCE{\kappa}{e_1}{T_1} \tand \typeCE{\kappa}{e_2}{T_2}.$$

The inductive hypothesis for the first predicate semantics derivation,
instantiated at $\sigma, \kappa \tand T_1$ gives us that
\cmpltIHd{}{_1}{T_1}{v_{p_1}}{_1}
Its assumptions already hold so we can obtain such witnesses. 
By the second inductive hypothesis, we know that 
\cmpltIHd{_1}{_2}{T_2}{v_{p_2}}{_1}
Since $\sigma_1|_{\sigma}$ and $\sigma(e_1) = e_{p_1}$, then $\sigma_1(e_1) =
e_{p_1}$. By preservation, we get that $\kappa_1$ is well typed. Finally,
narrowing only extends the typing environment (\autoref{lemma:narrowing_types})
and then by Unknown Invariance (\autoref{lemma:unknown_invariance}) we can
obtain the last assumption $\typeCE{\kappa}{e_1}{T_1}$ of the inductive
hypothesis.

We combine the results from the two inductive hypotheses to provide witnesses
for the existentials:
\[(v_1, v_2),\enskip\kappa_2,\enskip\sigma_2,\enskip q_1 * q_2 \tand t_1 \concat t_2.\]
By transitivity of restrict, we get that
$\sigma \equiv \sigma_2 |_{\sigma}$,
while the inclusion property $\sigma_2 \in \sem{\kappa_2}$ is satisfied
by the inductive hypothesis directly. To prove that
$\sigma_2((v_1,v_2)) = (v_{p_1}, v_{p_2})$
we just need to prove that $\sigma_2(v_1) = v_{p_1}$, but
that holds because $\sigma(e_1) = e_{p_1}$ and $\sigma$ is a restriction of
$\sigma_2$. Using the $\NPair$ constructor completes the proof.

\pcase{\PApp}
We know that
\[e_{p_0} \Downarrow v_{p_0},\,e_{p_1} \Downarrow v_{p_1}
\tand e_{p_2}[v_{p_0}/f, v_{p_1}/x] \Downarrow v_p,\]
for some $v_{p_0} = (\RecT{f}{x}{e_{p_2}}{T_1}{T_2})$.
Inversion on the substitution gives us only one possible $e$, since unknowns only
range over values: $e = (e_0 ~ e_1)$,
where $\sigma(e_0) = e_{p_0}$ and $\sigma(e_1) = e_{p_1}$.
Inversion of the typing premise gives us
$$\typeCE{\kappa}{e_0}{T_1' \rightarrow T_2'} \tand \typeCE{\kappa}{e_1}{T_1'}.$$

By the inductive hypothesis for the derivation of $e_{p_0}$ we get that
\cmpltIHd{}{_0}{T_1' \rightarrow T_2'}{v_{p_0}}{_0}
All its assumptions hold, so we can invert the last substitution to obtain that
$v_0 = (\RecT{f}{x}{e_2}{T_1}{T_2})$, where $\sigma_0(e_2) = e_{p_2}$.
By preservation, we know that the type of $v_0$ is the type of $e_0$ in
$\kappa_0$ and uniqueness of typing equates $T_1$ with $T_1'$ and $T_2$ with
$T_2'$.

By the second inductive hypothesis, we know that 
\cmpltIHd{_0}{_1}{T_1}{v_{p_1}}{_1}
As in \PPair{} we can discharge all of its assumptions. 

\newcommand{\esubbed}[3]{#1[#2/f, #3/x]}
Let $e_2' = \esubbed{e_2}{v_0}{v_1}$ and
$e_{p_2'} = \esubbed{e_{p_2}}{v_{p_0}}{v_{p_1}}$.
The last inductive hypothesis states that 
\cmpltIHd{_1}{_2'}{T_2}{v}{_2'}
The substitution premise can be discharged using the Substitution Interaction
Lemma (\autoref{lemma:substU_subst}), while the typing premise by repeated
applications of the Substitution Lemma and Unknown Invariance.  The proof
concludes by combining the probabilities and traces by multiplication and
concatenation respectively.

\pcases{\PInl, \PInr, \PFold, \PTil} These cases are similar to \PPair.

\pcase{\PCasePair}
From the predicate derivations we have that 
$$e_p  \Downarrow (v_{p_1}, v_{p_2}) \tand e_p'[v_{p_1}/x, v_{p_2}/y] \Downarrow v_p'.$$
Inverting the substitution premise leaves us with $\sigma(e) = e_p$ and
$\sigma(e') = e_p'$, while inverting the typing premise yields 
$$\typeCE{\kappa}{e}{T_1 + T_2} \tand \typed{\kappa}{(x\mapsto T_1,y\mapsto T_2)}{e'}{T}.$$

The inductive hypothesis for $e_p$ gives us that
\cmpltIHd{}{_1}{T_1 + T_2}{(v_{p_1}, v_{p_2})}{}
To decide which of \NCasePairP{} and \NCasePairU{} we will use, we invert the
substitution relation for $v$.
In the simple case, $v = (v_1, v_2)$ and the proof flows similarly to \PApp.

The interesting case is when $v$ is an unknown $u$, in which case we need to
``build up'' the derivation of \NCasePairU.
Let
\begin{gather*}
  \Cfresh{\kappa_1}{[T_1, T_2]}{[u_1, u_2]}{\kappa_{1a}}, \\
  \Cunify{\kappa_{1a}}{(u_1, u_2)}{u}{\kappa_{1b}}, \\
  \sigma_1' = \sigma_1 \oplus u_1 \mapsto v_1 \oplus u_2 \mapsto v_2, \\
  e'' = e'[u_1/x, u_2/y] \tand e_p'' = e_p'[v_{p_1}/x, v_{p_2}/y].
\end{gather*}
The second inductive hypothesis (instantiated at $\sigma_1', \kappa_{1b}$)
states that
\[
  \left.
    \begin{array}{l}
      \sigma_1' \in \sem{\kappa_{1b}} \\
      \vdash \kappa_{1b} \\
      \typeCE{\kappa_{1b}}{e''}{T} \\
      \sigma_1'(e'') = e_p''
    \end{array}
  \right\}
  \Rightarrow
  \begin{array}{l}
    \exists v' ~ \kappa_2 ~ \sigma_2 ~ q_2 ~ t_2 .\\
    \left\{
      \begin{array}{l}
        \sigma_2|_{\sigma_1'}\equiv \sigma_1'
        \enskip\wedge\enskip \sigma_2 \in \sem{\kappa_2} \\
        \sigma(v') = v_p' \\
        \narrow{e''}{\kappa_{1b}}{t_2}{q_2}{v'}{\kappa_2}
      \end{array}
    \right.
  \end{array}
  \rlap{.}
\]
To use this inductive hypothesis we need to discharge all of its assumptions
first.

To prove that $\sigma_1' \in \sem{\kappa_{1b}}$ we start with
$\sigma_1 \in \sem{\kappa_1}$ by the first induction hypothesis. By the
specification of \ii{fresh}, the denotation of $\kappa_1$ remains unchanged,
therefore $\sigma_1 \in \sem{\kappa_{1a}}$. Since $u_1, u_2$ are not in the
domain of $\kappa_{1a}$, the restriction $\sigma_1' |_{\ii{dom}(\kappa_{1a})}$
is $\sigma_1$. Therefore, by the specification of \ii{unify} we just need to show
that $\sigma_1'(u) = \sigma_1'((u_1, u_2))$. Indeed,
\begin{align*}
  \sigma_1'(u) = \sigma_1(u) = (v_{p_1},v_{p_2})
    &= (\sigma_1'(u_1), \sigma_1'(u_2)) \\
    &= \sigma_1'((u_1, u_2)),
\end{align*}
which concludes the proof of the first premise.

The fact that $\kappa_{1b}$ is well typed is a direct corollary of the typing
lemmas for \ii{fresh} (\autoref{lemma:fresh_types}) and \ii{unify}
(\autoref{lemma:unify_types}).

To prove that $\typeCE{\kappa_{1b}}{e''}{T}$ we apply the
Substitution Lemma twice. Then we need to prove that
\begin{gather*}
  \typed{\kappa_{1b}}{\emptyset}{u_i}{T_i} \;\text{ for } i=1,2\\
  \text{and } \typed{\kappa_{1b}}{(x\mapsto T_1, y\mapsto T_2)}{e'}{T}.
\end{gather*}
By the
specification of \ii{unify} we know that $U(\kappa_{1b}) = U(\kappa_{1a})$,
while from the specification of \ii{fresh} we obtain
\[U(\kappa_{1a}) = U(\kappa_1) \oplus u_1 \mapsto T_1 \oplus u_2 \mapsto T_2.\]
This directly proves
the former results for $u_1, u_2$, while Unknown Invariance (since
$U(\kappa_{1b})$ is an extension of $U(\kappa)$) proves the latter.

The final premise of the inductive hypothesis, $\sigma_1'(e'') = e_p''$,
is easily proved by applying the Substitution
Interaction lemma (\autoref{lemma:substU_subst}) twice.

Since we have satisfied all of its premises, we can now use the result of the
second inductive hypothesis. It provides most of the witnesses to completeness
($v$, $\kappa_2$ and $\sigma_2$), while, as usual, we combine the probabilities
and traces by multiplying and concatenating them.  The result follows by
transitivity of restrict and use of the \NCasePairU{} constructor.

\pcases{\PCaseInl, \PCaseInr, \PUnfold}
These cases are in direct correspondence with \PCasePair{}. The only difference
is that to choose between which \Cchoosename\ rule to follow we case analyze on
the satisfiability of the corresponding constraint set.

\pcase{\PBang}
We know that $e_p \Downarrow v_p$.  By the inductive hypothesis, we immediately
obtain that there exists some $v$, $\sigma_1$, $\kappa_1$, $q_1$ and $t_1$ such
that $\sigma_1(v) = v_p \tand \sigma_1 \in \sem{\kappa} \tand \sigma_1
|_{\sigma} \equiv \sigma$ and that $\narrow{e}{\kappa}{t_1}{q_1}{v}{\kappa_1}$.
By the completeness requirement of \Csamplename{} lifted to \CsampleVname, we
know that there exists some $q_2, t_2 \tand \kappa_2$ such
that $\CsampleV{v}{\kappa_1}{t_2}{q_2}{\kappa_2}$ and $\sigma_1 \in \sem{\kappa_2}$. The
result follows easily.

\pcase{\PInst}
We know that 
$$e_p \Downarrow v_p,\enskip e_{p_1} \Downarrow v_{p_1} \tand e_{p_2} \Downarrow
v_{p_2},$$ while $\sem{v_{p_1}} > 0$ and $\sem{v_{p_2}} > 0$.  Chaining the
induction hypothesis as in \PPair, we get that there exist $v, v_1,
v_2, \sigma', \kappa_1, \kappa_2, \kappa', q, q_1$, $q_2, t, t_1 \tand t_2$
such that
\[
  \sigma' \in \sem{\kappa'},
  \quad
  \begin{aligned}
    \sigma'|_{\sigma} &\equiv \sigma \\[-0.2em]
    \sigma'(v) &= v_p \\[-0.2em]
    \sigma'(v_i) &= v_{p_i}
  \end{aligned}
  \tand
  \begin{gathered}
    \narrow{e}{\kappa}{t}{q}{v}{\kappa_1} \\[-0.2em]
    \narrow{e_1}{\kappa_1}{t_1}{q_1}{v_1}{\kappa_2} \\[-0.2em]
    \narrow{e_2}{\kappa_2}{t_2}{q_2}{v_2}{\kappa_3}
  \end{gathered}
\]

By the lifted completeness requirement of \Csamplename{}, we know that there
exist $q_1', q_2', t_1', t_2', \kappa_4 \tand \kappa_5$ such that
$\sigma \in \sem{\kappa_5}$,
\[
  \CsampleV{v_1}{\kappa_3}{t_1'}{q_1'}{\kappa_4}
  \tand
  \CsampleV{v_2}{\kappa_4}{t_2'}{q_2'}{\kappa_5}
  \rlap{\;.}
\]
By definition,
$\Cnatdenote{\kappa_5}{v_1}{v_{p_1}}$ and $\Cnatdenote{\kappa_5}{v_2}{v_{p_2}}$.

Without loss of generality, assume that $v_p = \iinl~v_p'$ for some $v_p'$ and
let
$$\sigma'' = \sigma' \oplus u_1 \mapsto v_p' \tand 
\Cfresh{\kappa_5}{[T_1, T_2]}{[u_1, u_2]}{\kappa'}$$ and
$$\Cunify{\kappa'}{(\iinl~u_1)}{v}{\kappa_l} \tand 
\Cunify{\kappa'}{(\iinr~u_2)}{v}{\kappa_r}.$$

By transitivity of restrict, $\sigma''|_{\sigma} \equiv \sigma$. Moreover,
$\sigma''(v) = v_p$. The proof that $\sigma'' \in \sem{\kappa_l}$ is similar to
the proof that $\sigma_1' \in \sem{\kappa_{1b}}$ in \PPair. To conclude the
proof, we case analyze on whether $\kappa_r$ is satisfiable or not and choosing
which \ii{choose} derivation to follow accordingly.
\qed

\bigskip
\fi

\paragraph*{Properties of the Matching Semantics}

Before we proceed to the theorems for the matching semantics, we
need a specification for the \ii{union} operation.

\paragraph{Specification of union}

\[
\begin{array}{c}
  \left.
  \begin{array}{l}
    U(\kappa_1)|_{U(\kappa_1) \cap U(\kappa_2)} = U(\kappa_2)|_{U(\kappa_1) \cap U(\kappa_2)}\\
    \ii{union} ~ \kappa_1 ~ \kappa_2 = \kappa
  \end{array}
  \right\}\\
  \Rightarrow \\
  \left\{
  \begin{array}{l}
    U(\kappa) = U(\kappa_1) \cup U(\kappa_2) \\
    \sem{\kappa} = \sem{\kappa_1} \cup \sem{\kappa_2} \\
  \end{array}
  \right.
  \end{array}
\]
To take the $\ii{union}$ of two constraint sets, their typing maps must
obviously agree on any unknowns present in both.  The denotation of the
$\ii{union}$ of two constraint sets is then just the union of their corresponding
denotations.

\iffull
Similar lemmas concerning types and ordering can be proved about \ii{union}.

\lemma[union\_ordered]
\label{lemma:union_ordered}
\[
 \kappa' = \ii{union}~{\kappa_1}~{\kappa_2}
 \;\Rightarrow\; \kappa' \leq \kappa_1 \wedge \kappa' \leq \kappa_2
\]

\lemma[union\_types]
\label{lemma:union_types}
\[
\left.
\begin{array}{c}
 \kappa' = \ii{union}~{\kappa_1}~{\kappa_2}\\
 \vdash \kappa_1 \\
 \vdash \kappa_2 
\end{array}
\right\} \Rightarrow \enskip \vdash \kappa'
\]

\paragraph{Specification of rename}

The \ii{rename} function as introduced in the previous section can be
encoded in terms of \ii{fresh} and a function that renames a single unknown
to the result of fresh, iteratively.

\fi

The decreasingness property for the matching semantics is very similar to
the narrowing semantics: if the matching semantics yields $\just{\kappa'}$,
then $\kappa'$ is smaller than the input constraint set.

\thm[Decreasingness]
\label{thm:gen_decreasing}
\[
\geneval{e}{p}{\kappa}{q}{t}{\{\kappa'\}} \;\Rightarrow\; \kappa' \leq \kappa
\]
\iffull
\Proof
This is again the simplest proof: by induction on the derivation of matching
judgment, using the lemmas about ordering for fresh
(\autoref{lemma:fresh_ordered}), sample (\autoref{lemma:sample_ordered}) and
unify (\autoref{lemma:unify_ordered}) and repeated applications of the
transitivity of $\leq$. 
\qed

\medskip

\fi

\iffull
\medskip
Preservation is simpler than before since we only deal with a single output. We
still need a similar lemma about the effect of the matching semantics on types:

\lemma[Matching Effect on Types]
\label{lemma:match_types}
\[
\geneval{e}{p}{\kappa}{q}{t}{v}{\just{\kappa'}} \;\Rightarrow\; U(\kappa')|_{U(\kappa)} \equiv U(\kappa)
\]
\Proof
By induction on the derivation and transitivity of restrict.
\qed

\thm[Preservation]
\label{thm:gen_preservation}
\[
\left.
\begin{array}{l}
\geneval{e}{p}{\kappa}{q}{t}{\just{\kappa'}} \\
\typeCE{\kappa}{e}{\Tnf}\\
\typeCE{\kappa}{p}{\Tnf}\\
\vdash \kappa\\
\end{array}
\right\} \Rightarrow \\
 \vdash \kappa'
\]

\Proof

\pcase{\GVal}
Follows directly from the typing lemma of \ii{unify}
(\autoref{lemma:unify_types}).

\pcase{\GPair}
We know that
$$\geneval {e_1} {u_1} {\kappa_0} {q_1} {t_1} {\just{\kappa_1}} \tand \geneval
  {e_2} {u_2} {\kappa_1} {q_2} {t_2} {\just{\kappa_2}},$$
where
\begin{gather*}
  \Cfresh{\kappa}{[\Tnf_1,\Tnf_2]}{[u_1,u_2]}{\kappa'}, \\
  \Cunify {\kappa'}{(u_1, u_2)} {p} {\kappa_0}.
\end{gather*}

By inversion of the typing relation for $(e_1, e_2)$ we know that 
$$\typeCE{\kappa}{e_1}{\Tnf_1} \tand \typeCE{\kappa}{e_2}{\Tnf_2}.$$
Based on the specification of \ii{fresh}, 
$$U(\kappa') = U(\kappa) \oplus u_1 \mapsto \Tnf_1 \oplus u_2 \mapsto \Tnf_2 \tand \vdash \kappa',$$
while \ii{unify} preserves all type information. Therefore, $\kappa_0$ is well
typed and $\typeCE{\kappa_0}{u_1}{\Tnf_1}$ and
$\typeCE{\kappa_0}{u_2}{\Tnf_2}$. By Unknown Invariance (\autoref{lemma:unknown_invariance}) $e_1$ and $e_2$ are well typed in $\kappa_0$ as well. 

Now we can use the inductive hypothesis for the derivation of $u_1$ which gives
us that $\vdash \kappa_1$. To conclude the proof, we can use the other inductive
hypothesis; for that we just need to show that $\typeCE{\kappa_1}{u_2}{\Tnf_2}$
and $\typeCE{\kappa_1}{e_2}{\Tnf_2}$.  However, by the typing lemma for the
matching semantics (\autoref{lemma:match_types}) we known that $U(\kappa_1)
|_{U(\kappa_0)} \equiv U(\kappa_0)$. Unknown Invariance completes this case.

\pcase{\GCasePair}
We know that 
$$\geneval{e}{(u_1, u_2)}{\kappa_a}{q_1}{t_1}{\just{\kappa_b}}
\tand \geneval{e''}{p}{\kappa_b}{q_2}{t_2}{\just{\kappa'}},$$
where
\begin{gather*}
  e'' = e'[u_1/x, u_2/y], \\
  \Cfresh{\kappa} {[\Tnf_1, \Tnf_2]} {[u_{1}, u_{2}]} {\kappa_a}.
\end{gather*}

Like in the \GPair{} case, using the definition of \ii{fresh} we can obtain that
$U(\kappa') = U(\kappa) \oplus u_1 \mapsto \Tnf_1 \oplus u_2 \mapsto \Tnf_2$ as
well as $\vdash \kappa_a$ and therefore
$\typeCE{\kappa_a}{(u_1,u_2)}{\Tnf_1 \times \Tnf T_2}$.  We again can invert the
typing relation for the entire case to obtain that 
$$\typeCE{\kappa}{e}{\Tnf_1' \times \Tnf_2'} \tand \typed{\kappa}{x \mapsto \Tnf_1',
y \mapsto \Tnf_2'}{e'}{\Tnf},$$ while type uniqueness equates $\Tnf_i$ with
$\Tnf_i'$. Using Unknown Invariance we can propagate these typing relations to
$\kappa_a$. 

We can now use the inductive hypothesis on the matching derivation for $e$ to
obtain that $\kappa_b$ is well typed. By the typing lemma for the matching
semantics and Unknown Invariance we lift all typing relations to $\kappa_b$.  To
conclude the proof using the second inductive hypothesis we need only prove that
$\typeCE{\kappa_b}{e'[u_1/x, u_2/y]}{\Tnf}$, which follows by consecutive
applications of the Substitution Lemma.

\pcases{\GInlSat, \GInrSat, \GFold}
Follow similarly to \GPair.

\pcase{\GApp}
For some $v_0 = (\Rec{f}{x}{e_2})$, we have that
$$\narrow {e_0} {\kappa} {t_0} {q_0} {v_0} {\kappa_0}
\tand    \narrow{e_1}{\kappa_0}{t_1}{q_1}{v_1}{\kappa_1},$$
while
$$\geneval{e'[v_0/f, v_1/x]}{p}{\kappa_1}{q_2}{t_2}{\just{\kappa'}}.$$

By inverting the typing relation for $e_0 ~ e_1$ we get that
$\typeCE{\kappa}{e_0}{T \rightarrow \Tnf} \tand \typeCE{\kappa}{e_1}{T}$.  Using
the preservation theorem for the narrowing semantics
(\autoref{thm:narrow_preservation}) we know that $\kappa_0$ is well typed and
the lambda has the same type as $e_0$ in $\kappa$. That means that
$$\typed{\kappa_0}{(f \mapsto (T \rightarrow \Tnf), x \mapsto T)}{e_2}{\Tnf}.$$ The
typing lemma for the narrowing semantics (\autoref{lemma:narrowing_types}) and
Unknown Invariance allow us to lift type information to $\kappa_0$. We repeat
this process for the second narrowing derivation. To use the inductive
hypothesis and conclude the proof, we only need to apply the Substitution Lemma
twice as in $\GCasePair$.

\pcase{\GUnfold}
This case follows directly from the induction hypothesis.

\pcase{\GTil}
We know that 
$$\geneval{e_1}{p}{\kappa}{q_1}{t_1}{\just{\kappa_1}}
\tand \narrow{e_2}{\kappa_1}{t_2}{q_2}{v}{\kappa_2}.$$

As in \GPair, we invert the typing relation to obtain type information for $e_1$
and $e_2$. We the use the inductive hypothesis on the first derivation to obtain
that $\kappa_1$ is well typed. To conclude the proof, we need only apply the
preservation lemma for the narrowing semantics, and its premise that $e_2$ is
well typed is discharged as usual using the typing lemma for the matching
semantics and Unknown Invariance.

\pcases{\GPairFail, \GCasePairFail, \GTilFail, \GInlUnSat, \GInrUnSat, \GCaseFour, \GBangFail, \GInstFail}
These cases are vacuously true since no constraint set is returned.

\pcases{\GCasePairF, \GCaseInlF, \GCaseInrF}
Similar to \GApp.

\pcase{\GBang}
We know that 
$$\geneval{e}{p}{\kappa}{q_1}{t_1}{\just{\kappa_1}},$$
where $$\CsampleV{p}{\kappa_1}{t_2}{q_2}{\kappa'}.$$

By the inductive hypothesis we immediately get that $\kappa_1$ is well
typed. The specification of \Csamplename{} lifted to \CsampleVname{} yields the
result.

\pcase{\GInst}
We know that 
$$\geneval{e}{p}{\kappa}{q}{t}{\just{\kappa_a}}$$
and
$$\narrow{e_1}{\kappa_a}{t_1}{q_1}{v_1}{\kappa_b} 
\tand \narrow{e_2}{\kappa_b}{t_2}{q_2}{v_2}{\kappa_c}.$$
As in the previous cases, we use the inductive hypothesis and the preservation
lemma for the narrowing semantics to ensure all variables are appropriately
typed in $\kappa_c$.

Following the matching judgment, we proceed to \CsampleVname{} twice resulting
in a constraint set $\kappa_e$; as in \GBang, $\kappa_e$ is well typed. We then
generate two unknowns $u_1$ and $u_2$ with types $\Tnf_1$ and $\Tnf_2$ to obtain
a constraint set $\kappa_0$, that is well typed because of the specification
of \ii{fresh}. Finally, we \ii{unify} the pattern $p$ with the fresh unknowns
tagged $L$ or $R$, yielding $\kappa_l$ and $\kappa_r$ that are both well typed
because of the specification of \ii{unify}. Since all \ii{choose} does is pick
which of $\kappa_l$ and $\kappa_r$ to return, the result follows immediately.

\pcases{\GCaseOne, \GCaseTwo, \GCaseThree}
These cases flow similarly, using repeated applications of the inductive
hypotheses. The only case that hasn't been encountered in a previous rule is
for \GCaseOne, when both branch derivations yield some (well-typed) constraint
sets $\just{\kappa_a}$ and $\just{\kappa_b}$ that are combined
using \ii{union}. But by the typing lemma for \ii{union}, its result is also
well typed.
\qed

\medskip
\fi

Soundness is again similar to the matching semantics.

\thm[Soundness]
\label{thm:gen_soundness}
\[
  \left.
    \begin{array}{l}
      \geneval{e}{p}{\kappa}{q}{t}{\just{\kappa'}} \\
      \sigma'(p) = v_p \wedge \sigma' \in \sem{\kappa'} \\
      \forall u.~
        (u \in e \vee u \in p)
        \Rightarrow u \in \dom{\kappa}
    \end{array}\hspace{-0.1cm}
  \right\}
  \Rightarrow \exists \sigma ~ e_p.
  \left\{
    \begin{array}{l}
      \sigma' |_\sigma \equiv \sigma \\
      \sigma \in \sem{\kappa} \\
      \sigma(e) = e_p \\
      e_p \Downarrow v_p
    \end{array}
  \right.
\]
\iffull
\Proof
By induction on the matching derivation, following very closely the structure of
proof of soundness for the narrowing semantics: we use the inductive hypothesis
for every matching derivation in reverse order, obtaining witnesses for
valuations and expressions, while concluding the proof with the specifications
of constraint set operations and transitivity.

\pcase{\GVal}
In the base case, just like in the proof for the \NVal{} rule, the witnesses are
$\sigma'$ and $v_p$. The inclusion $\sigma' \in \kappa$ is a direct result of
the specification of \ii{unify}.

\pcase{\GPair}
We know that
$$\geneval {e_1} {u_1} {\kappa_0} {q_1} {t_1} {\just{\kappa_1}} \tand \geneval
  {e_2} {u_2} {\kappa_1} {q_2} {t_2} {\just{\kappa'}},$$
where
$$\Cfresh{\kappa}{[\Tnf_1,\Tnf_2]}{[u_1,u_2]}{\kappa_a}$$
and
$$\Cunify {\kappa_a}{(u_1, u_2)} {p} {\kappa_0}.$$

By the definition of \ii{fresh} and the fact that the domain is increasing, we
know that $u_2$ is in the domain of $\kappa'$. That means that there exists some
value $v_{p_2}'$ such that $\sigma'(u_2) = v_{p_2}'$. By the inductive
hypothesis for $\sigma'$ and $u_2$ we get that there exist some $\sigma_1$ and
$e_{p_2}$ such that $\sigma_1$ is a restriction of $\sigma'$ in $\kappa_1$, while 
\[
\sigma_1(e_2) = e_{p_2} \tand e_{p_2} \Downarrow v_{p_2}'.\\
\]

Using a similar argument to obtain a $v_{p_1}'$ such that $\sigma_1(u_1) =
v_{p_1}'$, we can leverage the inductive hypothesis again on the first
derivation gives us that there exists some $\sigma$ and $e_{p_1}$ such that
$\sigma$ is a restriction of $\sigma_1$ in $\kappa_0$ and
\[
\sigma(e_1) = e_{p_1} \tand e_{p_1} \Downarrow v_{p_1}'.\\
\]

Our soundness witnesses are $\sigma$ and $(e_{p_1}, e_{p_2})$.  By the
specification of \ii{unify} we know that $\sigma(p) = \sigma((u_1,u_2))$ and
decreasingness helps us conclude that $v_p = (v_{p_1}', v_{p_2}')$ which
concludes the proof of the pair case, along with transitivity of valuation
restriction.

\pcases{\GCaseOne, \GCaseTwo, \GCaseThree}
The only new rules are the case rules; however, the general structure of the
proof is once again similar. For \GCaseOne, we know that:

\[
    \Cfresh{\kappa}{[\Tnf_1, \Tnf_2]}{[u_1,u_2]}{\kappa_0},
\]
\[
    \geneval{e}{(\inl{\Tnf_1 + \Tnf_2}{u_1})}{\kappa_0}{q_1}{t_1}{\just{\kappa_1}},
\]
\[
    \geneval{e}{(\inr{\Tnf_1 + \Tnf_2}{u_2})}{\kappa_0}{q_2}{t_2}{\just{\kappa_2}},
\]
\[
    \geneval{e_1[u_1/x_l]}{p}{\kappa_1}{q_1'}{t_1'}{\kappa_a^?} \tand
    \geneval{e_2[u_2/y_r]}{p}{\kappa_2}{q_2'}{t_2'}{\kappa_b^?}, 
\]
while
\[	\kappa^? = \ii{combine} ~ \kappa_0 ~ \kappa_a^? ~ \kappa_b^?.    \]


If either of the non-union \ii{combine} cases fire, the proof is simple.  If
$\kappa_a^? = \kappa_b^? = \emptyset$, then there exists no $\kappa'$ such that
the result of the derivation is $\just{\kappa'}$.

Let's assume that $\kappa_a^?  = \just{\kappa_a}$ for some $\kappa_a$ and
$\kappa_b^? = \emptyset$ (the symmetric case follows similarly). Then we know
that $\sigma' \in \sem{\kappa_a}$ and from the inductive hypothesis for the
$e_1$ derivation we get that there exist $\sigma_1$ and $e_{p_1}$ such that
$\sigma \in \sem{\kappa_1}$ and $\sigma_1(e_1[u_1/x_l]) = e_{p_1}$.
As in the narrowing soundness proof, we can leverage the inverse substitution
interaction lemma (\ref{lemma:substU_subst}) to conclude that there exists some
$e_1'$ such that $\sigma_1(e_1) = e_1'$. An additional application of the
inductive hypothesis for the evaluation of $e$ against the $\inl{\Tnf_1
+ \Tnf_2}{u_1}$ gives us $\sigma$ and $e_p$ such that $\sigma \in \sem{\kappa}$
and $\sigma(e) = e_p$, which are also the soundness witnesses that conclude the
proof.

The more interesting case is when $\kappa_a^? = \just{\kappa_a}$ and $\kappa_b^?
= \just{\kappa_b}$ for some constraint sets $\kappa_a$ and $\kappa_b$. In that
case, $\sigma' \in \sem{\kappa_a}$ or
$\sigma' \in \sem{\ii{rename}~(U(\kappa_a) \hbox{-} U(\kappa_0))~\kappa_b}$. The
first case proceeds exactly like the one for $\kappa_b^? = \emptyset$. For the
latter, we need to push the renaming to $\sigma'$, obtaining some $\sigma_r$
which is an alpha-converted version of $\sigma'$ and then proceed
similarly. Since the alpha conversion only happens in the unknowns that are not
present in the original constraint set, the choice of these unknowns doesn't
matter for the final witness.
\qed

\bigskip
\fi

For the completeness theorem, we need to slightly strengthen its premise; since
the matching semantics may explore both branches of a {\em case}, it can fall
into a loop when the predicate semantics would not (by exploring a
non-terminating branch that the predicate semantics does not take). Thus, we
require that \emph{all} valuations in the input constraint set result in a
terminating execution.

\iffull 
Before we go to completeness we need an
auxiliary lemma that ensures there exists \emph{some} derivation that
returns a constraint set option if this requirement holds. This is
only necessary for the combining \GCaseOne.

\lemma[Termination]
\label{lemma:gen_termination}
\[
  \left.
    \begin{array}{l}
      \prsv{\kappa}{e}{\Tnf} \\
      \forall \sigma \in \sem{\kappa}. ~ \exists v'. ~ \sigma(e) \Downarrow v'
    \end{array}
  \right\}
  \Rightarrow
    \exists \kappa^? ~ q ~ t. ~ \geneval{e}{p}{\kappa}{q}{t}{\kappa^?}
\]

The proof of this lemma is almost identical to the completeness proof. Since it
doesn't require or enforce particular valuation memberships of the constraint
sets involved, every case can follow with the same argument. The only rules
where the difference matters is in the \ii{case} rules, where the lack of
assumptions allows to provide some termination witness without guaranteeing that
the resulting constraint set is not $\nothing$.
 
We also need another straightforward lemma regarding the completeness of values:

\lemma[Value Completeness]
\[
\left.
\begin{array}{l}
  \typeCE{\kappa}{e}{\Tnf} \\
  \vdash \kappa\\
  \sigma \in \sem{\kappa} \\
  \sigma(e) = v_p\\
  \sigma(p) = v_p\\
\end{array}
\right\}
\Rightarrow
\exists \kappa' ~ \sigma' ~ q ~ t. \\
\left\{
\begin{array}{l}
  \sigma' |_\sigma \equiv \sigma\\
  \sigma' \in \sem{\kappa'}\\
  \geneval{e}{p}{\kappa}{q}{t}{\just{\kappa'}}
\end{array}
\right.
\]
\Proof
By induction on $e$.

\pcase{$e = ()$ or $e = u$}
If $e$ was unit or an unknown, let $\kappa' = \ii{unify}~\kappa~e~p$. By the
specification of \ii{unify} $\sigma \in \sem{\kappa'}$. The witnesses to
conclude the case are $\kappa',~\sigma,~1$ and $\emptylist$ using the \GVal{} rule.

\pcase{$e = (e_1, e_2)$}
Following the \GPair{} rule, let 
$$\Cfresh{\kappa}{[\Tnf_1,\Tnf_2]}{[u_1,u_2]}{\kappa'}$$
and
$$\Cunify {\kappa'}{(u_1, u_2)} {p} {\kappa_0}.$$
We invert the substitution relation to obtain that $\sigma(e_1) = v_{p_1}$ and
$\sigma(e_2) = v_{p_2}$ for some $v_{p_1}, v_{p_2}$. Let $\sigma'
= \sigma \oplus u_1 \mapsto v_{p_1} \oplus u_2 \mapsto v_{p_2}$. Then since
$u_1$ and $u_2$ are fresh, $\sigma'|_{\sigma} \equiv \sigma$ and, by the
specification of \ii{unify}, $\sigma' \in \kappa'$.

By the inductive hypothesis for $e_1$ (inverting the typing relation for the
typing premise), there exist $\sigma_1,~ \kappa_1,~ q_1$ and $t_1$ such that
$\sigma_1|_{\sigma'} \equiv \sigma'$ and $\sigma_1 \in \sem{\kappa_1}$ and
$$\geneval{e_1}{u_1}{\kappa_0}{q_1}{t_1}{\kappa_1}.$$ 

Using Unknown Invariance we can apply the second inductive hypothesis to get
similar $\sigma_2, ~\kappa_2, q_2$ and $t_2$. We conclude the case by providing
the witnesses $\sigma_2$ and $\kappa_2$, while combining the probabilities and traces as usual ($q_1 * q_2$ and $t_1 \concat t_2$).

\pcases{$L$, $R$ or \ii{fold}}
The remaining cases are similar to the pair case, with only one inductive
hypothesis.

\qed

Finally, we will need to propagate the termination information across matching
derivations. For that we can prove the following corollary of decreasingness:

\cor{Termination Preservation}
\label{cor:termination_preservation}
\[
  \left.
    \begin{array}{l}
      \geneval{e}{p}{\kappa}{q}{t}{\kappa'} \\
      \forall \sigma \in \sem{\kappa}. ~\exists v. ~ \sigma(e) \Downarrow v
    \end{array}
  \right\}
  \Rightarrow\;
  \forall \sigma' \in \sem{\kappa'}. ~\exists v. ~\sigma'(e) \Downarrow v
\]
\Proof
By decreasingness, $\kappa' \leq \kappa$, which means that
$\sigma' |_{\sigma} \in \sem{\kappa}$.\linebreak Then, there exists $v$ such that
$\sigma'|_{\sigma}(e) \Downarrow v$ and the result follows.
\qed

\fi

\thm[Completeness]
\label{thm:gen_completeness}
\[
  \left.
    \begin{array}{l}
      e_p \Downarrow v_p \;\wedge\; \sigma \in \sem{\kappa} \\
      \prsv{\kappa}{e}{\Tnf} \\
      \sigma(e) = e_p \wedge \sigma(p) = v_p \\
      \forall \sigma' \in \sem{\kappa}. ~ \exists v'. ~\sigma'(e) \Downarrow v'
    \end{array}
  \right\}
  \Rightarrow
  \begin{array}{l}
    \exists \kappa' ~ \sigma' ~ q ~ t.\\
    \left\{
      \begin{array}{l}
        \sigma' |_\sigma \equiv \sigma \\
        \sigma' \in \sem{\kappa'} \\
        \geneval{e}{p}{\kappa}{q}{t}{\just{\kappa'}}
      \end{array}
    \right.
  \end{array}
\]
\iffull
\Proof
By induction on the predicate derivation.

\pcase{\PVal}
Follows directly from the completeness lemma for values.

\pcase{\PPair}
We have 
$$e_{p_1} \Downarrow v_{p_1}
\tand 
e_{p_2} \Downarrow v_{p_2}.$$
As in the narrowing proof, we invert the substitution of $e$ and get two cases.
In the simple case, $e$ is some unknown $u$ and $\sigma(u) = (e_{p_1},
e_{p_2})$. But then $e_{p_1}$ and $e_{p_2}$ must be values, and the proof
follows by the value completeness lemma.

In the more interesting case, $e$ is a pair $(e_1, e_2)$ and we know that
$\sigma(e_1) = e_{p_1}$ and $\sigma(e_2) = e_{p_2}$. Inverting the typing
relation gives us
$$\typeCE{\kappa}{e_1}{\Tnf_1} \tand \typeCE{\kappa}{e_2}{\Tnf_2}.$$
Following the \GPair{} rule, let 
$$\Cfresh{\kappa}{[\Tnf_1,\Tnf_2]}{[u_1,u_2]}{\kappa'}$$
and
$$\Cunify {\kappa'}{(u_1, u_2)} {p} {\kappa_0}.$$
As in the value completeness lemma, let $\sigma_0 = \sigma \oplus u_1 \mapsto
v_{p_1} \oplus u_2 \mapsto v_{p_2}$. 
By the inductive hypothesis for the derivation of $e_{p_1}$,
\[
  \left.
    \begin{array}{l}
      \typeCE{\kappa_0}{e_1}{\Tnf_1} \\
      \vdash \kappa_0 \enskip\wedge\enskip \sigma_0 \in \sem{\kappa_0} \\
      \sigma_0(e_1) = e_{p_1} \\
      \sigma_0(u_1) = v_{p_1} \\
      \forall \sigma' \in \sem{\kappa_0}. ~ \exists v'. ~ \sigma'(e_1) \Downarrow v'
    \end{array}
  \right\}
  \hspace{-0.2em}
  \Rightarrow
  \hspace{-0.5em}
  \begin{array}{l}
    \exists \kappa_1 ~ \sigma_1 ~ q_1 ~ t_1. \\
    \left\{
      \begin{array}{l}
        \sigma_1 |_{\sigma_0} \equiv \sigma_0 \\
        \sigma_1 \in \sem{\kappa_1} \\
        \geneval{e_1}{u_1}{\kappa_0}{q_1}{t_1}{\just{\kappa_1}}
      \end{array}
    \right.
  \end{array}
\]

Since $u_1$ and $u_2$ are fresh, we get that $\sigma_0 |_{\sigma} \equiv \sigma$
as well as $\sigma_0((u_1, u_2)) = \sigma_0(p)$. But then by the specification
of unify $\sigma_0 \in \sem{\kappa_0}$. Moreover, by the ordering lemmas
for \ii{fresh} and \ii{unify} (\autoref{lemma:fresh_ordered}
and \autoref{lemma:unify_ordered}) we know that $\kappa_0 \leq \kappa$ which
means that the termination assumption for valuations in $\kappa$ is preserved
and we can now use the above inductive hypothesis. 

The inductive hypothesis for the derivation of $e_{p_2}$ yields
\[
  \left.
    \begin{array}{l}
      \typeCE{\kappa_1}{e_2}{\Tnf_2} \\
      \vdash \kappa_1 \enskip\wedge\; \sigma_1 \in \sem{\kappa_1} \\
      \sigma_1(e_2) = e_{p_2} \\
      \sigma_1(u_2) = v_{p_2} \\
      \forall \sigma' \in \sem{\kappa_1}. ~ \exists v'. ~ \sigma'(e_2) \Downarrow v' \\
    \end{array}
  \right\}
  \hspace{-0.2em}
  \Rightarrow
  \hspace{-0.5em}
  \begin{array}{l}
    \exists \kappa_2 ~ \sigma_2 ~ q_2 ~ t_2. \\
    \left\{
      \begin{array}{l}
        \sigma_2 |_{\sigma_1} \equiv \sigma_1 \\
        \sigma_1 \in \sem{\kappa_2} \\
        \geneval{e_2}{u_2}{\kappa_1}{q_2}{t_2}{\just{\kappa_2}}
      \end{array}
    \right.
  \end{array}
\]

Like in the narrowing proof, we can discharge the typing hypothesis by using a
lemma similar to \autoref{lemma:narrowing_types} (which in turn is once again
a simple induction on the matching derivation) and Unknown Invariance, while the
termination assumption can be discharged using the Termination Preservation
corollary. 

Our final witnesses are $\sigma_2, ~ \kappa_2$ and the standard combinations of
probabilities and traces.

\pcase{\PApp} We know that 
\[
  e_{p_0} \Downarrow v_{p_0},\enskip e_{p_1} \Downarrow v_{p_1}
  \tand e_{p_2} \Downarrow v_p,
\]
where $v_{p_0}$ is of the form $(\RecT{f}{x}{e_{p_2}}{T_1}{T_2})$\linebreak
and $e_{p_2}' = e_{p_2}[v_{p_0}/f, v_{p_1}/x]$.
Inversion on the substitution gives us only one possible $e$,
since unknowns only range over values: $e = (e_0 ~ e_1)$,
where $\sigma(e_0) = e_{p_0}$ and $\sigma(e_1) = e_{p_1}$.
Inversion of the typing premise gives us
$$\typeCE{\kappa}{e_0}{T_1' \rightarrow \Tnf_2'} \tand \typeCE{\kappa}{e_1}{\Tnf_1'}.$$
Using preservation and type uniqueness we can equate $T_1$ with $T_1'$ as well
as $T_2$ with $\Tnf_2'$.

We can then turn to the completeness theorem for the narrowing semantics twice
to obtain witnesses such that:
$$ \narrow{e_0}{\kappa}{t_0}{q_0}{(\RecT{f}{x}{e_2}{T_1}{\Tnf_2})}{\kappa_0}$$
and $$\narrow{e_1}{\kappa_0}{t_1}{q_1}{v_1}{\kappa'}.$$
Completeness also guarantees that there exists $\sigma' \in \sem{\kappa'}$ such
that $\sigma' |_{\sigma} \equiv \sigma$, as well as
$\sigma'(e_1) \Downarrow \sigma'(v_1)$ and, through restriction to
$\dom{\kappa_0}$,
$\sigma'(e_0) \Downarrow (\RecT{f}{x}{\sigma(e_2)}{T_1}{\Tnf_2})$.

Using a Termination Preservation corollary for the narrowing semantics (that can
be proved identically to the one for the matching semantics), in addition to
Substitution Interaction as in the narrowing proof, we can use the inductive
hypothesis for the substituted $e'$ to complete the proof.

The rest of the cases follow using similar arguments, with the same overall
structure as the narrowing proof. The only cases that are interestingly
different (and where the termination assumption actually comes into play) are
the ones that necessitate use of the combining case rule \GCaseOne, which
are \PCaseInl{} and \PCaseInr.

\pcase{\PCaseInl}
Once again, the only interestingly different cases are the ones for the pattern
matching constructs. For \PCaseInl, we know that
$$e_p \Downarrow \inl{\Tnf_1 \Tsum \Tnf_2}{v_{p_1}} \tand e_{p_1}[v_{p_1}/x] \Downarrow v_{p_1}'.$$
Let $\Cfresh{\kappa}{[\Tnf_1, \Tnf_2]}{[u_1,u_2]}{\kappa_0}$ and 
$\sigma_0 = \sigma \oplus u_1 \mapsto v_{p_1}$.

As usual, we can immediately use the inductive hypothesis for the predicate
derivation of $e$ to obtain $\kappa_1$, $\sigma_1$, $q_1$ and $t_1$ such that
\begin{gather*}
  \sigma_1 |_{\sigma_0} \equiv \sigma_0, \quad
  \sigma_1 \in \sem{\kappa_1}\rlap{\enskip\text{and}}\\
  \geneval{e_1}{\inl{\Tnf_1 \Tsum \Tnf_2}{u_1}}{\kappa_0}{q_1}{t_1}{\just{\kappa_1}}.
\end{gather*}
However, we can't conclude that there exists a similar derivation for
$\inr{\Tnf_1 \Tsum \Tnf_2}{u_2}$ from some inductive hypothesis since we don't
have a corresponding derivation! That's where the termination assumptions comes
in: by the Termination Lemma (\autoref{lemma:gen_termination}) there exists some
$\kappa^?$ such that
$\geneval{e_1}{\inr{\Tnf_1 \Tsum \Tnf_2}{u_2}}{\kappa_0}{q_2}{t_2}{\kappa^?}$. 

We now do case analysis on $\kappa^?$. If it is equal to $\nothing$, then the
proof is straightforward following rule \GCaseThree, using the inductive
hypothesis for the other predicate derivation. 

If, on the other hand, $\kappa^? = \just{\kappa_2}$ for some $\kappa_2$, we face
a similar problem for the second derivation. We can obtain $\kappa_a$,
$\sigma_a$, $q_1'$ and $t_1'$ such that
\begin{gather*}
  \sigma_a |_{\sigma_1} \equiv \sigma_1, \quad
  \sigma_a \in \sem{\kappa_a}\rlap{\enskip\text{and}}\\
  \geneval{e_1[u_1/x]}{p}{\kappa_1}{q_1'}{t_1'}{\just{\kappa_a}}
\end{gather*}
by the
inductive hypothesis for the derivation of $e_{p_1}[v_{p_1}/x]$, but we have no
corresponding derivation for the other branch. Using the Termination Lemma once
again, we can obtain the there exists some such $\kappa_b^?$.

Once again we do case analysis on $\kappa_b^?$. If $\kappa_b^? = \nothing$ then
the branch of \ii{combine} that fires returns $\just{\kappa_a}$ and the result
follows directly. If $\kappa_b^? = \just{\kappa_b}$ for some $\kappa_b$, then,
by the specification of \ii{union}, $\sigma_a$ is contained in the denotation of
the combination and the result follows.
\qed
\bigskip
\fi

\section{Implementation}
\label{sec:source}

We next describe the Luck prototype: its top level, its treatment of
backtracking, and the implementation of primitive integers
instantiating the abstract specification presented in \autoref{sec:semantics}.

\paragraph{At the Top Level}
\label{sec-toplevel}

The inputs provided to the Luck interpreter consist of an expression $e$ of type
$\bool$ 
containing zero or more free unknowns $\vec{u}$ (but no free variables), and an
initial constraint set $\kappa$ providing types and finite domains%
\footnote{
This restriction to finite domains appears to be crucial for our technical
development to work, as discussed in the previous section. In practice, we have
not yet encountered a situation where it was important to be able to generate
examples of \emph{unbounded} size (as opposed to examples up to some large
maximum size).  We do sometimes want to generate structures containing large
numbers, since they can be represented efficiently, but here, too, choosing an
enormous finite bound appears to be adequate for the applications we've tried.
The implementation allows for representing all possible ranges of a
corresponding type up to a given size bound. Such bounds are initialized at the
top level, and they are propagated (and reduced a bit) to fresh unknowns created
by pattern matching before these unknowns are used as inputs to the interpreter.
}
for each
unknown in $\vec{u}$, such that their occurrences in $e$ are well typed
($\emptyset; U(\kappa) \vdash e : 1+1$).

The interpreter matches $e$ against \verb!True! (that is, $\inl{1+1}()$), to
derive a refined constraint set $\kappa'$:
\[ \geneval {e} {\inl{1+1}()} {\kappa} {q} {t} {\{\kappa'\}}\]
This involves random choices, and there is also the possibility that matching
fails (and the semantics generates $\emptyset$ instead of $\{\kappa'\}$). In
this case, a simple {\em global} backtracking approach could simply try the whole
thing again (up to an ad hoc limit).
While not strictly necessary for a correct
implementation of the matching semantics, some {\em local} backtracking allows
wrong choices to be reversed quickly and leads to an
enormous improvement in performance \cite{ClaessenDP15}. Our prototype
backtracks locally in calls to \ii{choose}: if {\em choose} has two choices
available and the first one fails when matching the instantiated expression
against a pattern, then we immediately try the second choice
instead. Effectively, this means that if $e$ is already known to be of the form
$\inl\_ \_$, then narrow will not choose to instantiate it using $\inr\_ \_$,
and vice versa.  This may require matching against $e$ twice, and our
implementation shares work between these two matches as far as possible.  (It
also seems useful to give the user explicit control over where backtracking
occurs, but we leave this for future work.)

After the interpreter matches $e$ against \lk{True}, all the resulting
valuations $\sigma \in \sem{\kappa'}$ should map the unknowns in $\vec{u}$ to
some values. However, there is no guarantee that the generator semantics will
yield a $\kappa'$ mapping every $\vec{u}$ to a unique values. The Luck top-level
then applies the $\ii{sample}$ constraint set function to each unknown in
$\vec{u}$, ensuring that $\sigma|_{\vec{u}}$ is the same for each $\sigma$ in
the final constraint set. The interpreter returns this common
$\sigma|_{\vec{u}}$ if it exists, and backtracks otherwise.

\paragraph{Pattern Match Compiler}
\ch{Give a standard reference to desugaring complex pattern matches in
  a functional language: \cite{Augustsson85}}

In Section~\ref{sec:examples}, we saw an example using a standard \lk{Tree}
datatype and instantiation expressions assigning different weights to each
branch. While the desugaring of simple pattern matching to core Luck syntax is
straightforward (\ref{sec:core}), nested patterns---as in
Fig.~\ref{fig:pattern-expand}---complicate things in the presence of
probabilities. We expand such expressions to a tree of simple case expressions
that match only the outermost constructors of their scrutinees.  However, there
is generally no unique choice of weights in the expanded predicate: a branch
from the source predicate may be duplicated in the result. We guarantee the
intuitive property that the {\em sum} of the probabilities of the clones of a
branch is proportional to the weights given by the user, but that still does not
determine the individual probabilities that should be assigned to these clones.

The most obvious way to distribute weights is to simply share the
weight equally with all duplicated branches.  But the probability of a
single branch then depends on the total number of expanded branches
that come from the same source, which can be hard for users to
determine and can vary widely even between sets of patterns that
appear similar.
Instead, Luck's default weighing strategy works as follows.  
\bcp{Don't understand the next sentence very well.}For any branch $B$ from
the source, at any intermediate case 
expression of the expansion, the subprobability distribution over the immediate
subtrees that contain at least one branch derived from $B$ is uniform.  This
makes modifications of the source patterns in nested positions affect the
distribution more locally.

In Figure~\ref{fig:pattern-expand}, the \lk{False} branch should have
probability $\frac13$. It is expanded into four branches, corresponding to
subpatterns \lk{Var \_\,}, \lk{Lam \_ \_\,}, \lk{App (Var \_) \_\,},
\lk{App (App \_ \_) \_\,}. The latter two are grouped under the
pattern \lk{App \_ \_\,}, while the former two are in their own groups.
These three groups receive equal shares of the total probability of the
original branch, that is $\frac19$ each. The two branches for
\lk{App (Var \_) \_} and \lk{App (App \_ \_) \_} split that further
into twice $\frac1{18}$. On the other hand, \lk{True} remains a single
branch with probability $\frac23$. The weights on the left of every pattern are
calculated to reflect this distribution.  \bcp{I'm reading fast, but I found
this example dense and hard to follow.}

\begin{figure}
\small
\begin{fancyluck}
data T = Var Int | Lam Int T | App T T

sig isRedex :: T -> Bool        -- Original
fun isRedex t =
  case t of
  | 2 
  | 1 

sig isRedex :: T -> Bool        -- Expansion
fun isRedex t =
  case t of
  | 1 
  | 1 
  | 7 
      case t1 of
      | 1 
      | 12 
      | 1 
\end{fancyluck}
\caption{Expanding case expression with a nested pattern and a wildcard.
Comments show the probability of each alternative.}
\label{fig:pattern-expand}
\end{figure}

\paragraph{Constraint Set Implementation}

Our desugaring of source-level pattern matching to core case expressions whose
discriminee $e$ is first narrowed means that rule $\GCaseOne$ is not executed for
datatypes; only one of the evaluations of $e$ against the $\ii{L}$ and $\ii{R}$
patterns will succeed and only one branch will be executed. This means that our
constraint-set representation for datatypes doesn't need to implement
$\ii{union}$. We leverage this to provide a simple and efficient implementation
of the unification constraints. For our prototype, the constraint solving behavior
of $\ii{case}$ is only exploited in our treatment of primitive integers,
which we detail at the end of this section.

The constraint set interface could be implemented in a variety of different
ways. The simplest would be to explicitly represent constraint sets as sets of
valuations, but this would lead to efficiency problems, since even unifying
two unknowns would require traversing the whole set, filtering out all valuations
in which the unknowns are different. On the other extreme, we could represent a
constraint set as an arbitrary logical formula over unknowns. While this is a
compact representation, it does not directly support the per-variable
sampling that we require.

For our prototype we choose a middle way, using a simple data structure we call {\em
orthogonal maps} to represent sets of valuations.
An orthogonal map is a map from unknowns to {\em ranges}, which have
the following syntax:
\begin{grammar}
r ::= () | u | (r, r) | \ii{fold} ~ r | \iinl ~ r | \iinr ~ r | \{\iinl ~ r , \iinr ~ r\}
\end{grammar}
Ranges represent sets of non-functional values: units, unknowns, pairs of
ranges, and $\iinl$ and $\iinr$ applied to ranges. We also include the option
for a range to be a pair of an $\iinl$ applied to some range and an $\iinr$
applied to another. For example, the set of all Boolean values can be encoded
compactly in a range (eliding folds and type information) as $\{\iinl () , \iinr
()\}$. Similarly, the set $\{0, 2, 3\}$ can be encoded as $\{\iinl (), \iinr
(\iinr \{ \iinl () , \iinr () \}) \}$, assuming a standard Peano encoding of
\iffull natural numbers\else naturals\fi.

However, while this compact representation can represent all sets of naturals,
not all sets of Luck non-functional values can be precisely represented.
For instance the set $\{(0,1), (1,0)\}$ cannot be represented using ranges, only
approximated to $(\{L(), R (L())\},$ $\{L(), R (L())\})$, which represents the
larger set $\{(0,0), (0,1),$ $(1, $ $0), (1,1)\}$.
This corresponds to a form of Cartesian abstraction, in which we lose any
relation between the components of a pair, so if one used ranges as an abstract
domain for abstract interpretation it would be hard to prove say sortedness of
lists.
Ranges are a rather imprecise abstract domain for algebraic
datatypes~\cite{Jensen97, KakiJ14, NielsonN85}.

We implement constraint sets as pairs of a typing environment and an optional
map from unknowns to ranges. The typing environment of a constraint set
($U(\cdot)$ operation), is just the first projection of the tuple. A constraint
set $\kappa$ is $\ii{SAT}$ if the second element is not $\emptyset$. The
$\ii{sample}$ primitive indexes into the map and collects all possible values
for an unknown.

The only interesting operation with this representation is \ii{unify}. It is
implemented by straightforwardly translating the values to ranges and unifying
those. For simplicity, unification of two ranges $r_1$ and $r_2$ in the presence
of a constraint set $\kappa$ returns both a constraint set $\kappa'$ where
$r_1$ and $r_2$ are unified and the unified range $r'$. If $r_1 = r_2 = ()$
there is nothing to be done. If both ranges have the same top-level
constructor, we recursively unify the inner subranges.
If one of the ranges, say $r_1$, is an unknown $u$ we index into $\kappa$ to
find the range $r_u$ corresponding to $u$, unify $r_u$ with $r_2$ in $\kappa$ to
obtain a range $r'$, and then map $u$ to $r'$ in the resulting constraint set
$\kappa'$. If both ranges are unknowns $u_1, u_2$ we unify their corresponding
ranges to obtain $r'$. We then pick one of the two unknowns, say $u_1$, to map to
$r'$, while mapping $u_2$ to $u_1$. To keep things deterministic we introduce an
ordering on unknowns and always map $u_i$ to $u_j$ if $u_i < u_j$. Finally, if
one range is the compound range $\{\iinl~ r_{1l}, \iinr~ r_{1r}\}$ while the
other is $\iinl~ r_2$, the resulting range is only $\iinl$ applied to the result
of the unification of $r_{1l}$ and $r_2$. 

It is easy to see that if we start with a set of valuations that is representable
as an orthogonal map, non-\ii{union} operations will result in constraint sets whose
denotation is still representable, which allows us to get away with this simple
implementation of datatypes.
The $\GCaseOne$ rule is used to model our treatement of integers. We introduce
primitive integers in our prototype accompanied by standard integer equality and
inequality constraints. 
In Section~\ref{sec:example-application} we saw how a recursive less-than
function can be encoded using Peano-style integers and \ii{case} expressions
that do {\em not} contain instantiation expressions in the discriminee. All
integer constraints can be desugared into such recursive functions with the
exact same behavior---modulo efficiency.

To implement integer constraints, we extend the codomain of the mapping in the
constraint set implementation described above to include a compact
representation of sets of intervals of primitive integers as well as a set of
the unknown's associated constraints. Every time the domain of an unknown $u$ is
refined, we use an incremental variant of the AC-3 {\em arc consistency}
algorithm~\cite{Mackworth77} to efficiently refine the domains of all the
unknowns linked to $u$, first iterating through the constrains associated with
$u$ and then only through the constraints of other ``affected'' unknowns.

\ch{Is the main catch here that integers are a very special datatype
  (that doesn't include pairs!) for which orthogonal maps are a
  precise representation? My impression is that the focus on the
  primitiveness of integers is completely obscuring this conceptual idea.}
\leo{The idea is that our desugaring of integers into the core bypasses the ``add a narrowing at every case'' of the toplevel match}
\ch{This too, but it would not work if integers used pairs and thus
  union, for which your precise primitive integer implementation and
  the approximate orthogonal maps model do not agree.}
\bcp{OK, now I understand Catalin's comments here, and I think I agree there
  there is something a bit fishy going on.  I wonder whether we can beg for
  another day or two to discuss it---there's no time to change anything or
  even add discussion tonight, but it would be nice (for me, at least) to
  have a clearer idea what's going on before we commit this to its archival
  form.  I suggest we go ahead and submit it on time, but also send an email
  asking if we can update it in a couple of days.}
\leo{I don't understand the pairs part of the comment. Having pairs has nothing
to do with the imprecision of the representation. The imprecision comes from the
representation being non-relational. If we switch our internal representation to
sets of orthogonal maps, everything works out --- but it is slow. Here is the
overview of the set up: We have arbitrary data types in the source
language. Elimination forms for arbitrary data types translate into uses of case
where the discriminee is narrowed (using instantiation expressions with
user-supplied weights). We also have integers in the source language. They are
translated into the same core language as ADTs. BUT their elimination forms
(again, case expressions) do NOT use instantiation expressions and leverage the
``execute both branches behavior''. Our source level language restricts the
subset of core Luck that is exposed to the programmer, for both simplicity and
efficiency of implementation.}
\bcp{OK, I really need to understand all this better when I'm fresh.  But
  the high-order bit that I'm getting is that (a) we've successfully defined
the system in such a way that primitive integers and encoded integers behave
exactly the same, but (b) encoded integers don't behave the same as 
other datatypes.  I don't think that's what we originally had in mind!}
\leo{In the core, everything behaves the same. In the source, integers have
special binary operators that can be defined in the core.  Currently, those
definitions can't be written in source Luck. They could, if we exposed a bigger
subset of core Luck to the users. Because we don't, we can get away with a
simpler representation for arbitrary datatypes. It is not clear what an
efficient representation for arbitrary datatypes that can handle the entirety of
Luck would be. There is a *simple* representation - sets of denotations. What we
have now is a strictly better story than what we had before with orthogonal
maps.}\bcp{OK, the last point is a good one.  But still, it seems that the
most interesting rule in the core calculus (the constraint-solving version
of case) is used only for encoding numbers, which the implementation doesn't
actually do.}
\section{Evaluation}
\label{sec:casestudies}


To evaluate the expressiveness and efficiency of Luck's hybrid approach to test
case generation, we tested it with a number of small examples and two
significant case studies: generating well-typed lambda terms and
information-flow-control machine states.  The Luck code is generally much
smaller and cleaner than that of existing handwritten generators, though the
Luck interpreter takes longer to generate each example---around $20\times$ to
$24\times$ for the more complex generators.  (Also, while this is admittedly a
subjective impression, we found it significantly easier to get the generators
right in Luck.)

\paragraph{Small Examples}

The literature on random test generation includes many small examples---list
predicates such as \lk{sorted}, \lk{member}, and \lk{distinct}, tree
predicates like BSTs (\autoref{sec:examples}) and red-black trees, and so
on. 
\iffull
In Appendix~\ref{app:examples}
\else
In the extended version 
\fi
we show the implementation of many such examples in Luck,
illustrating how we can write predicates and generators together
with minimal effort.  

\begin{figure}
\includegraphics[scale=0.625]{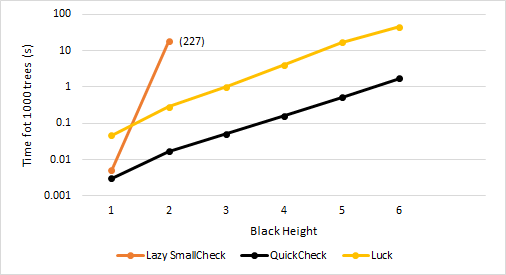}
\caption{Red-Black Tree Experiment}
\label{fig:rbt}
\end{figure}

We use red-black trees to compare the efficiency of our Luck interpreter to
generators provided by commonly used tools like QuickCheck
(random testing), SmallCheck (exhaustive testing) and Lazy
SmallCheck~\cite{RuncimanNL08}. Lazy SmallCheck leverages Haskell's laziness to
greatly improve upon out-of-the-box QuickCheck and SmallCheck generators in the
presence of sparse preconditions, by using partially defined inputs to explore
large parts of the search space at once.  Using both Luck and Lazy SmallCheck,
we attempted to generate 1000 red black trees with a specific black height
$bh$---meaning that the depth of the tree can be as large as $2 \cdot bh + 1$.
Results are shown in \autoref{fig:rbt}. Lazy SmallCheck was able to generate all
227 trees of black height 2 in 17 seconds, fully exploring all trees up to depth
5. When generating trees of black height 3, which required exploring trees up to
depth 7, Lazy SmallCheck was unable to generate 1000 red black trees within 5
minutes. At the same time, the Luck implementation lies consistently within an
order of magnitude of a very efficient handwritten QuickCheck generator that
generates valid Red-Black trees directly. Using rejection-sampling approaches by
generating trees and discarding those that don't satisfy the red-black tree
invariant (e.g., QuickCheck or SmallCheck's \lk{==>}) is prohibitively
costly: these approaches perform much worse than Lazy SmallCheck.


\paragraph{Well-Typed Lambda Terms}

Using our prototype implementation we reproduced the experiments of Pa{\l}ka
\ETAL~\cite{PalkaAST11}, who generated well-typed lambda terms in order to
discover bugs in GHC's strictness analyzer. We also use this case study to
indirectly compare to two narrowing-based tools that are arguably closer to
Luck and that use the same case study to evaluate their work: Claessen
\ETAL~\cite{ClaessenFLOPS14,ClaessenDP15} and Fetscher
\ETAL~\cite{FetscherCPHF15}.

We encoded a model of simply typed lambda calculus with polymorphism in Luck,
providing a large typing environment with standard functions from the Haskell
Prelude to generate interesting well-typed terms. The generated ASTs were then
pretty-printed into Haskell syntax and each one was applied to a partial list of
the form: \lk{[1,2,undefined]}. Using the same version of GHC (6.12.1), we
compiled each application twice: once with optimizations (\lk{-O2}) and once
without and compared the outputs.

A straightforward Luck implementation of a type system for the polymorphic
lambda calculus was not adequate for finding bugs efficiently. To improve its
performance we borrowed tricks from the similar case study of
Fetscher \ETAL~\cite{FetscherCPHF15}, seeding the environment with monomorphic
versions of possible constants and increasing the frequency of \lk{seq}, a basic
Haskell function that introduces strictness, to increase the chances of
exercising the strictness analyzer. Using this, we discovered bugs that seem
similar (under quick manual inspection) to those found by Pa{\l}ka \ETAL and
Fetscher \ETAL.

Luck's generation speed was slower than that of Pa{\l}ka's handwritten
generator. We were able to generate terms of average size 50 (internal
nodes), and, grouping terms together in batches of 100, we got a total time of
generation, unparsing, compilation and execution of around 35 seconds per
batch. This is a slowdown of 20x compared to that of Pa{\l}ka's.  However, our
implementation is a total of 82 lines of fairly simple code, while the
handwritten development is 1684 lines, with the warning ``...the code
is difficult to understand, so reading it is not recommended'' in its
distribution page~\cite{PalkaCodeUrl}.

The derived generators of Claessen \ETAL~\cite{ClaessenFLOPS14} achieved a 7x
slowdown compared to the handwritten generator, while the Redex
generators~\cite{FetscherCPHF15} also report a 7x slowdown in generation time
for their best generator. However, by seeding the environment with monomorphised
versions of the most common constants present in the counterexamples, they were
able to achieve a time per counterexample on par with the handwritten generator.

\paragraph{Information-Flow Control}
\label{sec:IFC}

For a second large case study, we re-implemented a method for generating
information-flow-control machine states~\cite{TestingNI}.  Given an abstract
stack machine with data and instruction memories, a stack, and a program counter,
one attaches {\em labels}---security levels---to runtime values, propagating
them during execution and restricting potential flows of information from {\em
high} (secret) to {\em low} (public) data.  The desired security property, {\em
termination-insensitive noninterference}, states that if we start with two
indistinguishable abstract machines \lk{s1} and \lk{s2} (\IE all their
low-tagged parts are identical) and run each of them to completion, then the
resulting states \lk{s1'} and \lk{s2'} are also indistinguishable.

Hri\c{t}cu \ETAL~\cite{TestingNI} found that efficient
testing of this property could be achieved in two ways: either by generating instruction
memories that allow for long executions and checking for indistinguishability at
each low step (called {\em LLNI}, low-lockstep noninterference), or by
looking for counter-examples to a stronger invariant (strong enough to {\em prove} noninterference),
generating two arbitrary indistinguishable states and then running for a single
step ({\em SSNI}, single step noninterference).  In both cases,
there is some effort involved in generating indistinguishable machines: for
efficiency, one must first generate one abstract machine \lk{s} and then {\em
vary} \lk{s}, to generate an indistinguishable one \lk{s'}. In writing such a
generator for variations, one must effectively reverse the indistinguishability
predicate between states and then keep the two artifacts in sync.

We first investigated the stronger property (SSNI), by encoding the
indistinguishability predicate in Luck and using our prototype to generate
small, indistinguishable pairs of states. In 216 lines of code we
were able to describe both the predicate and the generator for indistinguishable
machines. The same functionality required $>$1000 lines of complex Haskell code
in the handwritten version. The handwritten generator is reported to generate an
average of 18400 tests per second, while the Luck prototype generates \iffull an
average of \fi 1450 tests per second, around 12.5 times slower.

The real promise of Luck, however, became apparent when we turned to
LLNI. Hri\c{t}cu \ETAL~\cite{TestingNI} generate long sequences of instructions
using {\em generation by execution}: starting from a machine state where data
memories and stacks are instantiated, they generate the current instruction
ensuring it does not cause the machine to crash, then allow the machine to take
a step and repeat. While intuitively simple, this extra piece of generator
functionality took significant effort to code, debug, and optimize for
effectiveness, resulting in more than 100 additional lines of code.  The same
effect was achieved in Luck by the following 6 intuitive lines, where we just
put the previous explanation in code:
\begin{verbatim}
sig runsLong :: Int -> AS -> Bool
fun runsLong len st =
    if len <= 0 then True
    else case step st of
        | 99 % Just st' -> runsLong (len - 1) st'
        | 1 % Nothing  -> True
\end{verbatim}
We evaluated our generator on the same set of buggy
information-flow analyses as in
Hri\c{t}cu \ETAL~\cite{TestingNI}. We were able to find all of the same bugs,
with similar effectiveness (number of bugs found per 100 tests).
However, the Luck generator was 24 times slower (Luck: 150 tests/s, Haskell:
3600 tests/s). \bcp{IMO, the rest of this section should be moved to future
  work.  It doesn't really belong in Evaluation---not supported by data.}We expect to be able to improve this result (and the rest
of the results in this section) with a more efficient implementation that
compiles Luck programs to QuickCheck generators directly, instead of
interpreting them in a minimally tuned prototype.

The success of the prototype in giving the user enough flexibility to
achieve similar effectiveness with state-of-the-art generators, while
significantly reducing the amount of code and effort required,
suggests that the approach Luck takes is promising and
points towards the need for a real, optimizing implementation.

\section{Related Work}
\label{sec:relwork}

Luck lies in the intersection of many different topics in programming languages,
and the potentially related literature is huge. Here, we present just the
closest related work. \iffull Afterwards, we demonstrate how a Luck user can leverage Boltzmann
samplers to achieve uniformity guarantees.\bcp{Probably belongs in its own
  section, even if it's a short one, no?} \fi

\paragraph*{Random Testing}

The works that are most closely related to our own are the narrowing based
approaches of Gligoric \ETAL~\cite{GligoricGJKKM10},
Claessen \ETAL~\cite{ClaessenFLOPS14,ClaessenDP15} and
Fetscher \ETAL~\cite{FetscherCPHF15}.  Gligoric \ETAL use a ``delayed choice''
approach, which amounts to needed-narrowing, to generate test cases in
Java. Claessen \ETAL exploit the laziness of Haskell, combining a narrowing-like
technique with FEAT~\cite{Duregard12}, a tool for functional enumeration of
algebraic types, to efficiently generate near-uniform random inputs satisfying
some precondition. While their use of FEAT allows them to get uniformity by
default, it is not clear how user control over the resulting distribution could
be achieved.
Fetscher \ETAL~\cite{FetscherCPHF15} also use an algorithm that makes local
choices with the potential to backtrack in case of failure. Moreover, they add a
simple version of constraint solving, handling equality and disequality
constraints. This allows them to achieve excellent performance in testing GHC
for bugs (as in \cite{PalkaAST11}) using the ``trick'' of monomorphizing the
polymorphic constants of the context as discussed in the previous section. They
present two different strategies for making local choices: uniformly at random,
or by ordering branches based on their branching factor. While both of these
strategies seem reasonable (and somewhat complementary), there is no way of
exerting control over the distribution as necessary.

\paragraph*{Enumeration-Based Testing}

An interesting related approach appears in the inspiring work of
Bulwahn~\cite{Bulwahn12smartgen}. In the context of Isabelle's~\cite{Isabelle}
QuickCheck~\cite{Bulwahn12}, Bulwahn automatically constructs enumerators for a
given precondition via a compilation to logic programs using mode inference.
This work successfully addresses the issue of generating satisfying valuations
for preconditions directly and serves for exhaustive testing of ``small''
instances, significantly pushing the limit of what is considered ``small''
compared to previous approaches. Lindblad~\cite{Lindblad07} and
Runciman \ETAL~\cite{RuncimanNL08} also provide support for exhaustive testing
using narrowing-based techniques.
Instead of implementing mechanisms that resemble narrowing in standard
functional languages, Fischer and Kuchen~\cite{FischerK07} leverage the built-in
engine of the functional logic programming language Curry~\cite{Curry} to
enumerate tests satisfying a coverage criterion.  In a later, black-box approach
for Curry, Christiansen and Fischer~\cite{ChristiansenF08} additionally use {\em
level diagonalization} and randomization to bring larger tests earlier in the
enumeration order. While exhaustive testing is useful and has its own merits and
advantages over random testing in a lot of domains, we turn to random testing
because the complexity of our applications---testing noninterference or
optimizing compilers---makes enumeration impractical.

\paragraph*{Constraint Solving}

Many researchers have turned to
constraint-solving based approaches to generate random inputs satisfying
preconditions. In the constraint solving literature around SAT witness
generation, the pioneering work of Chakraborty \ETAL~\cite{Chakraborty2014}
stands out because of its efficiency and its guarantees of approximate
uniformity. However, there is no way---and no obvious way to add it---of
controlling distributions. In addition, their efficiency relies crucially on the
{\em independent support} being small relative to the entire space
(where the {\em support} $X$ of a boolean formula \lk{p} is the set of variables
appearing in \lk{p} and the \emph{independent support} is a subset $D$ of $X$
such that no two satisfying assignments for
\lk{p} differ only in $X \backslash D$).
While true for typical SAT instances, this is not the case for random testing
properties, like, for example, noninterference. In fact, a minimal independent
support for indistinguishable machines includes one entire machine state and the
high parts of another; thus, the benefit from their heuristics may be minimal.
Finally, they require logical formulae as inputs, which would require a rather
heavy translation from a high-level language like Haskell.

Such a translation from a higher-level language to the logic of a constraint
solver has been attempted a few times to support testing \cite{CarlierDG10,
GotliebICST09}, the most recent and efficient for Haskell being
Target~\cite{SeidelVJ15}. Target translates preconditions in the form of
refinement types, and uses a constraint solver to generate a satisfying
valuation for testing. Then it introduces the negation of the generated input to
the formula, in order to generate new, different ones. While more efficient than
Lazy SmallCheck in a variety of cases, there are still cases where a
narrowing-like approach outperforms the tool, further pointing towards the need
to combine the two approaches as in \Luck. Moreover, the use of an automatic
translation and constraint solving does not give any guarantees on the resulting
distribution, neither does it allow for user control.

Constraint-solving is also used in symbolic evaluation based techniques, where
the goal is to generate diverse inputs that achieve higher
coverage~\cite{DART:2005, CUTE:2005, KLEE:2008:OSDI, AvgerinosRCB14,
HybridConcolic:2007:Majumdar, CREST:2008, EXE:2006, GodefroidLM12,
ConcolicAssessment:ICSE11}.  Recently, in the context of
Rosette~\cite{TorlakB14}, symbolic execution was used to successfully find
bugs in the same information-flow control case study.

\paragraph*{Semantics for narrowing-based solvers}
Recently, Fowler and Hutton~\cite{FowlerH15} put needed-narrowing
based solvers on a firmer mathematical foundation. They presented an
operational semantics of a purely narrowing-based solver, named Reach,
proving soundness and completeness. In their concluding remarks, they
mention that native representations of primitive datatypes do not fit
with the notion of lazy narrowing since they are ``large, flat
datatypes with strict semantics.'' In Luck, we were able to exhibit
the same behavior for both the primitive integers and their datatype
encodings successfully addressing this issue, while at the same time
incorporating constraint solving into our formalization.

\paragraph*{Probabilistic programming}
\label{par:prob}

Semantics for probabilistic programs share many similarities with the semantics
of Luck~\cite{MilchMRSOK05,GoodmanMRBT08, GordonHNR14}, while the problem of
generating satisfying valuations shares similarities with probabilistic
sampling~\cite{MansinghkaRJT09,Latuszynski13,ChagantyAISTATS13, NoriAAAI14}. For
example, the semantics of \iffull the language\fi PROB in the recent probabilistic
programming survey of Gordon \ETAL~\cite{GordonHNR14} takes the form of
probability distributions over valuations, while Luck semantics can be viewed as
(sub)probability distributions over constraint sets, which induces a
distribution over valuations. Moreover, in probabilistic programs, observations
serve a similar role to preconditions in random testing, creating problems for
simplistic probabilistic samplers that use {\em rejection sampling}---\IE
generate and test. Recent advances in this domain, like the work on Microsoft's
R2 Markov Chain Monte Carlo sampler~\cite{NoriAAAI14}, have shown promise in
providing more efficient sampling, using pre-imaging transformations in
analyzing programs. An important difference is in the type of programs usually
targeted by such tools. The difficulty in probabilistic programming arises
mostly from dealing with a large number of complex observations, modeled by
relatively small programs. For example, Microsoft's TrueSkill~\cite{TrueSkill}
ranking program is a very small program, powered by millions of observations. In
contrast, random testing deals with very complex programs (\EG a type checker)
and a single observation without noise (\lk{observe true}).

We did a simple experiment with R2, using the following probabilistic
program to model the indistinguishability of \autoref{sec:examples},
where we use booleans to model labels:
\begin{verbatim}
    double v1 = Uniform.Sample(0, 10);
    double v2 = Uniform.Sample(0, 10);
    bool l1 = Bernoulli.Sample(0.5);
    bool l2 = Bernoulli.Sample(0.5);
    Observer.Observe(l1==l2 && (v1==v2 || l1));
\end{verbatim}
%
Two pairs of doubles and booleans will be indistinguishable if the
booleans are equal and, if the booleans are false, so are the
doubles. The result was somewhat surprising at first, since all the
generated samples have their booleans set to true. However, that is an
accurate estimation of the posterior distribution: for every ``false''
indistinguishable pair there exist $2^{64}$ ``true'' ones! Of course,
one could probably come up with a better prior or use a tool that
allows arbitrary conditioning to skew the distribution
appropriately. If, however, for such a trivial example the choices are
non-obvious, imagine replacing pairs of doubles and booleans with
arbitrary lambda terms and indistinguishability by a well-typedness
relation. Coming up with suitable priors that lead to efficient
testing would become an ambitious research problem on its own!

\iffull

\paragraph{Boltzmann Samplers}

Boltzmann samplers for algebraic datatypes work by making appropriately
weighted local choices based on a control parameter. In the binary tree case,
a Boltzmann sampler would look similar to a simplistic generator for trees:
flip a (potentially biased) coin, generating a \lk{Leaf} or a \lk{Node}
based on the outcome; then recurse (\autoref{fig:naivetreegen}). The
distribution this generator induces can be easily implemented by a
trivial Luck predicate, also in \autoref{fig:naivetreegen}.
A Boltzmann sampler is a slightly modified version of this simple
approach. First of all, the bias in the local choices must be systematically
calculated (or approximated with numerical methods). This bias depends on the
convergence of a generating function associated with the data type's recursive
specification. Then, we start sampling using the computed bias for each
choice. If at any point we reach a size bigger than $n(1+\epsilon)$, we stop
the generation and try again. If the generation stops with a term
$n(1-\epsilon)$, we throw away the generated term and try again. The theory
behind Boltzmann samplers guarantees that this approach will terminate in
expected linear time in $n$ (including discards!), and the result will be
uniformly selected among other elements of the same size.

\begin{figure}
\small
\begin{verbatim}
randomTree :: Gen Tree
randomTree =  frequency [(1, return Leaf)
                        ,(1, Branch <$> randomTree
                                    <*> randomTree)]

sig randomTree :: Tree -> Bool
fun randomTree tree =
    case tree of
    | 1 % Leaf -> True
    | 1 % Branch l r -> randomTree l && randomTree r
\end{verbatim}
\caption{Simple tree generator (Haskell vs Luck)}
\label{fig:naivetreegen}
\end{figure}

Using Boltzmann samplers is a natural fit for our setting and requires two
things: a way to compute the control parameter for the structure that is being
generated, and a way to reject samples that are outside the neighborhood of the
desired size. Both of these can be handled automatically, as shown by Canou
and Darrasse~\cite{CanouD09}.

The presence of additional constraints on generated data could skew the
posterior distributions, negating any uniformity guarantees. But if, every
time we reach an unsatisfiable constraint, we backtrack to the beginning of the
generation, uniformity is preserved since we are only concerned with
uniformity in the set of satisfying valuations. Unfortunately, throwing away
all progress is not efficient; it is usually a lot faster to backtrack to a
more recent choice instead of the beginning. Such an algorithm still gives us
some assurance about the distribution, similarly to
Claessen~\ETAL\cite{ClaessenFLOPS14}: the least likely value generated will be
at most a constant factor less likely than the most likely one, where the
factor is the amount of local backtracking allowed.
\fi

\section{Conclusions and Future Work}
\label{sec:concl}

We have presented \Luck, a language for writing generators in the form
of lightly annotated predicates. We presented the semantics of \Luck, combining
local instantiation and constraint solving in a unified framework and exploring
their interactions. We described a prototype implementation of
this semantics, which we used to replicate the results of state-of-the-art
handwritten random generators for two complex domains. The results showed the
potential of Luck's approach, allowing us to replicate the
generation presented by the handwritten generators with reduced code
and effort. The prototype was slower by an order of magnitude, but there is
still significant room for improvement.

In the future it will be interesting to explore compilation of Luck
into generators in a language like Haskell
to improve the performance of our interpreted prototype. Another way to improve
performance would be to experiment with other domain representations.%
\ch{Here is where SMT solvers could also come into the picture}
%
We also want to investigate Luck's equational theory, showing, for
instance, that the encoded conjunction, negation, and disjunction
satisfy the usual logical laws.
Finally, the backtracking strategies in our implementation can be
abstractly modeled on top of our notion of choice-recording trace;
Gallois-Wong~\cite{diane-report} shows promising preliminary results
using Markov chains for this.

Another potential direction for future work is automatically deriving smart
shrinkers. Shrinking, or delta-debugging, is crucial in
property-based testing, and it can also require significant user effort and
domain 
specific knowledge to be efficient~\cite{RegehrCCEEY12}. It would be
interesting to see if there is a counterpart to narrowing or constraint
solving that allows shrinking to preserve desired properties.

\iffull
Finally, we would like to see if transferring ideas from \Luck to
generate inductive datatypes for testing in the Coq proof
assistant~\cite{QuickChickCoq14, itp2015} can ease the formalization
effort, allowing users to discover flaws early in the proof process as
in other proof assistants~\cite{Bulwahn12, ChamarthiDKM11}.
\fi

\ifanon\else
\sloppy

\section*{Acknowledgments}
We are grateful to
Maxime D\'en\`es,
Nate Foster,
Thomas Jensen,
Gowtham Kaki,
George Karachalias,
Micha{\l} Pa{\l}ka,
Zoe Paraskevopoulou,
Christine Rizkallah,
Antal Spector-Zabusky,
Perdita Stevens,
and Andrew Tolmach
for useful comments.
\ch{The people at the PPS workshop too?}
This work was supported by
NSF awards \#1421243, {\em Random Testing
  for Language Design}
and \#1521523, \emph{Expeditions in Computing: The Science of Deep
  Specification}.
  This work was, in part, supported by the
  \href{https://erc.europa.eu/}{European Research Council}
  under \href{https://secure-compilation.github.io/}{ERC
    Starting Grant SECOMP (715753).}

\fussy
\fi


\iffull
\pagebreak

\appendix

\section{Appendix}
\label{sec:appendix}

\bcp{These last little bits should just be brought into the main body of the
(full version of the) paper.  No need for appendices.}\leo{They really
add very little and would disrupt the flow of the evaluation section}

\subsection{Luck examples}
\label{app:examples}

In this appendix we present the Luck programs that serve as
both predicates and generators for the small examples of
\autoref{sec:casestudies}.
\begin{verbatim}
sig sorted :: [Int] -> Bool
fun sorted l =
    case l of
    | (x:y:t) -> x < y && sorted (y:t)
    | _ -> True
    end

sig member :: Int -> [Int] -> Bool
fun member x l =
    case l of
    | h:t -> x == h || member x t
    | _ -> False
    end

sig distinctAux :: [Int] -> [Int] -> Bool
fun distinctAux l acc =
    case l of
    | []  -> True
    | h:t -> not (member h acc) !h
             && distinctAux t (h:acc)
    end

sig distinct :: [Int] -> Bool
fun distinct l = aux l []
\end{verbatim}

In order to obtain lists of a specific size, we could skew the distribution
towards the \lk{cons} case using numeric annotations on the branches, or, we
can use the conjunction of such a predicate with the following simple length
predicate (which could be greatly simplified with some syntactic sugar).

\begin{verbatim}
sig length :: [a] -> Int -> Bool
fun length l n =
    if n == 0 then
      case l of
      | [] -> True
      | _  -> False
      end
    else case l of
         | h:t -> length t (n-1)
         | _ -> False
         end
\end{verbatim}

Finally, the Luck program that generates red black trees of a specific height
is:

\begin{verbatim}
data Color = Red | Black
data RBT a = Leaf | Node Color a (RBT a) (RBT a)

fun isRBT h low high c t =
  if h == 0 then
      case (c, t) of
        | (_, Leaf) -> True
        | (Black, Node Red x Leaf Leaf) ->
          (low < x && x < high) !x
        | _ -> False
      end
  else case (c, t) of
         | (Red, Node Black x l r) ->
              (low < x && x < high) !x
              && isRBT (h-1) low x  Black l
              && isRBT (h-1) x high Black r
         | (Black, Node Red x l r) ->
              (x | low < x && x < high) !x
              && isRBT h low x  Red l
              && isRBT h x high Red r
         | (Black, Node Black x l r) ->
              (x | low < x && x < high) !x
              && isRBT (h-1) low x  Black l
              && isRBT (h-1) x high Black r
         | _ -> False
       end
\end{verbatim}
\fi

\ifcamera\balance\fi
\bibliographystyle{plainurl}
\bibliography{quick-chick}

\end{document}